\documentclass[]{emulateapj}

\def\lya{Ly$\alpha$ }

\def\hMsun{h^{-1}M_\odot}
\def\hMpc{h^{-1}{\rm Mpc}}
\def\hkpc{h^{-1}{\rm kpc}}
\def\VhMpc{h^{-3}{\rm Mpc}^3}
\def\denhMpc{h^3{\rm Mpc}^{-3}}
\def\kms{{\rm km\, s^{-1}}}
\def\Lint{L_{\rm intrinsic}}
\def\Lapp{L_{\rm apparent}}

\slugcomment{Draft, 20101010}

\begin{document}

\title{Radiative Transfer Modeling of Lyman Alpha Emitters. II. New Effects in Galaxy Clustering}
\author{
Zheng Zheng\altaffilmark{1},
Renyue Cen\altaffilmark{2},
Hy Trac\altaffilmark{3,4},
and
Jordi Miralda-Escud\'e\altaffilmark{5,6}
}
\altaffiltext{1}{Yale Center for Astronomy and Astrophysics, Yale University,
                 New Haven, CT 06520; zheng.zheng@yale.edu
}
\altaffiltext{2}{Department of Astrophysical Sciences, Princeton University,
                 Peyton Hall, Ivy Lane, Princeton, NJ 08544
}
\altaffiltext{3}{Harvard-Smithsonian Center for Astrophysics,
                 Cambridge, MA 02138
}
\altaffiltext{4}{Department of Physics, Carnegie Mellon University, Pittsburgh,
                 PA 15213
}
\altaffiltext{5}{Instituci\'o Catalana de Recerca i Estudis Avan\c cats,
                 Barcelona, Catalonia
}
\altaffiltext{6}{Institut de Ci\`encies del Cosmos, Universitat de Barcelona,
                 Barcelona, Catalonia
}

\begin{abstract}

We study the clustering properties of $z\sim 5.7$ \lya emitters (LAEs) 
in a cosmological reionization simulation with a full \lya radiative transfer 
calculation. \lya radiative transfer substantially modifies the intrinsic
\lya emission properties, compared to observed ones, depending on the
density and velocity structure environment around the \lya-emitting galaxy. 
This environment-dependent \lya selection introduces new features in LAE
clustering, suppressing (enhancing) the line-of-sight (transverse) density 
fluctuations and giving rise to scale-dependent galaxy bias. In real space, 
the contours of the three-dimensional two-point correlation function 
of LAEs appear to be prominently elongated along the line of sight on large
scales, an effect that is opposite to and much stronger than the linear
redshift-space distortion effect. The projected two-point correlation
function is greatly enhanced in amplitude by a factor of up to a few, 
compared to the case without the environment-dependent selection effect.
The new features in LAE clustering can be understood with a simple,
physically motivated model, where \lya selection
depends on matter density, velocity, and their gradients. 
We discuss the implications and consequences of the effects on galaxy
clustering from \lya selection in interpreting clustering measurements
and in constraining cosmology and reionization from LAEs.

\end{abstract}


\section{Introduction}

\lya emitters (hereafter LAEs) are galaxies with strong \lya emission
lines. Owing to the strong \lya line feature, LAEs can be efficiently detected 
through narrowband imaging or with integral-field-units (IFU) spectroscopy,
which makes them natural targets for searches of high-redshift galaxies.
Large samples of high-redshift LAEs are expected with ongoing and forthcoming
LAE surveys. In this paper, we investigate the clustering of $z\sim 5.7$ LAEs 
in a cosmological reionization simulation, focusing on the effects of \lya 
radiative transfer on their clustering properties.

Large samples of LAEs would provide exciting opportunities to probe the high 
redshift universe. The resonance nature of \lya line makes LAEs a sensitive 
probe of the high-redshift intergalactic medium (IGM), especially across the 
reionization epoch (e.g., \citealt{Miralda98a,Miralda98,Haiman99,
Santos04,Haiman05,Wyithe07,Malhotra04,Kashikawa06,Furlanetto06,Dijkstra07a,
Dijkstra07b,McQuinn07,Mesinger08,Iliev08,Dayal08,Dayal09,Dayal10}). 
In particular, the 
clustering of LAEs at redshift $z>6$ can potentially put tight constraints on 
the ionization status of the IGM 
\citep[e.g.,][]{Furlanetto06,McQuinn07,Iliev08}.

Large samples of LAEs should enable accurate measurements of their clustering.
As with galaxy clustering in general, the clustering of LAEs encodes useful
information of galaxy formation and evolution. We expect to learn about the
relation between these young galaxies and dark matter halos, and to obtain 
insights on the early stage of structure formation. 

Galaxy clustering also encodes important cosmological information.
The fluctuation power spectrum of galaxies is related to that of 
matter, usually differing by a constant, multiplicative galaxy bias factor on 
large scales. Because nonlinearity is weaker at higher redshift, the galaxy 
power spectrum may be used down to smaller spatial scales for constraining 
cosmology, tightening constraints on parameters like the neutrino mass and 
providing tests of inflation \citep[e.g.,][]{Takada06}. The efficient 
detection of LAEs at high redshift makes them attractive candidates in this 
endeavor. Large-volume surveys of LAEs, such as the Hobby--Eberly Telescope 
Dark Energy Experiment (\citealt{Hill08}), might also enable the 
detection of the baryon acoustic oscillations feature
\citep[e.g.,][]{Eisenstein05} in the LAE power spectrum.
Baryon acoustic oscillations can be used
to measure the expansion history of the universe,
contributing to constraints on the evolution of dark energy and the 
curvature of the universe. 

At present, measurements of LAEs clustering are very limited by
small survey volumes and small samples of LAEs. 
\citet{Ouchi03} present the angular two-point correlation function
(hereafter, 2PCF) 
of 87 $z=4.86$ LAEs in a 0.15deg$^2$ narrowband survey in the Subaru Deep 
Field. The correlation length is estimated to be $(3.5\pm 0.3) \hMpc$ (and 
$(6.2\pm 0.5)\hMpc$ if a maximum contamination correction is applied), larger
than that of $z\sim 4$ Lyman break galaxies. 
With a set of LAEs found in a larger area ($\sim$0.3 deg$^2$) around
the same field, \citet{Shimasaku04} also report strong clustering
for 41 $z=4.86$ LAEs, but they find almost no clustering for 51
$z=4.79$ LAEs. The strong clustering of $z=4.86$ LAEs is difficult to
reproduce with a simple model that relates LAEs to dark matter halos
\citep{Hamana04}.
With 151 $z=4.5$ LAEs in a 0.36deg$^2$ field of the narrowband Large Area 
Lyman Alpha (LALA) survey, \citet{Kovac07} estimate a correlation length of
$3.2\pm 0.4\hMpc$ ($4.6\pm 0.6\hMpc$ with contamination correction). The 
clustering can be reproduced if these LAEs reside in halos more massive
than (1--2)$\times 10^{11}\hMsun$. 
\citet{Gawiser07} measure the angular 2PCF of 162 $z=3.1$ 
LAEs discovered in a 0.28deg$^2$ field of the MUltiwavelength Survey by 
Yale-Chile (MUSYC). They find a moderate clustering with a correlation 
length of $2.5_{-0.7}^{+0.6}\hMpc$, corresponding to that of halos with a 
minimum mass of $\sim$2.8$\times 10^{10}\hMsun$ (and a median mass of 
$5.6\times 10^{10}\hMsun$). The 261 $z=2.1$ LAEs in the same MUSYC field have 
a correlation length of $3.2\pm 0.6\hMpc$ \citep{Guaita10}, corresponding to 
a median halo mass of $1.8\times 10^{11}\hMsun$. 
From the clustering of $z=3$--7 LAEs in the 1deg$^2$ Subaru/{\it XMM-Newton} 
Deep Survey (SXDS), \citet{Ouchi10} infer that the average host halo mass 
is $10^{10}$--$10^{11}M_\odot$.
Understanding the LAE clustering results needs to take into account the 
differences in LAE samples and redshifts, as well as sample variance caused
by small survey volumes. The development of a physically based 
theoretical model of LAEs and their clustering is also needed.

In models of LAEs, the \lya flux of an LAE is usually computed from the
emitted ionizing photons in the galaxy residing in a dark matter halo,
assuming Case B 
recombination \citep{Osterbrock89}. The theoretical modeling of LAE clustering
depends on how LAEs and dark matter halos are connected,
and how the observed \lya luminosity is determined from the intrinsic
\lya luminosity. Generally speaking, there are two scenarios considered
in LAE models, the duty cycle scenario and the \lya escape fraction scenario.
In the duty cycle scenario, LAEs are short-lived and at any given time only a 
fraction of all galaxies are active as LAEs. In the \lya escape fraction 
scenario, it is assumed that only a fraction of \lya photons can 
escape from the source, and therefore the observed \lya luminosity is a 
fraction of the intrinsic one. Either scenario can make the predicted \lya 
luminosity function (LF) match the observation. 
For a given number density of LAEs, the masses of host halos in the 
duty cycle scenario would be on average lower than those in the escape 
fraction scenario. As a consequence, the clustering of LAEs would be different
in the two scenarios, with a stronger clustering in the escape fraction 
scenario. 

\citet{Nagamine10} predict LAE clustering based on cosmological
smoothed particle hydrodynamic (SPH) simulation and consider both scenarios. 
They find that LAE clustering 
measurements from observations are in favor of their duty cycle scenario. 
\citet{Tilvi09} present an LAE model in which \lya luminosity or star 
formation rate (SFR) is related to the halo mass accretion rate, rather than
halo mass, and the model naturally gives rise to the duty cycle of LAEs.
Their model predict correlation lengths of LAEs in agreement with the 
observations. \citet{Orsi08} combine a semi-analytic model of galaxy formation 
with a large $N$-body simulation to predict the clustering of LAEs. They 
adopt the scenario of \lya escape fraction and assume the escape fraction to 
be constant (2\%) and independent of galaxy properties. By accounting for the 
large sample variance, the model is found to reproduce the angular clustering 
measurements from current surveys of LAEs. \citet{McQuinn07} develop a model
of LAEs using reionization simulations with cosmological volume (see also
\citealt{Iliev08}) and discuss the effect of reionization on LAE clustering. 
Their model computes the escape fraction based on a simplified treatment of 
\lya radiative transfer (hereafter, RT) that consists of multiplying
the intrinsic line profile by $\exp(-\tau_\nu)$, where $\tau_\nu$ is the 
optical depth at frequency $\nu$ along the line of sight.
The $\exp(-\tau_\nu)$ model is reasonably accurate in the case of studying
the absorption feature of \lya photons passing through some neutral regions
along the line of sight, especially for absorption caused by
the damping wing or small optical depth.
For studying the observed \lya emission, the model is not accurate.
Even if one limits
the $\exp(-\tau_\nu)$ model to study the transfer outside of a radius much 
larger than halo size by assuming a \lya line profile at that radius, it is 
not clear what radius to use, what line profile to assume, and what angular 
distribution of \lya emission at that radius to adopt. The model neglects the 
spatial and frequency diffusion of \lya photons, which is important because 
of the scattering, not the absorption nature of the transfer of \lya photons.

\citet{Zheng10} (hereafter Paper I) present a simple physical model of LAEs, 
where \lya RT is the primary physical process transforming 
intrinsic \lya emission properties to observed ones.  For the first time, 
a full RT calculation of \lya photons \citep{Zheng02}
in gas halos around LAEs
is performed in a self-consistent fashion with the radiation-hydrodynamic 
reionization simulations \citep{Trac08}.
While in this model the number of \lya photons in the IGM is 
correctly conserved, only a fraction of them can be observed, those included in
the central part of the extended \lya emission with
high enough surface brightness.
The model predicts a broad distribution of apparent (observed) \lya luminosity 
at fixed intrinsic \lya luminosity or ultraviolet (UV) luminosity, 
a consequence of a variable intergalactic environment of LAEs and 
the environment-dependent RT of \lya photons. 
Therefore, the model predicts an effective \lya escape fraction that is
not constant, but has a broad distribution and is correlated with the
environment.
This simple physical model is able to 
explain an array of observed properties of $z\sim$5.7 LAEs in \citet{Ouchi08}, 
including \lya spectra, morphology, and apparent \lya LF. The broad 
distribution of apparent \lya luminosity at fixed UV luminosity provides a 
natural explanation for the observed UV LF, especially the turnover toward
the low-luminosity end. The model also reproduces the observed distribution 
of \lya equivalent width (EW) and explains the deficit of UV bright, high-EW 
sources. 

In this paper, we investigate the clustering of LAEs within the model presented
in Paper I. As we will show, the environment-dependent \lya RT 
introduces new and significant effects in the clustering of galaxies selected 
by \lya emission, a real physical effect that has not been properly taken into 
account in previous studies. We first present the environment dependence of 
the \lya selection and the dependence on halo mass in 
Section~\ref{sec:environ}. In Section~\ref{sec:clustering}, 
we present the results of LAE clustering from our model, in terms of the
2PCFs.
Following an intuitive interpretation of the features seen in LAE clustering, 
we provide a simple physical model to further aid our understanding of LAE 
clustering. In Section~\ref{sec:hod}, we show the environment dependence of the 
halo occupation distribution (HOD) of LAEs. In Section~5, we summarize our main 
findings and discuss the implications. In the appendices, we provide an 
extended simple physical model of LAE clustering, present the power spectrum 
of LAEs, make comparisons to the LAE clustering in the $\exp(-\tau_\nu)$ 
model, and present tests on factors that may mask the new clustering effects.

Throughout the paper, we adopt the same cosmological model as in the 
reionization simulation \citep{Trac08} used in our RT 
calculation. It is a spatially flat $\Lambda$CDM cosmological 
model with Gaussian initial density fluctuations, and the cosmological 
parameters are consistent with the {\it Wilkinson Microwave Anisotropy Probe}
5 year data \citep{Dunkley09}: $\Omega_m=0.28$, $\Omega_\Lambda=0.72$,
$\Omega_b=0.046$, $h=0.70$, $n_s=0.96$, and $\sigma_8=0.82$.
Our \lya RT calculation is based on the $z=5.7$ output of
the simulation, which has a box size of $100\hMpc$ on a side. In our 
calculation, a 768$^3$ grid is used to represent the neutral hydrogen density, 
temperature, and peculiar velocity fields in the simulation box.
The Hubble flow is added to the velocity field. LAEs are assumed to 
reside in dark matter halos with positions and velocities from the halo 
catalog. To reduce source blending in the \lya image and spectra, \lya photons
are collected with a finer spatial resolution, a $6144^2$ grid for the image 
of the whole box, corresponding to 16.3$\hkpc$ (comoving) or 0.58\arcsec per 
pixel. The spectral resolution and range are 0.1\AA (25$\kms$) and 24\AA in 
rest frame, respectively. We divide the whole simulation box into three layers 
so that the depth of each layer approximates that from the width of the 
narrowband filter used in searching for $z\sim 5.7$ LAEs \citep{Ouchi08}. The 
calculation result for each layer is saved in an IFU-like datacube of dimension 
6144$\times$6144$\times$240. We refer the readers to Paper I for more details 
about the characteristics of the simulation and calculation.

\begin{figure*}
\plotone{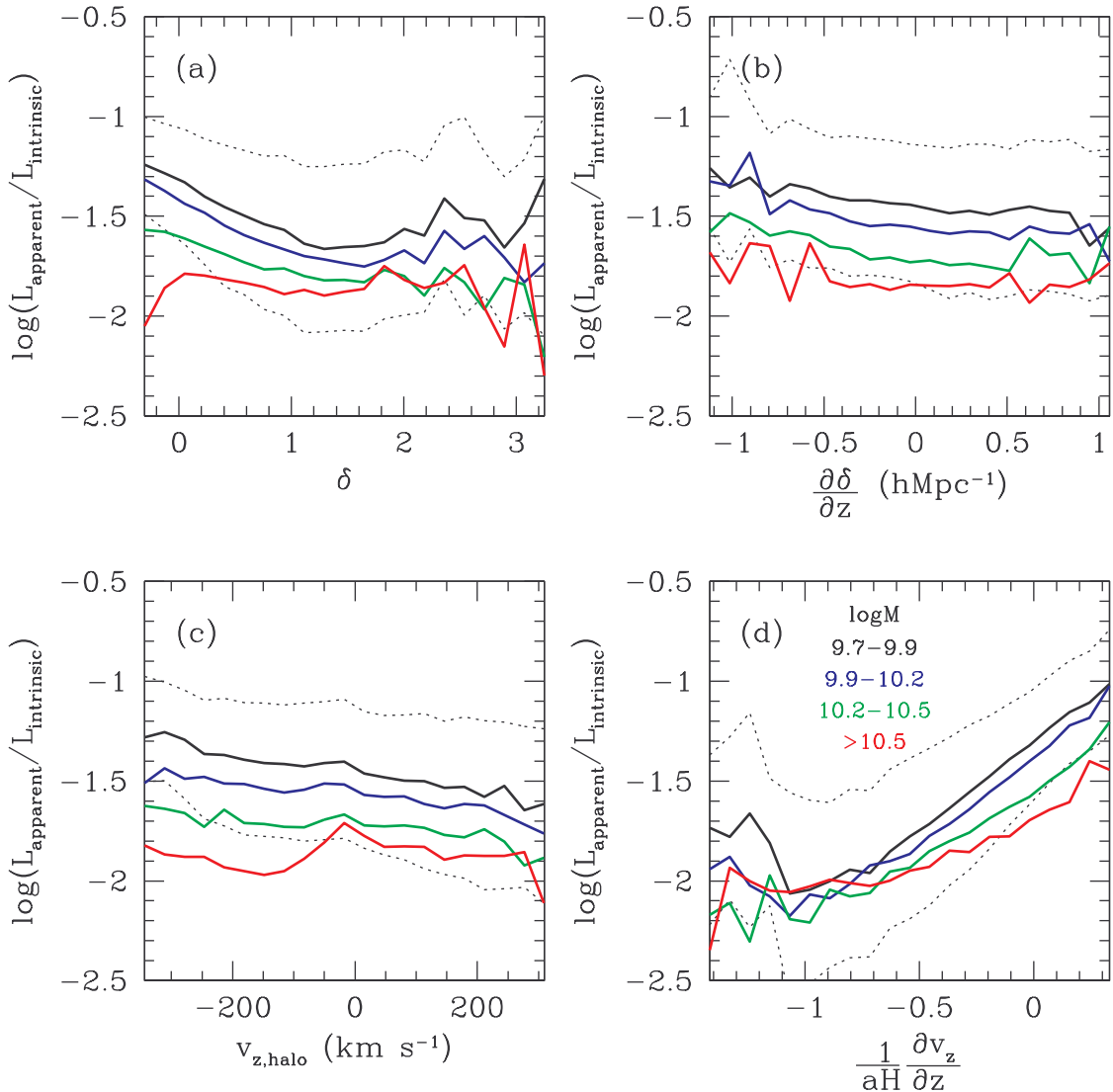}
\caption[]{
\label{fig:denvelgrad_mass}
Dependence of \lya flux suppression of LAEs on density and peculiar velocity,
as a function of halo mass.
The suppression is characterized by the ratio of the apparent (observed) and
intrinsic \lya luminosity $L_{\rm apparent}/L_{\rm intrinsic}$.
(a) Dependence on the smoothed overdensity field at the source
position. The overdensity field is smoothed with a 3D top-hat filter of radius
2$\hMpc$ (comoving).
(b) Dependence on the density gradient along the $Z$-direction.
The derivative is with respect to comoving coordinate.
(c) Dependence on the host halo velocity.
(d) Dependence on the linear peculiar velocity gradient along the
$Z$-direction. The linear peculiar velocity is obtained from the smoothed
overdensity field based on the continuity equation (see the text for details).
The velocity gradient is put in units of the Hubble parameter.
Different colors are for LAE host halos of different masses, as labeled in
panel~($d$). The median of the ratio is plotted as
a solid curve. The two dotted curves delineate the upper and the lower 
quartiles, and for clarity we only plot those for the lowest mass range. 
Note that the line-of-sight direction (from the observer to sources) is along 
the $-Z$-direction, which matters for interpreting the results in panels~($b$) 
and ($c$). See Section~\ref{sec:environ} for further discussion.
}
\end{figure*}

\begin{figure*} 
\epsscale{1.1}
\plottwo{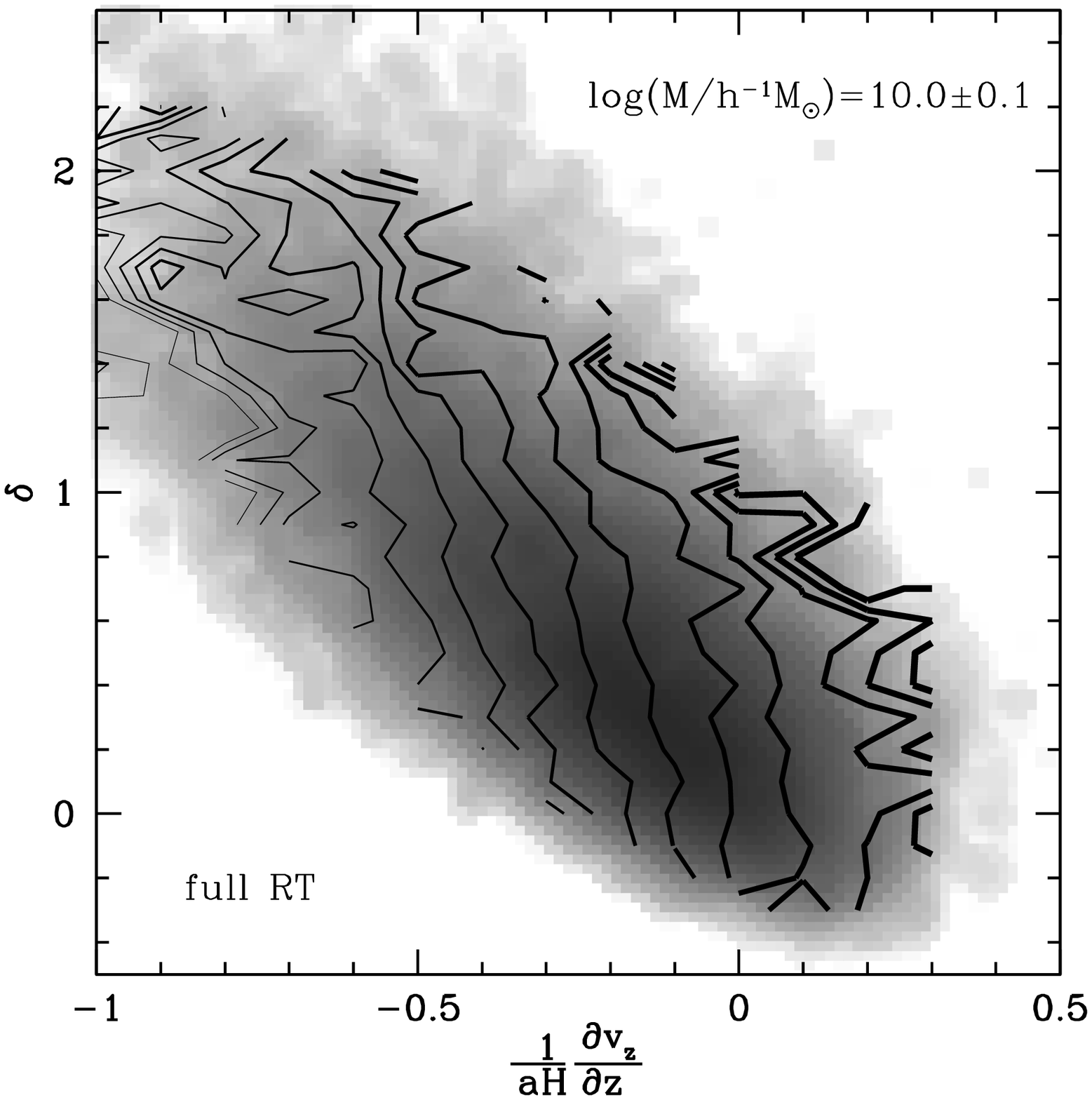}{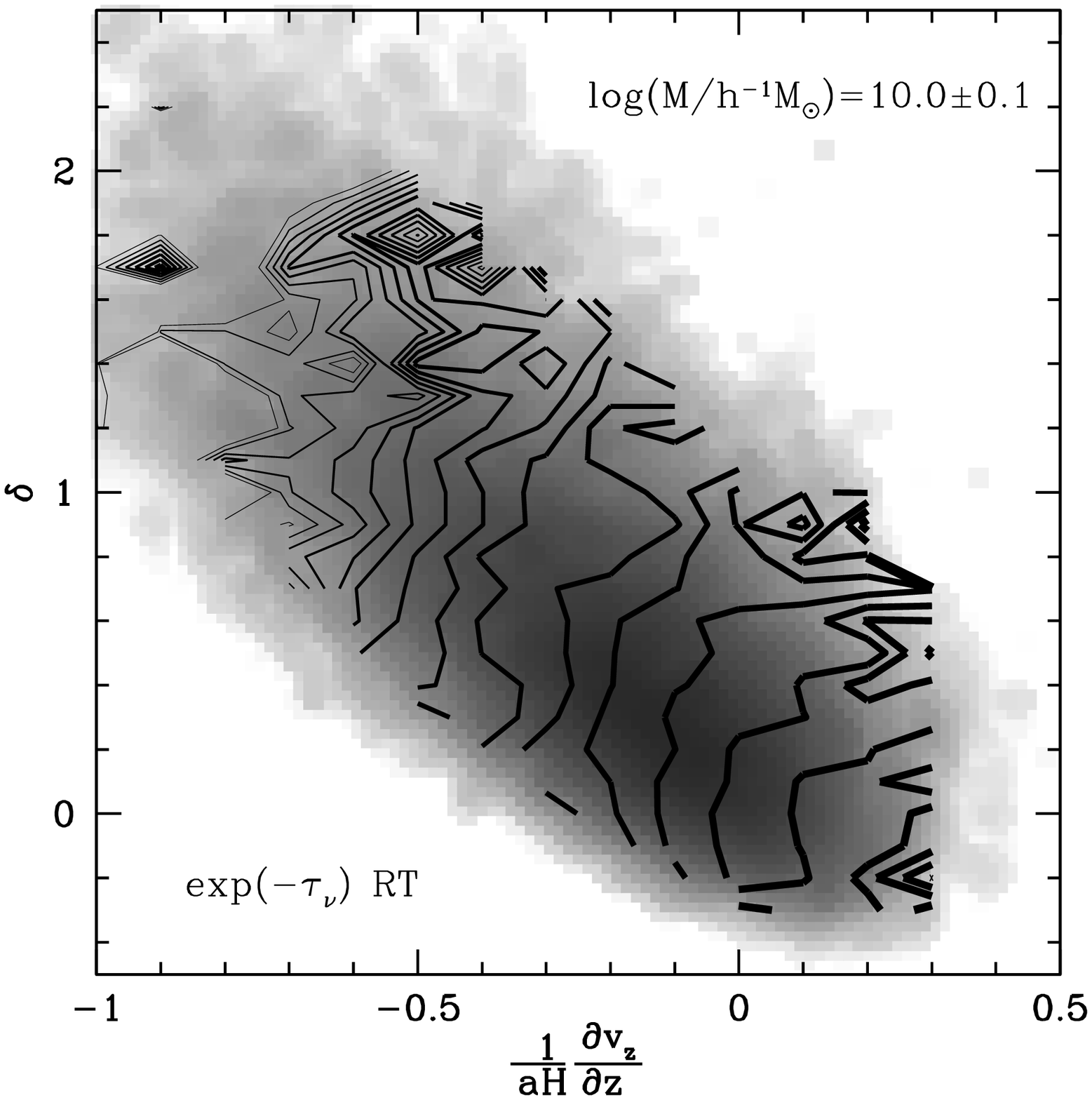}
\caption[]{
\label{fig:jointdep}
Joint dependence of \lya flux suppression of LAEs on environments in the 
full \lya RT model (left) and the $\exp(-\tau_\nu)$ model (right).
In each panel, the gray scale shows the distribution of halos 
($\sim10^{10}\hMsun$) in the plane of density and line-of-sight velocity 
gradient, darker for higher probability density. 
Contours indicate the median \lya flux suppression, defined as the ratio of 
observed to intrinsic \lya luminosity, for LAEs residing in these halos. 
Thicker contours correspond to higher ratios, and adjacent contours differ by 
0.1 dex in contour levels. Note the difference in the trend of flux ratios at 
fixed velocity gradient in the two models.
See the text for details.
}
\epsscale{1.0}
\end{figure*}

\begin{figure}
\epsscale{1.2}
\plotone{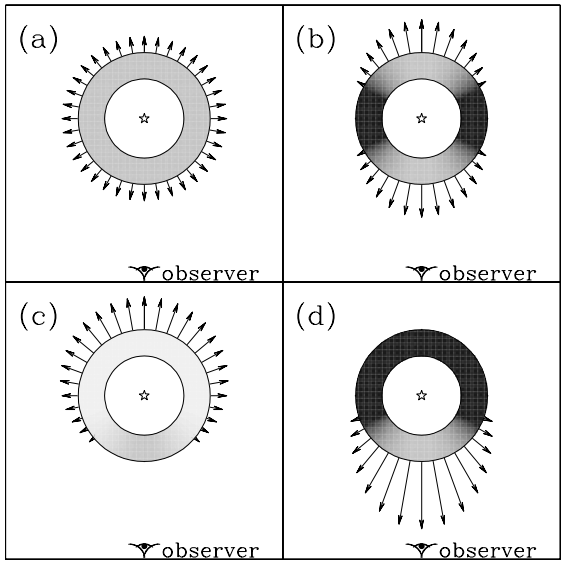}
\caption[]{
\label{fig:schematic}
Schematic illustration of RT effect on the observed 
flux of \lya emission. In each panel, \lya photons are emitted from the
central source and scattered by neutral gas surrounding it. The gas is
shown as the shaded region, and the gray scale indicates the \lya 
optical depth along each radial direction with darker for higher optical 
depth. The flux of escaped \lya photons along each direction is illustrated
by the arrows with length proportional to flux. Although the optical depths
along the line-of-sight direction (from the observer to the source) are the 
same in all four panels, we expect to have different observed flux because 
of different distribution of optical depth in other directions.
}
\epsscale{1.0}
\end{figure}

\begin{figure*}
\epsscale{1.15}
\plotone{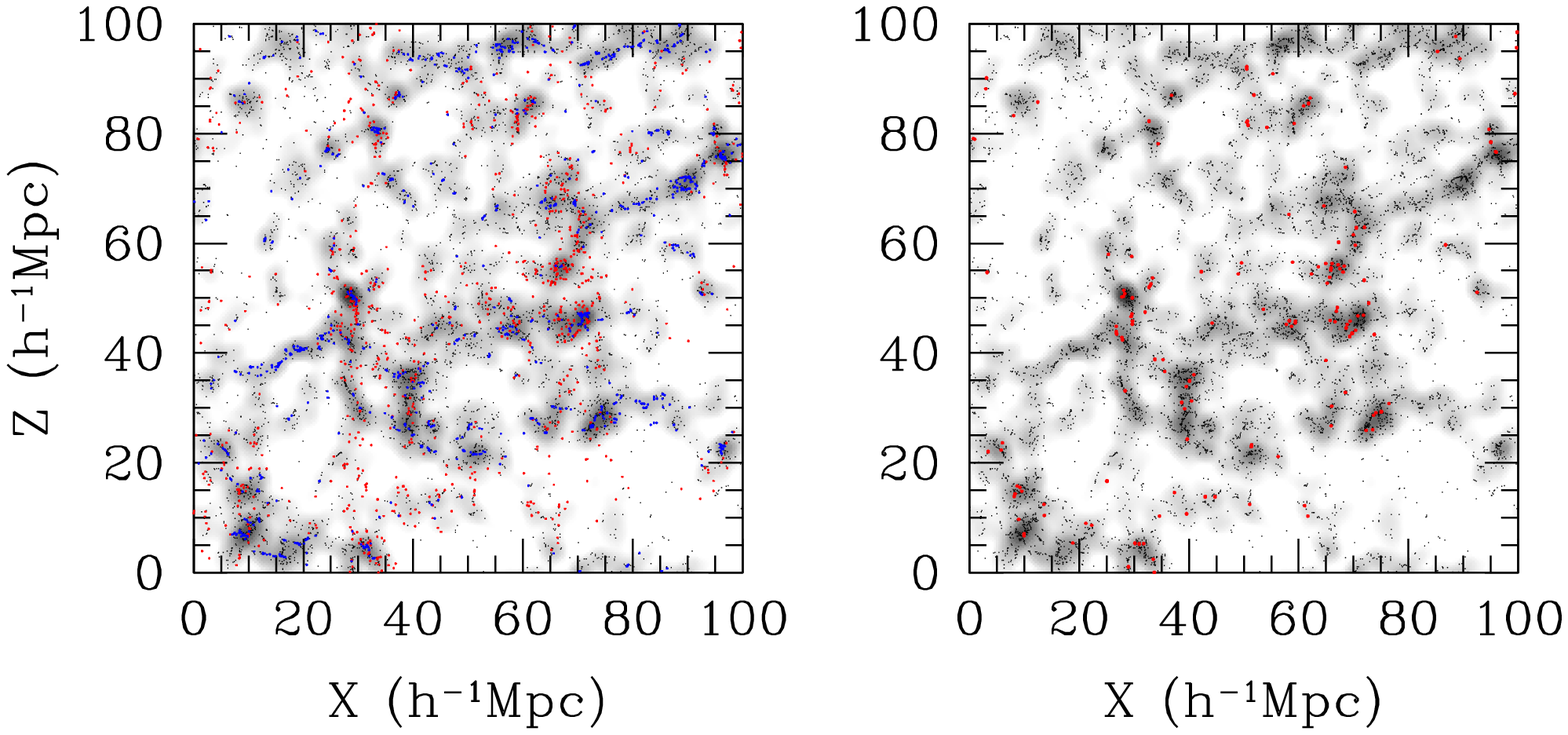}
\epsscale{1.0}
\caption[]{
\label{fig:slice}
Spatial distribution of LAEs. The slice is $5\hMpc$ thick. The gray scale map 
shows the matter density distribution (smoothed at 2$\hMpc$ scales; darker for
higher density) to delineate the large-scale filamentary structures. 
Points represent halos above $5\times 10^9 \hMsun$. 
Left: distribution of LAEs as a 
function of the ratio of the apparent (observed) and intrinsic \lya luminosity 
$L_{\rm apparent}/L_{\rm intrinsic}$. Red (blue) points are for sources with 
\lya luminosity weakly (strongly) suppressed, with the apparent to intrinsic 
luminosity ratio higher than 14\% (lower than 0.6\%), which approximately 
corresponds to the top (bottom) 10\% of the luminosity ratio distribution.
Right: distribution of LAEs (red points)
above a threshold in apparent (observed) luminosity. The luminosity threshold
corresponds to a LAE number density of $3.8\times 10^{-3}\denhMpc$.
Note that the line-of-sight direction (from the observer to sources) is along
the $-Z$-direction. That is, the distant observer observes the sources from 
the top of the panels, which matters for interpreting the relation between \lya 
flux suppression and the local environment (see the text).
}
\end{figure*}

\section{Environment Dependence of L\protect\lowercase{\rm y}$\alpha$ Radiative Transfer}
\label{sec:environ}

As shown in Paper I, a point \lya-emitting source becomes extended in the end 
as a result of spatial diffusion caused by \lya RT. Only the
central, high surface brightness part of the extended source can be observed
as an LAE. Therefore, the observed or apparent \lya luminosity ($\Lapp$) is
reduced with respect to the intrinsic \lya luminosity ($\Lint$). The 
suppression of \lya emission, characterized by the ratio of observed to 
intrinsic \lya luminosity, depends on the environments in the vicinities of 
LAEs. The environments here are interpreted broadly as the matter density and 
peculiar velocity. In Paper I, we identify a few environment variables. In 
general, the suppression is weaker in regions of lower density;
the suppression depends on the sign of the density 
gradient along the line of sight, in the sense that sources located in the 
near side (with respect to the observer) of an overdense region have a lower
suppression; relative to the Hubble flow, sources moving away from the 
observer have a lower suppression because of the additional redshift in 
\lya frequency; sources with larger line-of-sight gradient of the line-of-sight 
peculiar velocity are more easily observed because of an effectively larger
Hubble expansion rate.

In Figure~\ref{fig:denvelgrad_mass}, we show the environmental dependence of 
\lya RT outcome in terms of the observed-to-intrinsic \lya
luminosity as a function of four physical variables ---
overdensity $\delta$, its gradient along the line of sight, line-of-sight
halo velocity, and peculiar velocity gradient along the line of sight, 
separately for four subsets of halos divided according to their halo mass. 
The overdensity field is smoothed with a three-dimensional (hereafter, 3D) 
top-hat filter of radius $2\hMpc$ (comoving). 
It is seen that the overall dependence becomes weaker for 
sources in halos of higher mass, probably indicating that \lya scatterings 
encountered in virialized regions or their immediate surroundings,
which are less affected by environments, becomes more important for higher mass halos.

Gravitational evolution of the cosmic structure leads to correlations
among the above environment variables, which complicates the 
isolation of the role of each variable on the \lya RT.
To have a better understanding of the interplay of different environment
variables, we study in detail the joint dependence of \lya RT
on two variables, the matter density and the line-of-sight velocity 
gradient, which appear to be the major factors in the effects 
shown in the clustering of LAEs (as discussed in the following section).

The gray scale in Figure~\ref{fig:jointdep} shows the joint distribution of
local matter overdensity $\delta$ and $\partial v_z/\partial z$ for halos 
of mass $\sim 10^{10}\hMsun$. Clearly, the two quantities
are anti-correlated: $\partial v_z/\partial z$ is, on average, 
larger (i.e., less negative) in less dense region, albeit with a large 
dispersion, which means that there is an effective higher Hubble constant in 
less dense regions than in higher density regions.
This is easily understood using linear theory, where
the linear velocity field is related to overdensity by
$\delta \propto - \nabla\cdot{\mathbf v} $
from the continuity equation. Therefore, on average or in the spherical case,
a higher density corresponds to a lower value of velocity gradient.
Complex non-spherical geometry of density perturbations around sources causes 
the large dispersion seen in the correlation.

At each position in this
$\partial v_z/\partial z$--$\delta$ plane, we compute the median value of 
the ratio of the observed to intrinsic \lya luminosity. The contours show
the distribution of the median ratio, with thicker contours for higher values.
In the model with full \lya RT calculation (left panel), at a fixed velocity 
gradient, \lya luminosity appears to be {\it less} 
suppressed in regions of {\it higher} density. 
This may appear to be counter-intuitive and in contradiction with 
Figure~\ref{fig:denvelgrad_mass}(a). 
Naively, we expect more 
\lya scatterings and larger spatial diffusion in denser regions, which 
would lower the ratio of observed-to-intrinsic \lya luminosity. This is 
opposite to what is seen in Figure~\ref{fig:jointdep}, which implies that 
the density effect is masked by a stronger effect. 

To understand this, we need to note that the linear velocity field is related 
to overdensity, $\delta \propto - \nabla\cdot{\mathbf v} = 
-(\partial v_x/\partial x + \partial v_y/\partial y + \partial v_z/\partial z)$.
Therefore, at fixed $\partial v_z/\partial z$ a higher density corresponds to 
lower values of velocity gradients in the transverse directions. Since 
velocity gradient is dominant in determining observed-to-intrinsic 
\lya luminosity ratio (see Section~7 in Paper I), lower values of velocity 
gradients in the transverse 
directions indicate a lower probability for \lya photons to escape along the 
transverse directions. This is expected from the \lya RT --- we can think of 
this as that the scatterings of \lya photons enable them to probe the optical 
depth in all directions and they prefer to travel the path with the least 
resistance.

Figure~\ref{fig:schematic} gives a schematic illustration of the dependence of 
\lya RT on the surroundings. \lya photons are emitted from a central source 
(marked as a star in each panel). The gray scale of the ring around the source 
indicates \lya optical depth in each direction, darker for higher optical 
depth. Scatterings of \lya photons tend to make them escape in directions with 
lowest optical depth, leading to higher flux in these directions. The lengths
of arrows illustrate the magnitude of flux in each direction. Although 
the line-of-sight optical depth is the same and the total flux over all 
directions is the same in the four cases shown in Figure~\ref{fig:schematic}, 
the observed fluxes are not. For example, in panel~(d), the optical depths
are high in directions other than the line-of-sight direction, and \lya
photons are all likely to be reflected toward the observer. 
While in panel~(c), the optical depths are low in directions other than the 
line-of-sight direction, and the observer can only see a very low flux of 
\lya photons. {\it The observed 
flux is therefore not purely determined by the line-of-sight optical depth, but 
the line-of-sight optical depth relative to those in all other directions.} 
This is a fundamental difference between the $\exp(-\tau_\nu)$ model and 
the full \lya RT model, with the former only using the
line-of-sight information. In the $\exp(-\tau_\nu)$ model, at fixed 
line-of-sight velocity gradient, the optical depth should increase with 
density, which is opposite to the full RT model and is seen in the right 
panel of Figure~\ref{fig:jointdep}. The relative amplitude of the optical
depth is the zeroth-order, primary effect on the directional propagation of 
\lya photons. Of course, the overall amplitude of the optical depth (e.g.,
in the spherically symmetric case) determines the spatial extent of the \lya 
photon diffusion and the overall degree of suppression of the observed \lya
luminosity.

While at fixed $\partial v_z/\partial z$ the observed-to-intrinsic \lya 
luminosity ratio increases with density, it decreases with density if averaged
over environments, as seen in Figure~\ref{fig:denvelgrad_mass}(a). The apparent
contradiction is resolved by noticing that the latter is driven by the 
strong dependence of the \lya luminosity ratio on $\partial v_z/\partial z$
(Figure~\ref{fig:denvelgrad_mass}(d) and the anti-correlation between density 
and $\partial v_z/\partial z$ (gray scale in Figure~\ref{fig:jointdep}).

To have a visual impression of the environmental dependence of \lya flux 
suppression in the simulation, we show a slice 
(5$\hMpc$ thick) of the spatial distribution of LAEs in Figure~\ref{fig:slice}. 
In our \lya RT calculation, the observational direction is
along the $-Z$-direction, i.e., the observer is assumed to observe from the
top of the plot. Points denote all the halos more massive than 
$9\times 10^9\hMsun$. In the left panel, red points represent sources with
weak \lya flux suppression (high ratio of $\Lapp/\Lint$) and blue points are for
sources with strong \lya flux suppression (low ratio of $\Lapp/\Lint$). One can 
see the environmental dependence in the sense that sources in underdense
regions (larger line-of-sight velocity gradient) have a higher probability of 
being less suppressed in \lya flux, e.g., sources around the void at 
($X$, $Z$) = (35, 20). One can also see the line-of-sight density gradient 
effect. Sources on the near side of an overdense region are more likely 
to be less suppressed in \lya flux, e.g., sources above and below the 
overdense region near ($X$, $Z$) = (30, 40).

In the right panel of Figure~\ref{fig:slice}, red points represent LAEs
above a threshold in the observed \lya luminosity. The luminosity threshold 
corresponds to a sample of LAEs with number density of 
$3.8\times 10^{-3}\denhMpc$. The red points therefore show the spatial
distributions of observed LAEs. They appear to be strongly clustered with
large voids between them. In the next section, we study the clustering 
of LAEs in more detail.

\section{The Clustering of LAE\protect\lowercase{\rm s}}
\label{sec:clustering}

\begin{figure*}
\epsscale{1.15}
\plotone{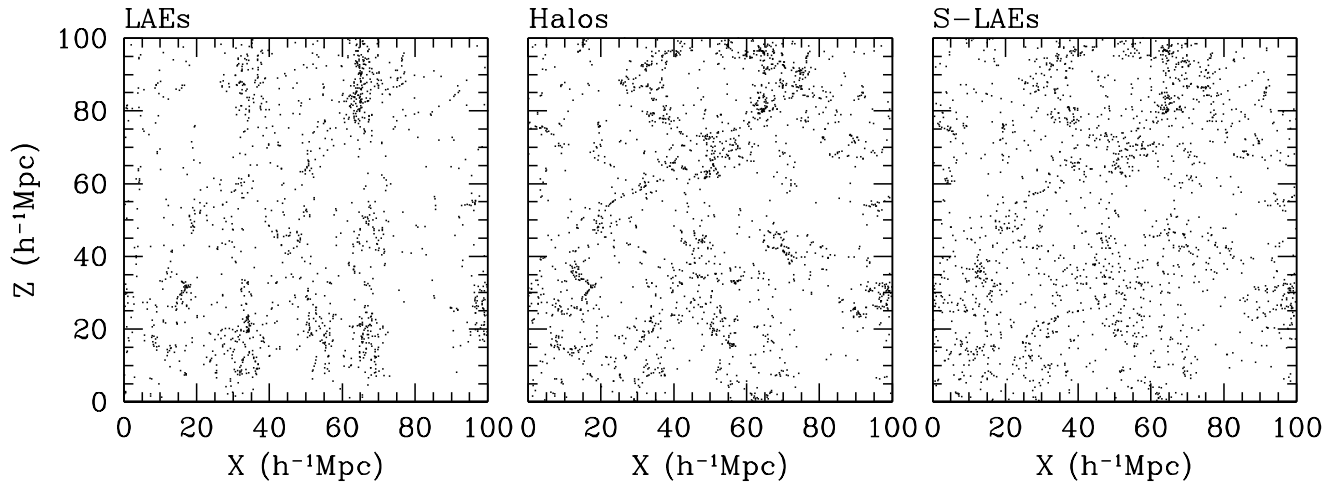}
\caption[]{
\label{fig:slice_los}
Spatial distributions of LAEs and control sources in real space, with one
spatial direction being along the line of sight. For LAEs (left panel), the 
observer is supposed to be on the top of the panel, observing toward the 
$-Z$-direction. The slice in each panel is from the same part of the box 
projected onto the $X$--$Z$ plane, with a thickness of 20$\hMpc$. The two 
control samples are the halo sample (middle panel) and the shuffled LAE 
sample (S-LAEs; right panel), with the same number density ($10^{-2}\denhMpc$)
as the LAE sample.
The LAE/S-LAE and halo samples are defined by thresholds in observed
\lya luminosity and halo mass, respectively. The shuffled LAE sample is 
expected to eliminate the effect of environmental dependence of \lya RT. 
Compared with the two control samples, the LAE sample shows a prominent 
pattern of elongated distribution along the line of sight, a result of 
environment-dependent \lya RT. See the text for details.
}
\epsscale{1.0}
\end{figure*}

\begin{figure*}
\epsscale{1.15}
\plotone{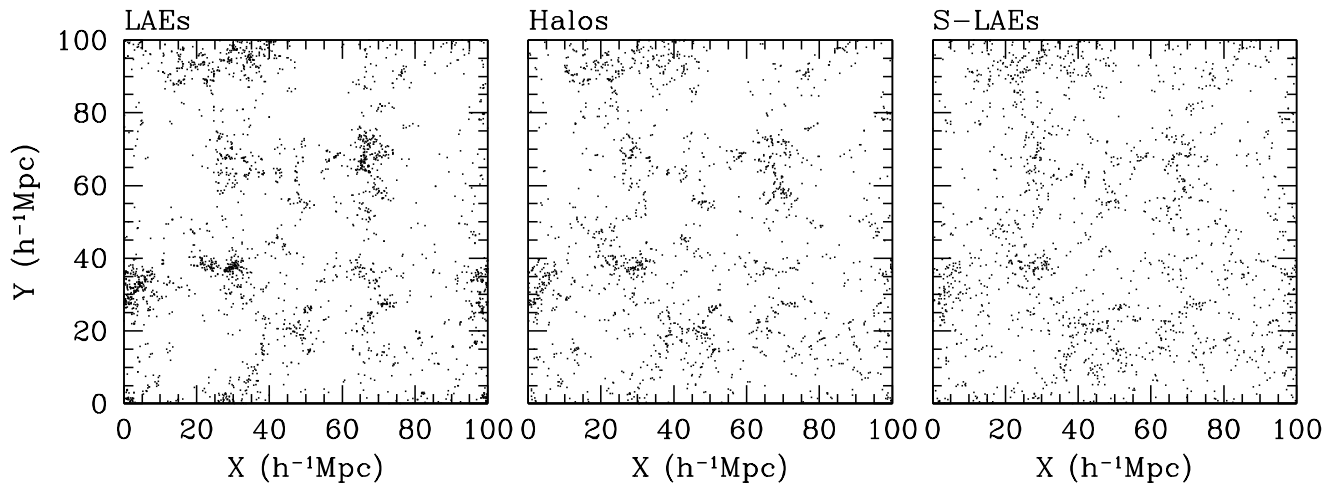}
\caption[]{
\label{fig:slice_trans}
Same as Figure~\ref{fig:slice_los}, but for distributions in the transverse
plane perpendicular to the line of sight. In the transverse direction, the 
LAE sample appears to be more strongly clustered than the two control samples, 
which is a result of environment-dependent \lya RT.
See the text for details.
}
\epsscale{1.0}
\end{figure*}

\begin{figure*}
\epsscale{1.0}
\plotone{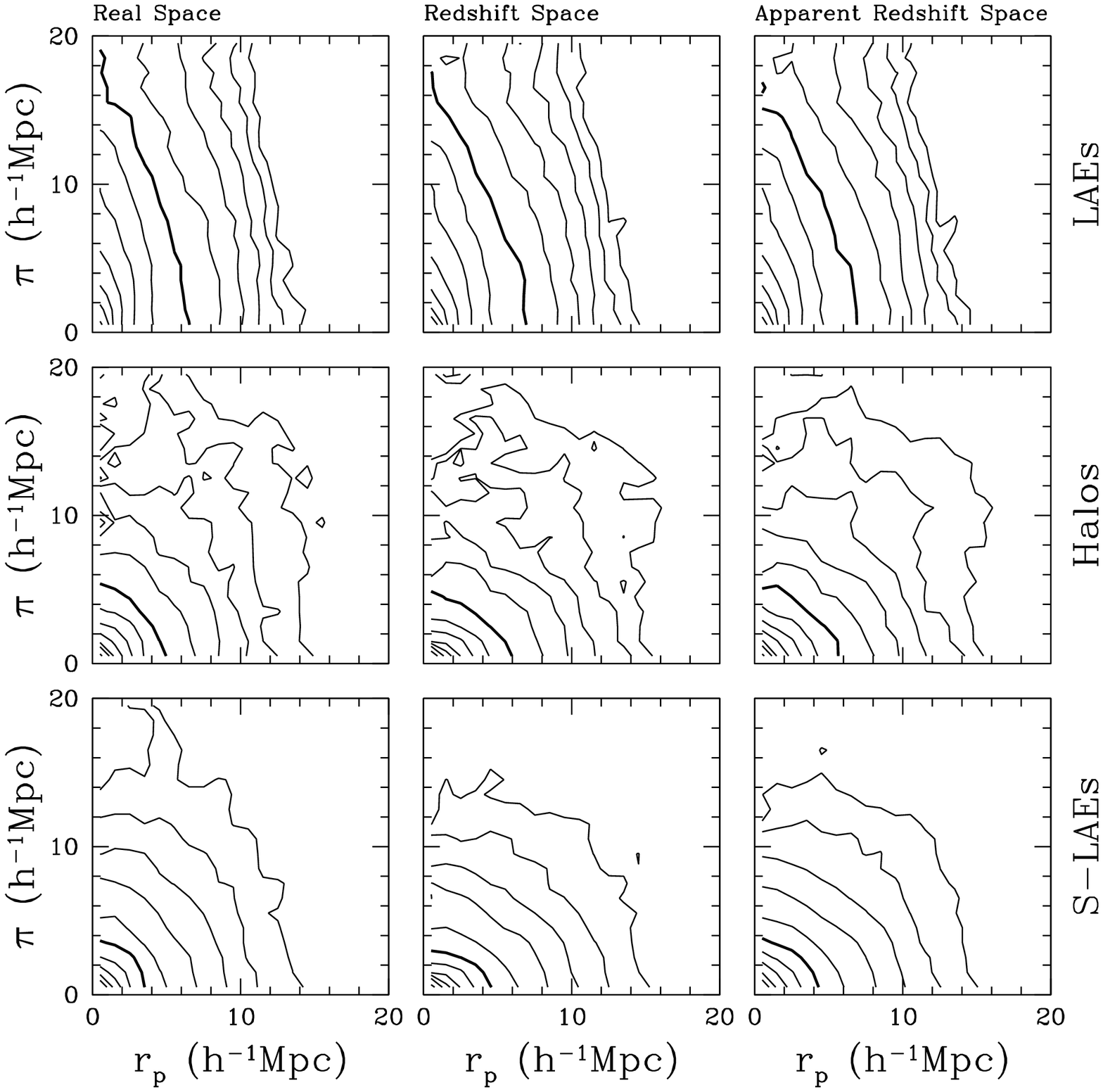}
\epsscale{1.0}
\caption[]{
\label{fig:xi_laehalo}
Three-dimensional (3D) two-point correlation functions of LAEs and halos
as a function of the line-of-sight separation ($r_p$) and the transverse
separation ($\pi$).
From top to bottom are those of LAEs, halos, and shuffled LAEs (S-LAEs).
All samples are threshold samples with the same number density 
($10^{-2}\denhMpc$), with the thresholds being the apparent (observed) \lya 
luminosity for LAEs/S-LAEs samples and halo mass for the halo sample,
respectively. From left to right are the two-point correlation functions 
in real space, redshift space, and apparent redshift space. In real space,
the position of a source is just its comoving position. In redshift space,
the position is determined by the source's comoving position and peculiar 
velocity. The apparent redshift is defined from the observed peak in \lya 
spectra, which has the additional shift caused by \lya RT 
(see the text). It is used to determine a source's position in apparent
redshift space. The solid contour in each panel denotes a contour level of 
unity, and the adjacent contours differ by 0.2dex in contour levels.
Note the prominent elongation pattern along the line of sight in the 
clustering of LAEs in all three spaces considered here. See the text.
}
\epsscale{1.0}
\end{figure*}

In this section, we start with the clustering of LAEs in one sample and the 
associated control samples to show a few new features in galaxy clustering 
introduced by \lya RT. Then we present a simple physical model to aid our
understanding of these new features. With this model in hand, we study LAE
clustering in detail, including the projected/angular clustering and
the dependence on observed \lya luminosity. We present the clustering results
from our simulation in terms of the 2PCFs. The fluctuation power spectra for 
a couple of LAE samples are shown in Appendix~\ref{sec:powerspectrum}.
A comparison between the predictions 
of LAE clustering in the full \lya RT and the $\exp(-\tau_\nu)$ models is 
presented in Appendix~\ref{sec:clustering_expmtau}.

\subsection{Real-space and Redshift-space 2PCFs of LAEs}

To mimic observations, we 
construct LAE samples based on the apparent ({\it observed}) \lya luminosity,
with each sample consisting of all LAEs with observed \lya luminosity above
a threshold. For such luminosity-threshold samples, luminosity threshold is 
uniquely mapped onto sample number density. Hereafter, we denote different 
samples by the number densities of LAEs.

To better illustrate the environment-induced effects, we also construct 
control samples that are supposed to be free of environment effects induced by
\lya RT. We
generate two types of control samples for each LAE sample. The first type 
is a mass-threshold sample of halos that has the same number density as the LAE 
sample. 
This sample corresponds to a model of LAEs that relates observed LAEs 
to halos through a one-to-one map of observed \lya luminosity onto halo mass 
(a.k.a. abundance matching). 

The second type of control sample is called the shuffled LAE sample (S-LAEs).
The sample is constructed as follows. We first sort halos by mass and divide
them into narrow mass bins with typical bin width of 0.02--0.04 dex in 
$\log M$. The number of halos in each bin is a few thousands (hundreds) around
$10^{10}\hMsun$ ($10^{11}\hMsun$). For halos in each narrow mass bin, 
the \lya emission properties (apparent/intrinsic \lya luminosity, \lya 
spectra, and even the UV luminosity) as a whole are randomly shuffled 
among them. Such a shuffling algorithm does not 
lead to any change in the \lya LF, UV LF, and the relation between observed 
to intrinsic \lya properties, which keeps the interpretations to the observed 
properties of LAEs presented in Paper I unchanged.
However, the shuffling is supposed to get rid of any correlation between 
observed \lya properties and environments, with the latter encoded in halo 
positions. From this shuffled catalog, we form the S-LAE sample above the same
threshold in observed \lya luminosity as in the LAE sample. By construction,
the S-LAE sample has the same number density as the LAE sample. For the S-LAE
sample, the dependence of \lya RT results on environment is
completely eliminated while the statistical properties of \lya luminosity
is kept, making it a more suitable control sample than the halo sample. 
Similar shuffling method is applied in studying environment effect on halo 
clustering (assembly bias; e.g., \citealt{Yoo06,Croton07,Zu08}). However,
we emphasize that what we intend to study here is not the effect of 
environments on halo clustering, which is small for galaxy-host halos at high 
redshifts \citep[e.g.,][]{Wechsler06,Gao07}, but the effect of the 
environment-dependent \lya RT on LAE clustering.

We perform nine realizations of shuffling with different random seeds. For each 
LAE sample, nine S-LAE samples are constructed from these realizations. The 
statistical results for S-LAEs (e.g., correlation functions) presented in 
this paper are always the average over the nine realizations. 

In Figure~\ref{fig:slice_los}, the real-space spatial distributions in 
the $X$--$Z$ plane of the LAE sample and the two corresponding control samples 
are compared. All the three samples have number density of $10^{-2}\denhMpc$.
By the term ``real space'', we mean to use the comoving coordinates of 
sources, with no peculiar velocity displacements.
Each slice has a thickness of 20$\hMpc$. The line-of-sight direction is along 
$-Z$. Unlike sources in the control samples, the spatial distribution of LAEs 
shows a distinct pattern stretching along the line of sight. The elongation
resembles the fingers-of-God (FoG) effect seen in redshift-space galaxy 
clustering. However, it clearly differs from the FoG effect. First, the
elongation effect for the LAE sample shows up in real space. Second, while the
FoG effect is on scales of virialized halos, the elongation effect for the 
LAE sample exists up to much larger scales. 

For the two control samples, there is no elongation pattern. The clustering 
of the S-LAE sample appears to be weaker than the halo sample. Although the 
two samples have the same number density, there are S-LAEs residing in halos 
of mass lower than the mass threshold of the halo control sample (see 
Section~\ref{sec:hod}). Since low-mass halos are less clustered than high-mass 
halos, the S-LAE sample displays a weaker clustering pattern. 

In Figure~\ref{fig:slice_trans}, the real-space spatial distributions of 
sources in the $X$--$Y$ plane (perpendicular to the line of sight) are compared
for the three samples. For the two control samples, the trend is the same
as in Figure~\ref{fig:slice_los}, as expected. The LAE samples show much 
stronger clustering than the control S-LAE sample. For this particular set of 
samples, it is also true that LAEs are more strongly clustered than halos in 
the halo control sample. 

The different effects of \lya selection on the line-of-sight
and transverse distributions of LAEs means that the spatial distribution 
(hence the clustering) of LAEs is no longer isotropic even in real space. 
We measure the 3D 2PCFs of the LAE and control samples using the Landy--Szalay 
estimator \citep{Landy93}. The real-space 3D 2PCF 
$\xi(r_p,\pi)$ for the above LAE sample (top-left panel of 
Figure~\ref{fig:xi_laehalo}) clearly shows the anisotropy in LAE clustering,
where $r_p$ and $\pi$ are the transverse and line-of-sight separations of
galaxy pairs, respectively. 
The contours are elongated along the line-of-sight separation. The scales 
range from sub-Mpc to $>10\hMpc$, where the 2PCF can be accurately measured.
The 2PCFs of sources in the halo and S-LAE control samples (middle-left and
bottom-left panels) appear to be isotropic, manifested by contours being 
circular. Note that the deviation from circular contours at large separation 
for the control samples is caused by statistical fluctuations (sample 
variance) with the finite simulation box. 

In redshift space, there is a well-known effect (Kaiser effect; 
\citealt{Kaiser87}) that introduces anisotropy in galaxy clustering on linear 
scales, which makes the 2PCF contours squashed along the line of sight.
The linear Kaiser effect is clearly seen for the control samples in redshift
space (the two lower panels in the middle column of 
Figure~\ref{fig:xi_laehalo}). However, for the LAE sample, the prominent 
elongation pattern along the line of sight is preserved in redshift space,
with a slight reduction in the degree of the elongation. 

In the above discussion, the redshift-space source positions are calculated
as the sum of comoving coordinates and the apparent position change 
introduced by the line-of-sight peculiar velocity. In practice, if redshifts 
of LAEs are determined from \lya line peak in spectra, there would be an 
additional shift in the apparent position. As shown in Paper I, the observed
\lya line peak generally shifts redward as a consequence of RT
and the shift is correlated with the environment. We name the redshift 
measured from \lya line peak as apparent redshift and the corresponding 
redshift space as apparent redshift space. The 2PCFs in the apparent redshift 
space (right column in Figure~\ref{fig:xi_laehalo}) look similar to those in 
the (true) redshift space (middle column in Figure~\ref{fig:xi_laehalo}). 
Since at a fixed (true) redshift the apparent redshift has a distribution,
the 2PCF contours in apparent redshift space are slightly smoothed. {\it In 
all the following discussions, when we refer to redshift-space measurements, 
we use apparent redshifts for LAE and S-LAE samples and (true) redshifts for 
halo samples, unless mentioned explicitly.}

We see that our RT model of LAEs predicts that the clustering 
of observed LAEs differ 
significantly from those of halos and S-LAEs, which are constructed to be
immune to the environment effect on \lya RT. Before 
presenting more results in LAE clustering, it is useful to gain some
understanding of the basic features in the clustering of LAEs, for which we 
give an intuitive picture and introduce a simple physical model.

\subsection{Understanding the Clustering of LAEs}
\label{sec:toy_model}

\begin{figure*}
\plotone{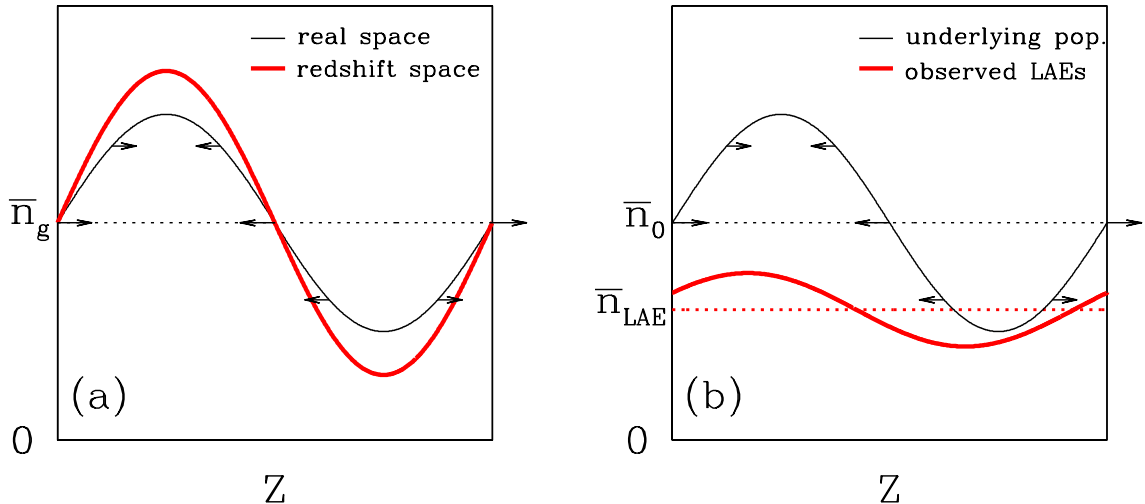}
\caption[]{
\label{fig:toy_illustration}
Illustration of effects of redshift-space distortion and \lya 
RT selection on the observed density fluctuation alone the line of 
sight. (a): Effect of redshift-space distortion. The solid 
black/thin curve 
is the real-space density distribution with the dotted line denoting the mean.
Arrows represent the linear peculiar velocity with the length proportional to
the amplitude. The red/thick curve represents the density distribution in 
redshift 
space. (b): Effect of \lya RT selection. The solid 
black/thin curve is the real-space density distribution of a population of 
galaxies. The observed LAEs are a fraction of these galaxies, with a selection 
function imposed by the \lya RT process. The density 
distribution of the observed LAEs in real-space is represented by the solid
red/thick curve. The selection favors sources in regions with low density and 
positive line-of-sight velocity gradient and to a smaller degree, in regions
with receding line-of-sight velocity that are on the near-side (far-side)
of overdense (underdense) regions. The observer is assumed to lie on the left
side of the panel. See the text for details.
}
\end{figure*}

\begin{figure*}
\epsscale{1.1}
\plotone{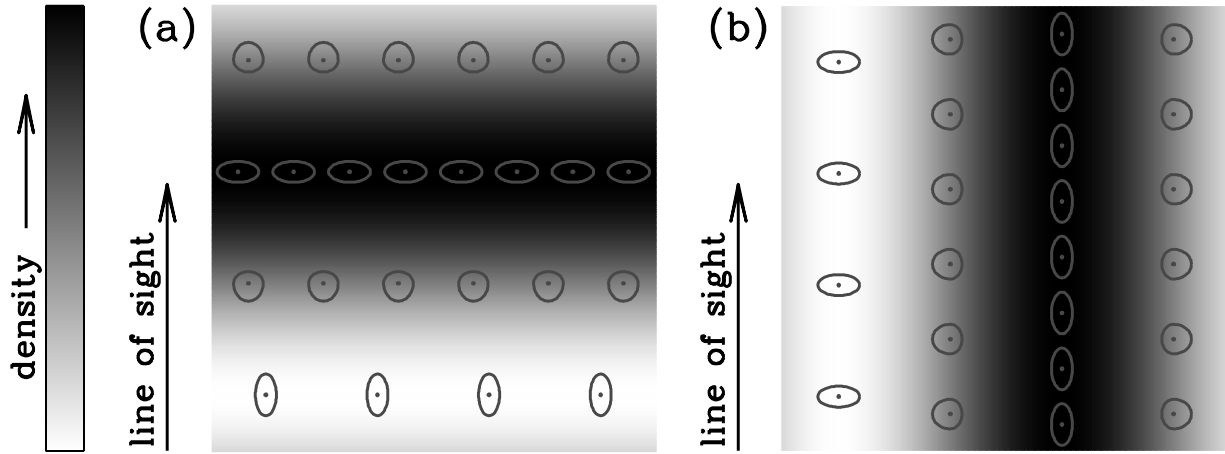}
\caption[]{
\label{fig:angle_dep}
Illustration of the \lya RT selection effect in the cases with 
fluctuation modes parallel and perpendicular to the line of sight.
Panel (a) shows the case of density fluctuation along the line of sight.
The gray scale delineates the overdensity across one wavelength of a plane wave,
darker for higher overdensity. Each point represents a \lya emitting source, 
and the ring 
around it illustrates the angular distribution of escaped \lya emission. In the 
underdense region, \lya emission preferentially comes out along the 
line-of-sight direction (and its opposite direction), mainly because of
the effect of velocity gradient on the RT (see the text). Sources
in the underdense (overdense) region have a higher (lower) probability to 
be observed than those in the overdense region. Therefore, the \lya 
RT selection causes a suppression of the fluctuation along the line of 
sight. Panel (b) is a $90^\circ$ rotation of panel (a), corresponding to the 
case of density fluctuation perpendicular to the line of sight. In this case, 
the \lya RT selection causes an enhancement in the fluctuation.
See the text for details.
}
\epsscale{1.0}
\end{figure*}

\subsubsection{An Intuitive Picture}

The new features appearing in LAE clustering are related to the selection
imposed by \lya RT, which is closely related to the environment
around the LAE sources.  Since the main feature in the 2PCF is a 
line-of-sight distortion, bearing some analogy to redshift-space distortion,
we start from an intuitive understanding of LAE clustering by making 
comparisons to redshift-space distortion.

The effect of redshift-space distortion in the linear regime \citep{Kaiser87} 
is illustrated in Figure~\ref{fig:toy_illustration}(a). 
The solid black curve shows the real-space line-of-sight density distribution 
of a sample of galaxies, with one mode of the plane-wave fluctuation 
considered. The arrows indicate peculiar velocities of galaxies with the length
representing the amplitude. In general, galaxies fall into overdense regions 
and stream out of underdense regions. When mapped onto redshift space, the 
line-of-sight density distribution of galaxies changes because of the 
additional redshifts caused by peculiar velocities. As a result, in redshift 
space, overdense regions become more overdense and underdense regions become 
more underdense, as shown by the red curve. Redshift-space distortion only 
takes effect along the line of sight, which leads to an enhancement of the
line-of-sight fluctuations. In terms of the 3D 2PCF, the effect shows up as 
contours squashed along the line of sight.

For LAEs, the dependence of \lya RT on environments leads to
a selection function that determines whether a source can be detected as
LAEs given the flux limit in observation. 
Figure~\ref{fig:toy_illustration}(b) illustrates such a selection effect. 
The black curve is the real-space
line-of-sight density distribution of an underlying population of LAEs (e.g., 
all the halos that could be detected as LAEs given the flux limit in 
observation, if there were no \lya RT effect). As shown in 
Section~\ref{sec:environ}, the line-of-sight velocity gradient has the 
strongest 
effect in determining the observed-to-intrinsic \lya luminosity ratio, larger 
gradient for higher ratio. For the plane-wave fluctuation in 
Figure~\ref{fig:toy_illustration}(b), the velocity gradient reaches
its maximum and minimum at the trough of the underdense region and the peak
of the overdense region, respectively. As a result, sources in the underdense
region have a higher probability to be observed as LAEs than those in the
overdense region, making the overdense region less overdense and
the underdense region less underdense (i.e., suppressing the line-of-sight 
fluctuation). The density distribution of LAEs that can
be observed is shown as the solid red curve (note that we also add a phase 
shift to reflect the dependence of the selection on density and velocity 
gradients). The dotted red line is the mean
density of LAEs, and the difference with respect to the dotted black line 
is an overall selection effect caused by \lya RT. 
The illustration shows that the \lya selection effect is similar to 
redshift-space distortion, but it originates in real space with an opposite 
sign. 

Figure~\ref{fig:angle_dep} illustrates how to understand 
the selection effect from the point view of angular distribution of \lya 
emission. In Figure~\ref{fig:angle_dep}(a), the grays cale delineates
a plane-wave density fluctuation along the line of sight, darker for higher
density. Each dot is an underlying LAE source. The ring around each
source shows the angular distribution of the \lya emission that can be 
observed. For example, in the underdense region (white part of the gray scale
map), we have a large velocity gradient along the line of sight and zero
along the transverse direction. As a consequence, \lya photons preferentially 
escape from the directions parallel and anti-parallel to the line of sight,
leading to higher surface brightness in these directions. While in the
overdense region, the situation changes to the opposite. For an LAE survey set 
by a \lya flux limit, sources in the underdense region have a higher 
probability to be observed as LAEs. Again we reach the same conclusion as 
above --- the line-of-sight density fluctuation is suppressed for LAEs.
Note that except for sources residing at the peak and trough of the 
(plane-wave) density fluctuations, the angular distribution of \lya emission
generally does not have a parity symmetry along the line of sight. For 
example, \lya photons in sources near the boundary between the overdense and 
underdense regions in Figure~\ref{fig:angle_dep} have their \lya photons 
preferentially escape toward the direction of underdense region, as a result 
of the density gradient and receding velocity effect. 

For the selection effect in the case of a transverse fluctuation, we only
need to rotate Figure~\ref{fig:angle_dep}(a) by $90^\circ$, which leads
to Figure~\ref{fig:angle_dep}(b). In the underdense region, \lya photons 
appear to preferentially escape in the direction perpendicular to the line 
of sight. So sources in the underdense region now have a lower probability 
to be detected as LAEs. This leads to the interesting result that the density 
fluctuation in the transverse direction is enhanced for LAEs.

Note that large-scale density filaments are a common feature of gravitational
dynamics. Filaments may be described by Figure~\ref{fig:angle_dep}.
The net result is that filaments oriented along the line of sight are 
preferentially observed, whereas those perpendicular to the line of sight are 
suppressed. This feature is clearly seen in the left panel of 
Figure~\ref{fig:slice_los}.

Based on the above discussion, the main selection effects in galaxy clustering
caused by environment-dependent \lya RT can be summarized as
the suppression of the fluctuation in the line-of-sight direction and 
the enhancement of the fluctuation in the transverse direction. In what 
follows, we present a simple physical model to describe these effects.

\subsubsection{A Simple Physical Model}

For the simple physical model, we limit our discussion to the linear regime. 
Given that the effects of \lya selection share some similarities with 
the Kaiser effect, we first review the formalism of redshift-space distortion 
in galaxy clustering (see, e.g., \citealt{Hamilton98}). We denote quantities 
in redshift space with a superscript ``s''. 

The linear density fluctuation 
$\delta_g({\mathbf r})$ of galaxies in real space may be related to that of
matter $\delta_m({\mathbf r})$ by
$\delta_g({\mathbf r})=b\delta_m({\mathbf r})$, where $b$ is the large-scale
galaxy bias factor that in general may be scale dependent.
From conservation of galaxy pairs, the redshift-space
density fluctuation $\delta^s_g({\mathbf s})$ satisfies 
\begin{equation}
\label{eqn:pairconserv}
[1+\delta^s_g({\mathbf s})]d^3s=[1+\delta_g({\mathbf r})]d^3r,
\end{equation}
with ${\mathbf s}={\mathbf r}+v_z\hat{z}/(Ha)$, where $v_z$ is the 
line-of-sight peculiar velocity and $H$ is the Hubble constant at the time
when the scale factor is $a$. Equation~(\ref{eqn:pairconserv}) reduces to  
\begin{equation}
\label{eqn:zspace_config}
\delta^s_g=\delta_g-\frac{1}{Ha}\frac{\partial v_z}{\partial z}.
\end{equation}
The relation is easier to study in Fourier space by noticing that 
$\delta_m = \sum_{\mathbf k} 
\delta_{m,{\mathbf k}}\exp(i{\mathbf k}\cdot{\mathbf r})$
and
${\partial v_z}/{\partial z}=-\sum_{\mathbf k} fHa(k_z^2/k^2)
\delta_{m,{\mathbf k}}\exp(i{\mathbf k}\cdot{\mathbf r})$, 
\begin{equation}
\delta^s_{g,{\mathbf k}}=\delta_{g,{\mathbf k}}+f\mu^2\delta_{m,{\mathbf k}}
                        =\left(1+\beta\mu^2\right)\delta_{g,{\mathbf k}}.
\end{equation}
In the above expression, $\mu=k_z/k$ is the cosine of the angle between the
line of sight and the wave vector of the Fourier mode of interest,  
$f=d\ln G/d\ln a$ is the derivative of the linear growth factor $G$, and 
$\beta\equiv f/b$ is the redshift-distortion parameter. The underlying
picture of the above equation is that galaxies fall into overdense regions
and stream out of underdense regions,
enhancing the line-of-sight density fluctuation in redshift space,
as illustrated in Figure~\ref{fig:toy_illustration}(a).
In terms of the power spectrum, we have
\begin{equation}
\label{eqn:zdistortion}
P^s_g({\mathbf k})=(1 + \beta\mu^2)^2 P_g({\mathbf k}).
\end{equation}

In general, whenever the detection of an object or the measurement of a 
quantity in redshift space is affected or altered by the gradient of peculiar 
velocity, there are two different bias factors, one related to 
density and the other related to peculiar velocity. The $\beta$ factor in 
the Kaiser formula may not have its usual meaning. \citet{McDonald00} and 
\citet{McDonald03} discuss this for the case of \lya forest correlation, where 
the optical depth or transmitted flux is affected by peculiar velocity 
gradient. In our case, the selection function of the galaxies varies with the 
velocity gradient.

For LAEs, the density and line-of-sight peculiar velocity 
and their line-of-sight gradient, as the primary environmental factors,
affect \lya RT, which imposes a
selection function for the appearance of LAEs. For the simple physical model presented 
here, we only consider the two main factors, line-of-sight peculiar velocity 
gradient and density. As shown below, including them is able to explain the 
main effects in LAE clustering. For completeness, in 
Appendix~\ref{sec:extendedtoy}, we add the other two 
factors into the model and show that their main effect is to introduce 
scale-dependent bias.

We consider a simple case that the selection function is a linear function 
of the line-of-sight peculiar velocity gradient $\partial v_z / \partial z$
and the matter density $\delta_m$. In this model, the real-space density 
of LAE galaxies is related to the matter density modified by the selection 
function,
\begin{equation}
\label{eqn:lae_den_sel}
\bar{n}_g(1+\delta_g) = q\bar{n}_0(1+b\delta_m) \times  
          \left[ 1 + \alpha_1 \delta_m
                   + \alpha_2 \frac{1}{Ha}\frac{\partial v_z}{\partial z}
          \right],
\end{equation}
where $b$ is the bias factor for the underlying galaxy population (with mean 
number density $\bar{n}_0$) before the 
\lya selection is imposed, $\bar{n}_g$ is the mean number density of galaxies 
that are selected as LAEs, $q$ is the overall fraction of galaxies 
that are selected as LAEs, and $\alpha_i$ ($i$=1 and 2) are dimensionless 
coefficients (assumed to be constant). 
Based on discussions related to Figure~\ref{fig:jointdep}, the $\delta_m$ 
term in the selection function represents a combined effect of the dependence 
of \lya RT on $\delta_m$ and the transverse peculiar velocity 
gradient, and both coefficients $\alpha_1$ and $\alpha_2$ are expected to be 
positive. Although it is more appropriate to separate
the terms of density and the transverse peculiar velocity gradient (see
Appendix~\ref{sec:extendedtoy}), we choose to use a positive $\alpha_1$ to 
denote the combined effect here for simplicity.

Keeping the first order terms in Equation~(\ref{eqn:lae_den_sel}) and noticing 
that $\bar{n}_g=q\bar{n}_0$, we have
\begin{equation}
\delta_g =  (b+\alpha_1) \delta_m 
               +\alpha_2 \frac{1}{Ha}\frac{\partial v_z}{\partial z}.
\end{equation}
In Fourier space, the relation becomes
\begin{equation}
\label{eqn:delta_realspace}
\delta_{g,{\mathbf k}} 
 =    \left[ (b+\alpha_1)-\alpha_2 f\mu^2
                \right]\delta_{m,{\mathbf k}}.
\end{equation} 
Since $\partial v_z /\partial z$ is the major variable in shaping the 
observed-to-intrinsic \lya luminosity ratio, the coefficient $\alpha_2$ 
is expected to be greater than $\alpha_1$ (also see Figure~\ref{fig:jointdep}).
Equation~(\ref{eqn:delta_realspace}) describes that the effect of \lya 
RT selection from the velocity gradient term ($\alpha_2$
term) is to suppress the fluctuation along the line of sight, as long as
$\alpha_2f$ is not so large to reverse the phase of the fluctuation.

Adding the Kaiser effect to the above \lya RT selection effect, in
redshift space, the relation for the total effect reads
\begin{equation}
\label{eqn:delta_zspace}
\delta^s_{g,{\mathbf k}} 
 =     \left[ (b+\alpha_1)+(1-\alpha_2)f\mu^2
                \right]\delta_{m,{\mathbf k}}.
\end{equation}

With Equation~(\ref{eqn:delta_zspace}), the power spectrum of LAEs in redshift 
space is then
\begin{equation}
\label{eqn:PLAE_zspace}
P^s_g({\mathbf k}) 
 = \left[ \left(1+\frac{\alpha_1}{b}\right)
              +(1-\alpha_2)\beta\mu^2\right]^2 
   b^2P_m({\mathbf k}).
\end{equation}
This full expression of power spectrum includes both redshift distortion and
\lya selection effects. The usual redshift space linear power spectrum is
a special case with the coefficients $\alpha_i$ set to zero. The 
dependence of \lya RT on the velocity gradient leads to the
same form of angular dependence as the redshift distortion in the anisotropic
power spectrum, but with an opposite sign. 

The power spectrum in Equation~(\ref{eqn:PLAE_zspace}) can be decomposed 
into monopole, quadrupole, and hexadecapole moments,
\begin{eqnarray}
\label{eqn:monopole}
P_0(k) & = &
\left[
 \left(1+\frac{\alpha_1}{b}\right)^2
+\frac{2}{3}\left(1+\frac{\alpha_1}{b}\right)(1-\alpha_2)\beta
\right. \nonumber\\
& & \left.
+ \frac{1}{5}(1-\alpha_2)^2\beta^2
 \right] b^2P_m(k),
\end{eqnarray}
\begin{equation}
P_2(k)   =  
\left[
 \frac{4}{3}\left(1+\frac{\alpha_1}{b}\right)(1-\alpha_2)\beta
+\frac{4}{7}(1-\alpha_2)^2\beta^2
 \right] b^2P_m(k),
\end{equation}
and
\begin{equation}
P_4(k)  = 
\frac{8}{35}(1-\alpha_2)^2\beta^2
b^2P_m(k).
\end{equation}
The changes in these multipole moments with respect
to the redshift-distortion-only case ($\alpha_i=0$) can be clearly seen.
The monopole and quadrupole are affected by both \lya selection factors,
while the hexadecapole only has additional contributions from the dependence
on velocity gradient.

The coefficients in Equation~(\ref{eqn:PLAE_zspace}) represent the magnitude
of the \lya selection effect. As shown in Figure~\ref{fig:xi_laehalo}, the
elongation pattern along the line of sight is clearly seen in redshift space.
Since this is mainly caused by the velocity gradient term, it means that
$\alpha_2>1$. The $\alpha_1$ term represents the combined effect of 
density and transverse velocity gradient. We can imagine that there are 
cases that the dependence on density itself is strong. For example, at the 
reionization stage when ionizing bubbles do not percolate yet, there are 
isolated ionized and neutral regions. If the clustering of sources in 
overdense regions makes these regions ionize earlier, we expect a strong 
dependence of \lya observability on density (a large $\alpha_1$), which would 
substantially enhance the clustering of LAEs (Equation~(\ref{eqn:PLAE_zspace})).
We will examine this in more detail in a subsequent investigation of \lya RT
at higher redshifts where the neutral fraction of the IGM is higher
and its fluctuation larger.

In practice, it is hard to measure the 3D clustering of LAEs, since it needs 
a large spectroscopic sample of LAEs. The 2D clustering, in particular the 
angular 2PCF, is relatively easy to measure with narrow-band surveys.
The angular 2PCF is closely related to the projected 2PCF $w_p$, which is 
the 3D 2PCF $\xi(r_p,\pi)$ integrated over the line-of-sight separation,
\begin{equation}
w_p(r_p)=\int_{-\infty}^{+\infty} \xi(r_p,\pi)d\pi.
\end{equation}
The projected 2PCF is the 2D Fourier transform of the power spectrum with the
line-of-sight wave vector set to zero (see Equation (4) of \citealt{Zheng04}). 
From
equation~(\ref{eqn:PLAE_zspace}), the projected 2PCF of LAEs corresponds to 
the following 2D power spectrum
\begin{equation}
\label{eqn:projectedP}
P_g({\mathbf k_p}) = \left(1+\frac{\alpha_1}{b}\right)^2 
   b^2P_m({\mathbf k_p}),
\end{equation}
where ${\mathbf k_p}$ is the wave vector in the plane perpendicular
to the line of sight. Since $\alpha_1>0$, Equation~(\ref{eqn:projectedP}) means
that the transverse fluctuation is enhanced for LAEs, as schematically shown
in Figure~\ref{fig:angle_dep}.

This highly simplified model enables us to see how different environment 
factors related to \lya selection affect the power spectrum of LAEs. A 
realistic model
is expected to be more complicated. For example, the simple model assumes that
the selection function is a linear combination of environment variables.
From Figure~\ref{fig:denvelgrad_mass}, we see that the
selection does not have a simple linear dependence on any of the factors. 
In fact, the observed-to-intrinsic \lya luminosity ratio approximately shows an
exponential dependence on environmental variables. Furthermore, the 
selection depends on halo mass. The simple model is nevertheless useful for 
understanding features in the clustering of LAEs, and it is able to describe
the main effects of \lya selection on LAE clustering. 
The form of the power spectrum in Equation~(\ref{eqn:PLAE_zspace}) 
and in Appendix~\ref{sec:extendedtoy} may be a starting point for a fitting 
formula. In what follows we study LAE clustering in more detail and use the 
simple model to aid our understanding.

\subsection{3D 2PCFs of LAEs}

\begin{figure*}
\epsscale{1.00}
\plotone{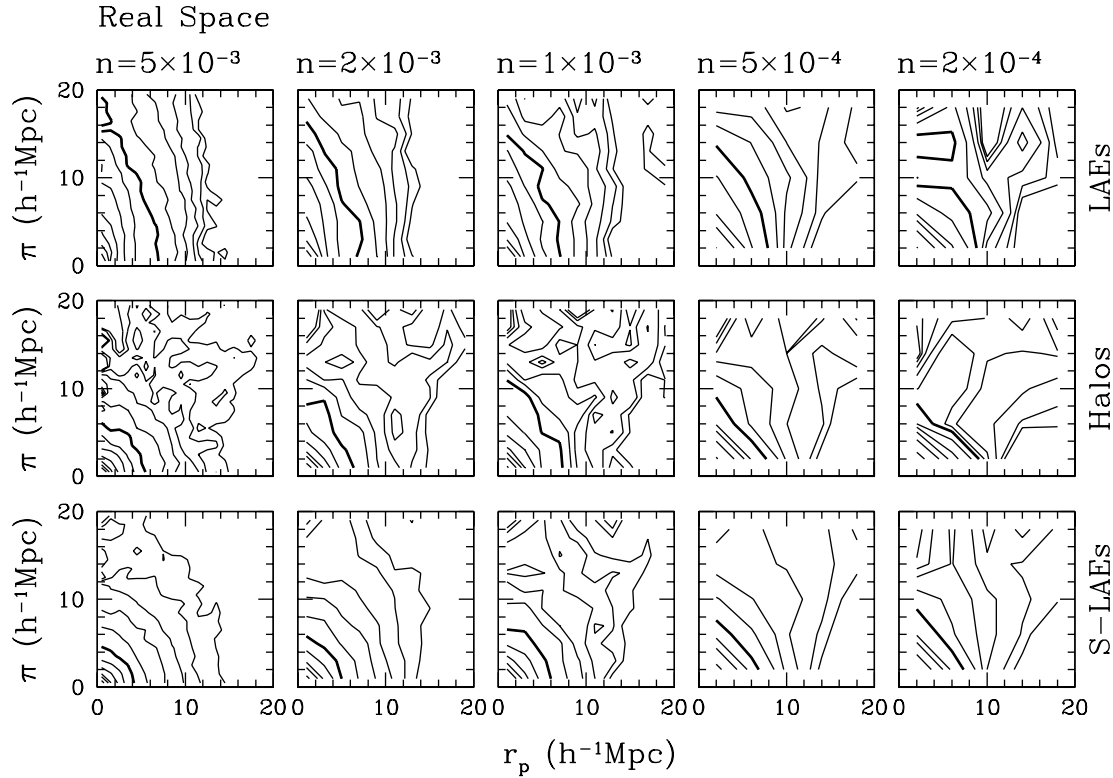}
\caption[]{
\label{fig:xi3d_realspace}
Real-space 3D two-point correlation functions $\xi(r_p,\pi)$ for threshold
samples. The top column shows the case for LAE samples. For each LAE sample, 
the middle and bottom columns show the cases for two types of control samples 
--- halos and shuffled LAE samples (S-LAEs), both having the same number 
density as the corresponding LAE sample. The shuffled LAE sample is expected
to eliminate the effect of environmental dependence of \lya RT
(see the text for details).
For LAE and S-LAE samples, thresholds 
in observed \lya luminosity are used to select sources, while for halos, the 
thresholds are in halo mass. The number density (in units of 
$h^3{\rm Mpc}^{-3}$) of each set of samples is labeled at 
the top of each column.
}
\end{figure*}

\begin{figure*}
\epsscale{1.00}
\plotone{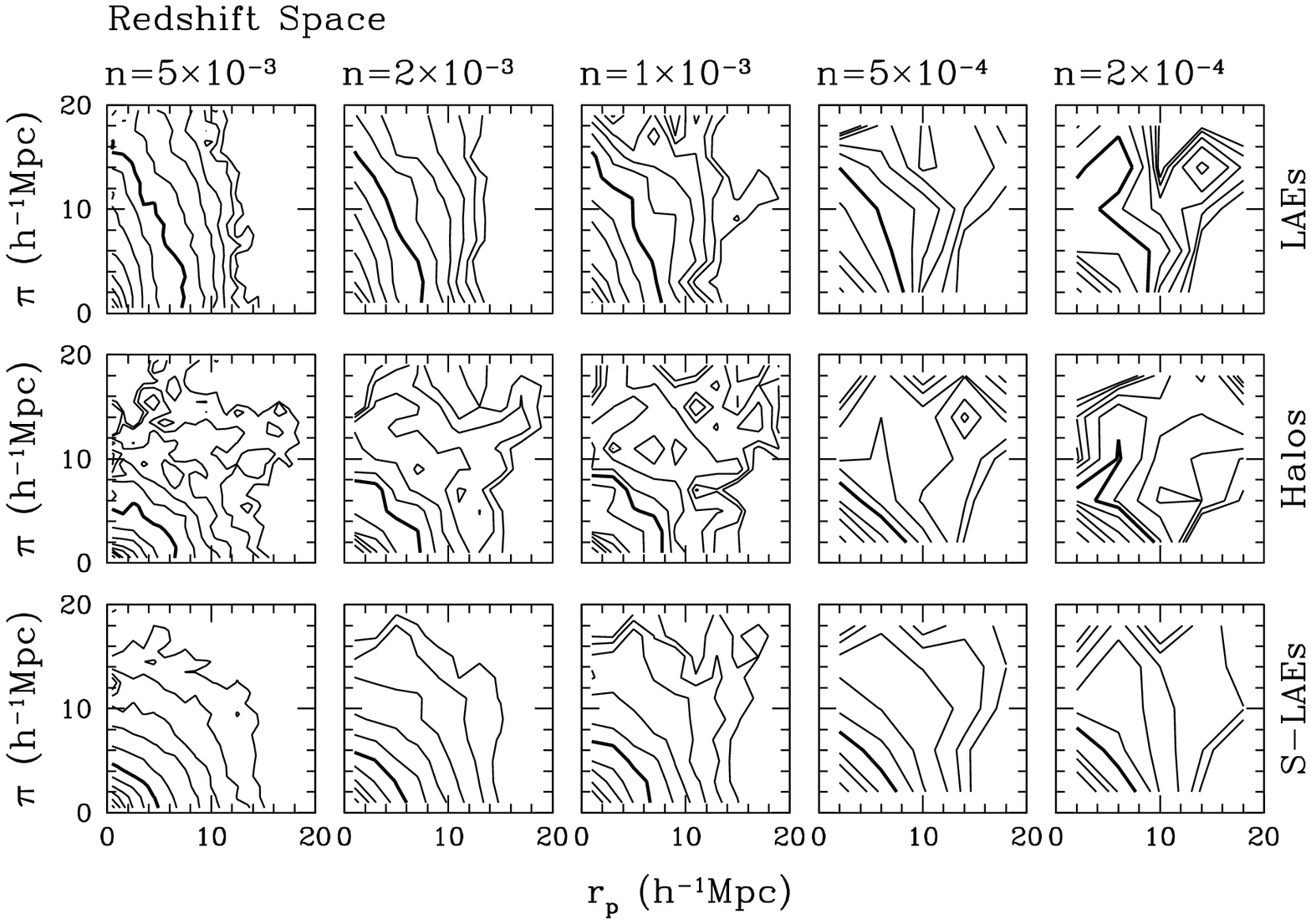}
\caption[]{
\label{fig:xi3d_zspace}
Same as in Figure~\ref{fig:xi3d_realspace}, but for redshift-space 3D two-point
correlation functions.
}
\end{figure*}

In this section, we present the 3D 2PCFs for threshold LAE samples. For each 
LAE sample, we compare the results to those of the two threshold control 
samples (halos and S-LAEs) of the same number density. 

Figure~\ref{fig:xi3d_realspace} shows the real-space 3D 2PCFs as a function
of sample number density. The 2PCFs of all LAE samples clearly show the 
elongation pattern along the line-of-sight direction, even though that of the 
$n=2\times 10^{-4}\denhMpc$ sample is somewhat noisy. The contours of the 2PCFs 
for the control samples appear to be circular except for the statistical
fluctuation at large scales. 

In redshift space, the elongation pattern in the 3D 2PCFs of LAEs is still 
preserved (Figure~\ref{fig:xi3d_zspace}), implying that the \lya selection 
effect is much stronger than the redshift-space distortion effect. For control 
samples, the contours of 2PCFs are still close to circular. Control samples
are only affected by the redshift-space distortion effect. Halos or S-LAEs 
at $z=5.7$ are highly biased, so the redshift distortion parameter 
$\beta\approx\Omega_m^{0.6}/b\sim 1/b$ is small 
(see Equation~(\ref{eqn:zdistortion})), which explains the weak redshift 
distortion seen in the 2PCFs of control samples.

For all the samples considered here, the 3D 2PCFs of LAEs are characterized
by elongation contours on scales ranging from $\sim 1$Mpc to a few tens of Mpc, a 
completely new phenomenon in galaxy clustering predicted by our model. If the 
anisotropic 3D clustering can be measured from LAE surveys, it will be a 
direct test of our model and the \lya RT selection effect.
Since the volume of the simulation we use is $10^6\VhMpc$, 
Figure~\ref{fig:xi3d_zspace} implies that a few hundred LAEs with 
spectroscopic redshifts are needed to have decent measurements of the 3D 2PCFs.
Current LAE surveys, such as the LALA survey \citep[]{Kovac07,Wang09} and 
SXDS \citep[]{Ouchi08}, are approaching this requirement.

\begin{figure*}
\plotone{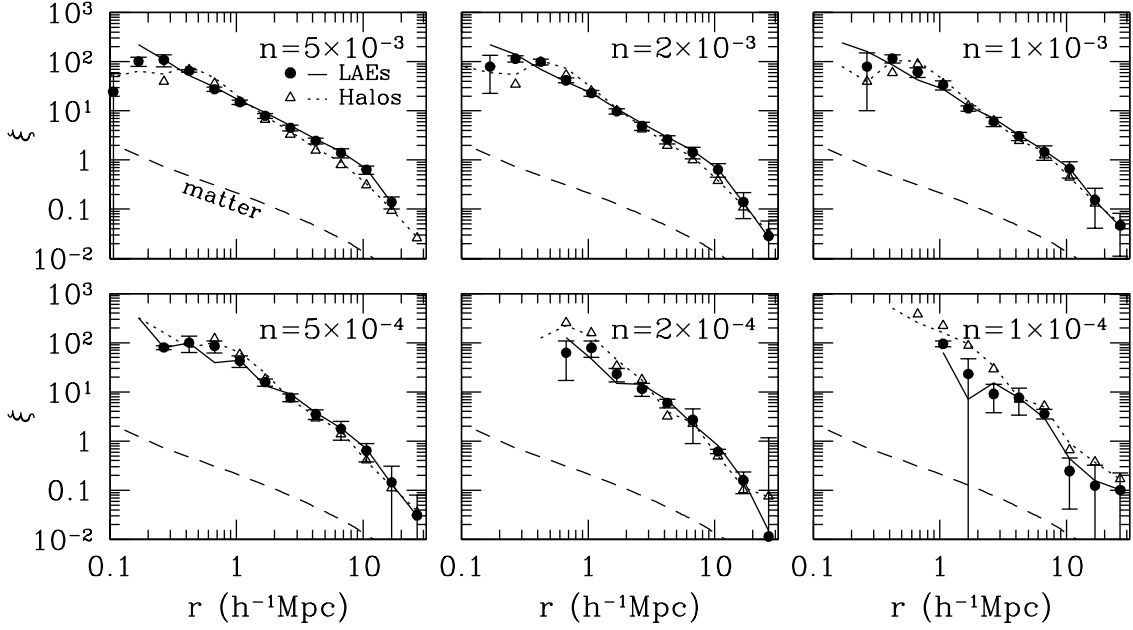}
\caption[]{
\label{fig:xi_lae_halo}
Spherically averaged two-point correlation functions of LAEs as a function
of number density. In each panel, filled circles are the real 
space two-point correlation function of LAEs, and open triangles are that
for the halo control sample with the same number density. 
LAE and halo samples are defined by thresholds in observed
\lya luminosity and halo mass, respectively. The number density
(in units of $h^3{\rm Mpc}^{-3}$) is marked in the panel. The solid and dotted
curves are from redshift space for LAEs and halos, respectively. The dashed
curve is the two-point correlation function of matter. 
}
\end{figure*}

\begin{figure*}
\plotone{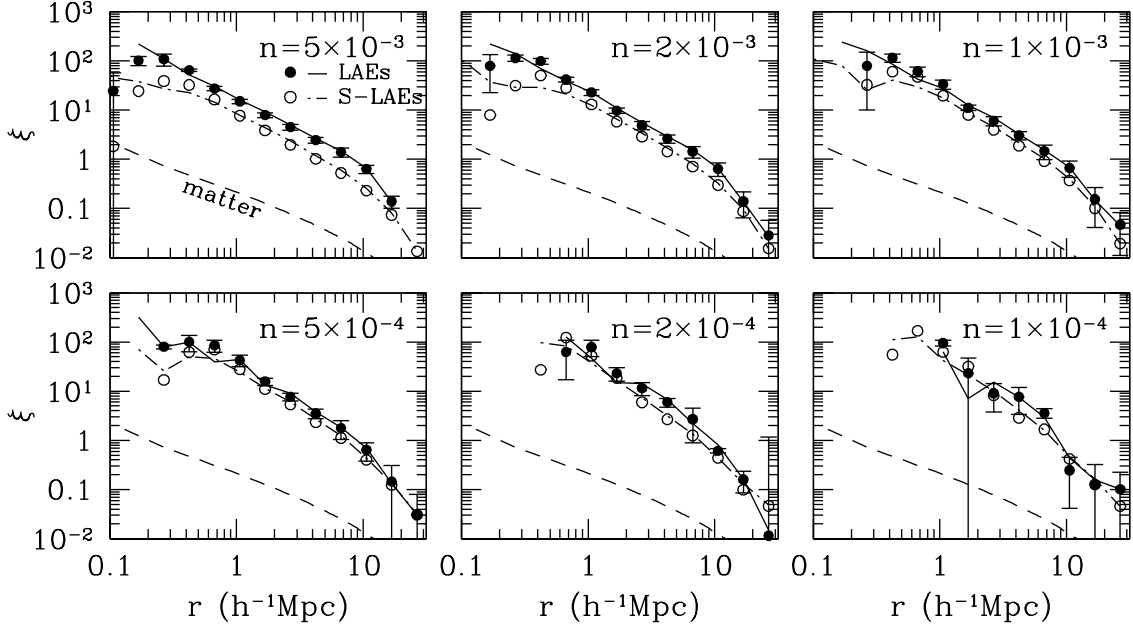}
\caption[]{
\label{fig:xi_lae_slae}
Same as Figure~\ref{fig:xi_lae_halo}, with control samples changed to 
shuffled LAE samples (S-LAEs; open circles and dot-dashed curves). 
In each panel, LAE and S-LAE sample are defined by the same thresholds 
in observed \lya luminosity. Shuffled LAE samples are expected
to eliminate the effect of environmental dependence of \lya RT
(see the text for details).
}
\end{figure*}

Figures~\ref{fig:xi_lae_halo} and \ref{fig:xi_lae_slae} show the spherically
averaged 3D 2PCFs of LAEs and control samples as a function of number density. 
The error bars are obtained with the jackknife method, whereby we divide the 
simulation box into eight octants and measure the 2PCFs by excluding one 
octant at a time.
The control samples are halos and S-LAEs in the two figures, respectively.
These plots allow a comparison of the overall amplitudes of 2PCFs. 
In each panel, points are the spherically averaged real-space 3D 2PCFs
and curves are averaged from redshift-space 3D 2PCFs. The difference 
between the real-space and redshift-space average is small, which again
demonstrates that the redshift-distortion effect is weak for sources at 
$z=5.7$. 

The spherically averaged 2PCFs of halos appear to be slightly steeper than
those of LAEs (Figure~\ref{fig:xi_lae_halo}). On large scales 
($\sim 1-10\hMpc$), LAEs are more clustered than halos for high number density
samples, a consequence of the enhancement of LAE clustering by \lya 
selection. At number density $10^{-4}\denhMpc$, the amplitude of the 2PCF of 
halos exceeds that of LAEs. The trend of the relative amplitude between LAE 
and halo samples reflects the competition of the \lya selection effect and 
halo bias. At fixed number density, LAEs can populate into halos of mass 
lower than the mass threshold that defines the halo control sample. One would
naively expect that LAEs should be less clustered than halos. However, the
selection effect from \lya RT boosts the clustering of LAEs
(Equation~(\ref{eqn:PLAE_zspace}) and Equation~(\ref{eqn:monopole})). 
So we see that LAEs
are more clustered than halos for samples with high number density. To explain 
the relative amplitude at the low number density end, we need to note that 
halo bias factor is a steeply rising function toward low number density (i.e.,
toward more massive and rare halos). At 
very low number density, the boost from \lya selection in LAE clustering 
can no longer catch up with the steep increase in halo bias factor, which
results in the reversal of the relative 2PCF amplitude.

The S-LAEs, on the other hand, seem to always have a lower amplitude in
the spherically averaged 2PCF than the LAEs of the same number density
(Figure~\ref{fig:xi_lae_slae}). S-LAE samples are fairer control samples
than halo samples, since they share the same statistical properties of 
\lya emission with the LAE samples except for the environmental effect on
\lya RT. The LAE and S-LAE samples of the same number density
have the same underlying population. Setting the coefficients $\alpha_1$ and 
$\alpha_2$ to zero in the simple physical model (e.g., 
Equation~(\ref{eqn:PLAE_zspace}) and Equation~(\ref{eqn:monopole})) 
corresponds to the cases of S-LAEs. From the
S-LAE and matter 2PCFs in Figure~\ref{fig:xi_lae_slae}, the bias factor $b$ 
is of the order of 10, so the redshift distortion parameter $\beta$ is at the 
level of 0.1. The monopole of the power spectrum 
(Equation~(\ref{eqn:monopole})) is 
likely to be dominated by the first term, which is from the density dependence.
The LAE sample therefore always has enhanced clustering with respect to the 
corresponding S-LAE sample. As the bias factor $b$ in 
Equation~(\ref{eqn:monopole})
increases with decreasing sample number density, we expect the enhancement 
effect to become weaker at lower number density, consistent with what is seen
in Figure~\ref{fig:xi_lae_slae}. For $n=10^{-4}\denhMpc$, the 2PCFs of the 
LAE and S-LAE samples are quite close to each other.

\subsection{Projected/Angular 2PCFs of LAEs}

\begin{figure*}
\plotone{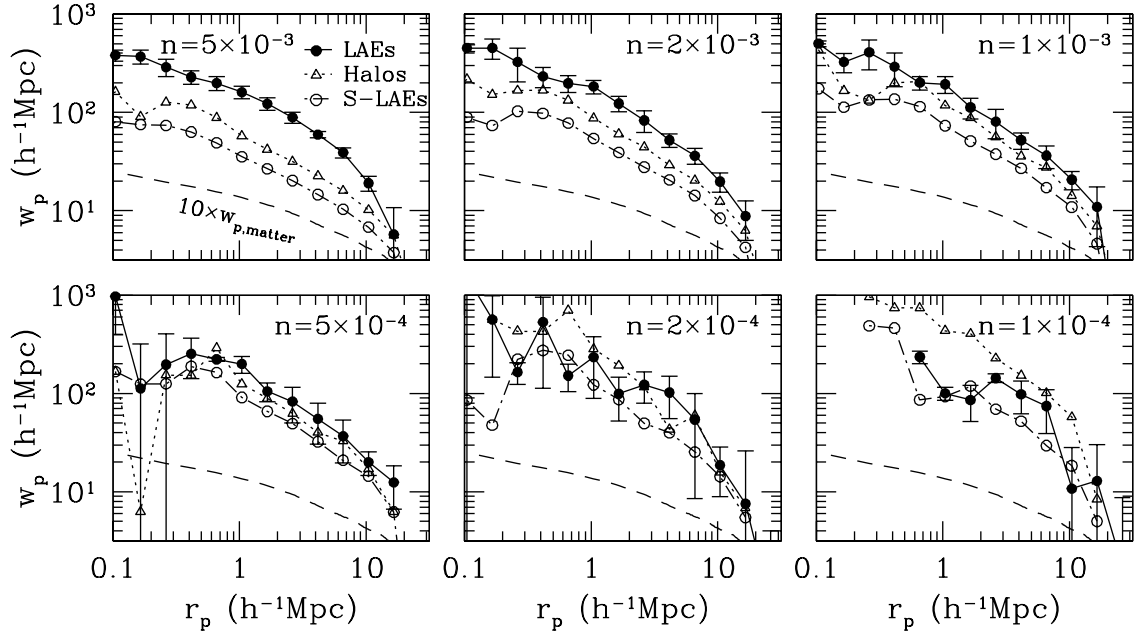}
\caption[]{
\label{fig:wp}
Projected two-point correlation functions $w_p$ of LAEs as a function
of number density. In each panel, besides the projected two-point 
correlation function of LAEs (filled circle), we also plot those for
the halo control sample (open triangles) and the shuffled LAE sample 
(S-LAEs; open circles) of the same number density (marked in the
panel in units of $h^3{\rm Mpc}^{-3}$). LAE/S-LAE and halo samples are defined 
by thresholds in observed \lya luminosity and halo mass, respectively. 
The dashed curve is the projected two-point correlation function of matter,
scaled by a factor of 10.
}
\end{figure*}

\begin{figure*}
\plottwo{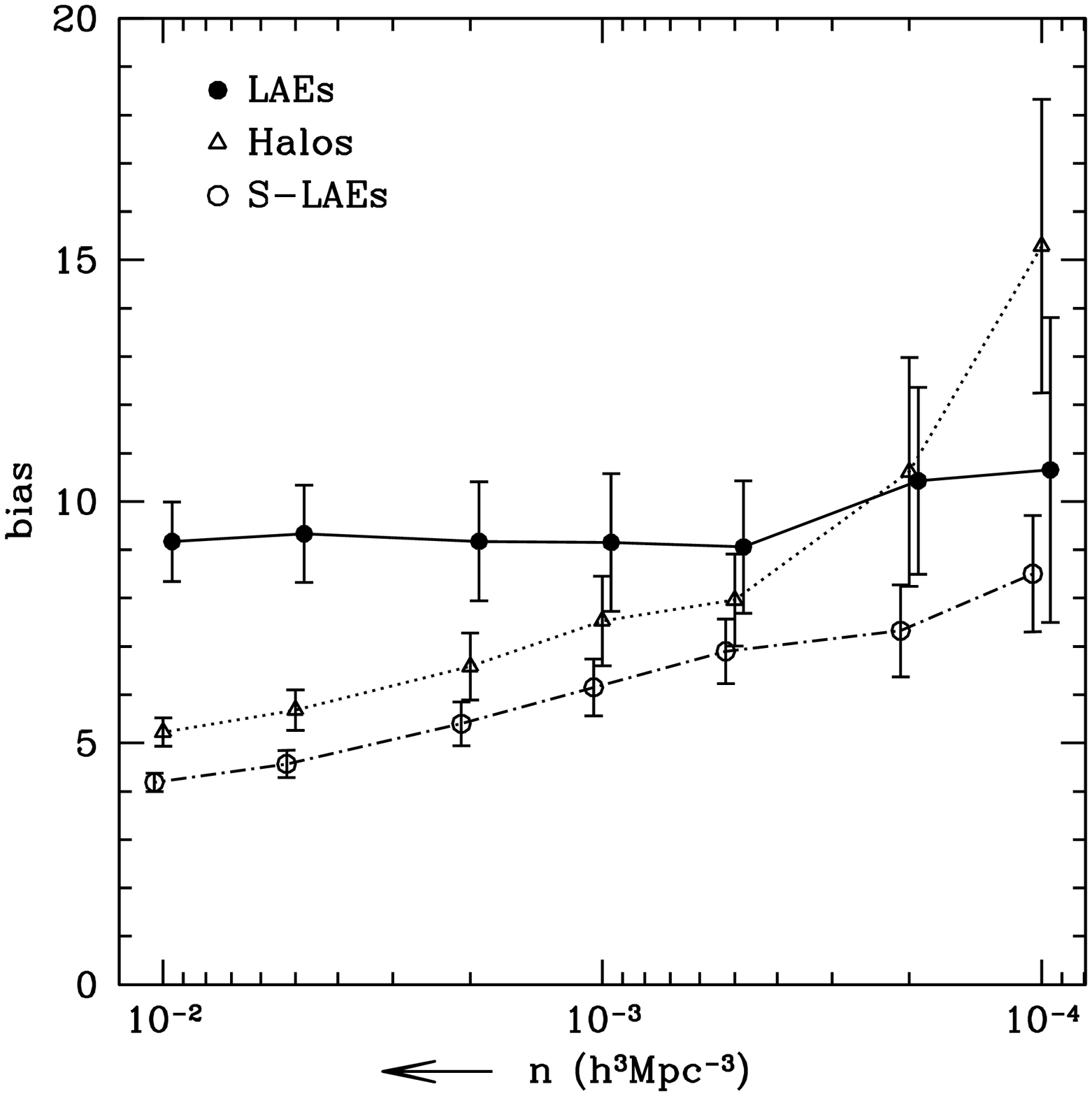}{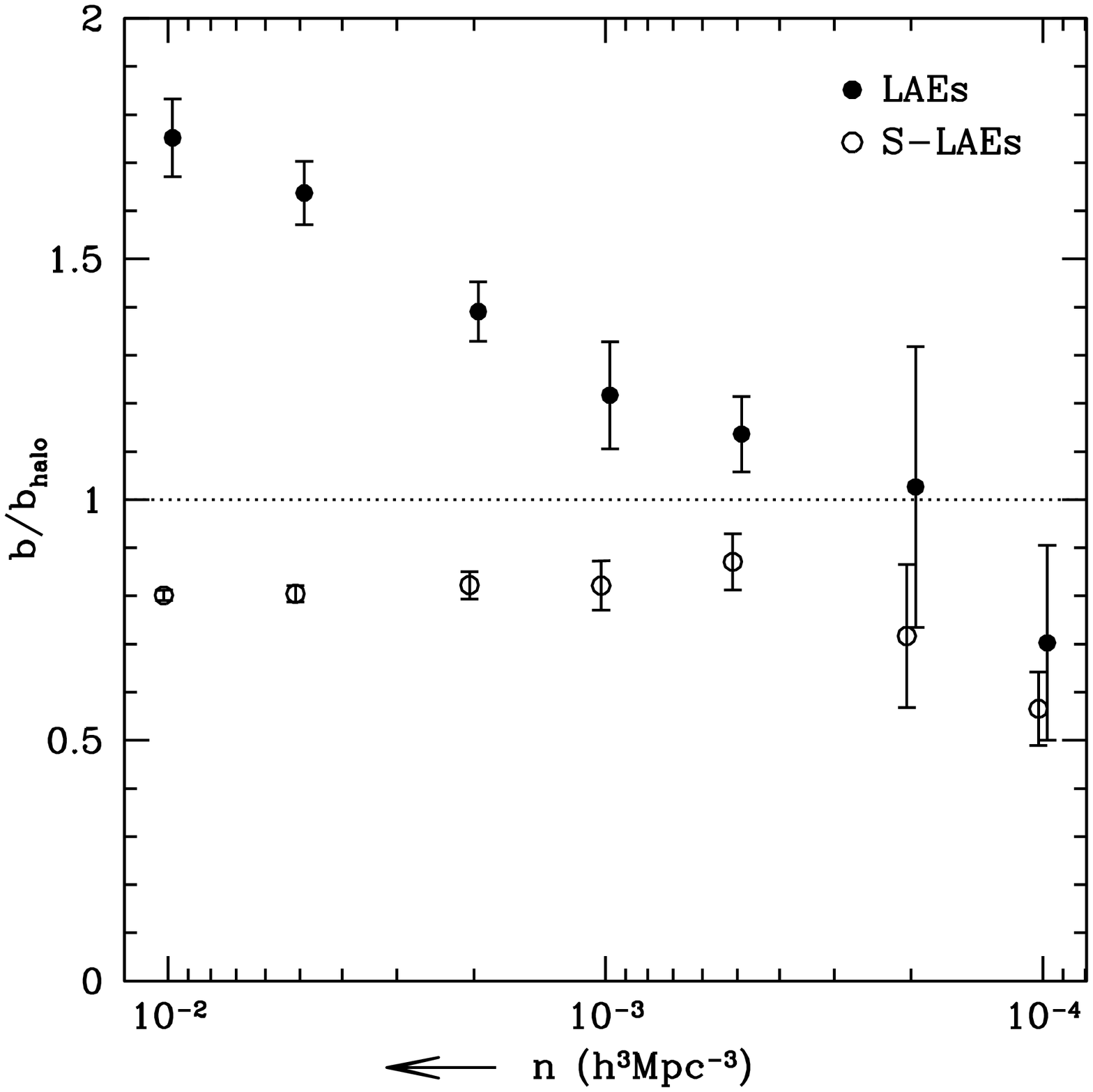}
\caption[]{
\label{fig:biasfactor}
Bias factors of LAEs, shuffled LAEs (S-LAEs), and halos as a function of 
sample number density.
Left: absolute bias factor. Right: bias factor relative to halos.
The relative bias factor is calculated as the square root of 
the ratio of the projected two-point correlation functions $w_p$ in 
Figure~\ref{fig:wp} of LAEs/S-LAEs and halos, averaged over the range of 
1--10$\hMpc$. The absolute bias factor is similarly calculated by using the 
projected matter 2PCF in Figure~\ref{fig:wp}. For clarity, the points for 
bias factors of LAEs and S-LAEs are slightly shifted horizontally. 
}
\end{figure*}

\begin{figure}
\plotone{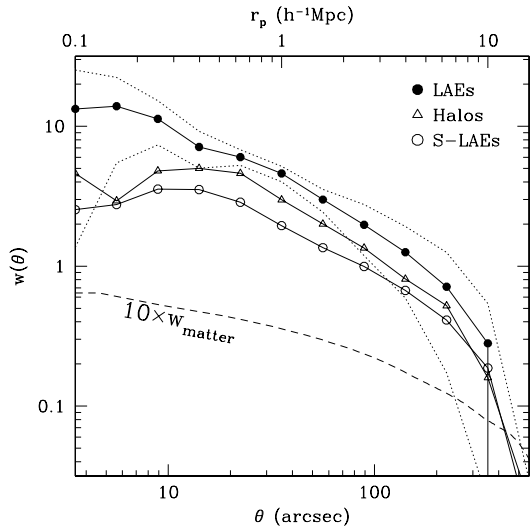}
\caption[]{
\label{fig:subaru1}
Prediction of the angular two-point correlation function for the $z\sim$5.7 
LAE sample in \citet{Ouchi08}. The threshold sample has a number density of 
$1.2\times10^{-3}\denhMpc$.  The predicted angular two-point correlation
function of this sample of LAEs is shown as filled circles. The two dotted 
curves represent the 1$\sigma$ scatter expected in a survey with SXDS-like 
volume. Triangles and open
circles are the two-point correlation functions of halos and shuffled
LAEs with the same number density as the LAE sample. The dashed curve
is the matter correlation function, scaled by a factor of 10.
}
\end{figure}

\begin{figure}
\plotone{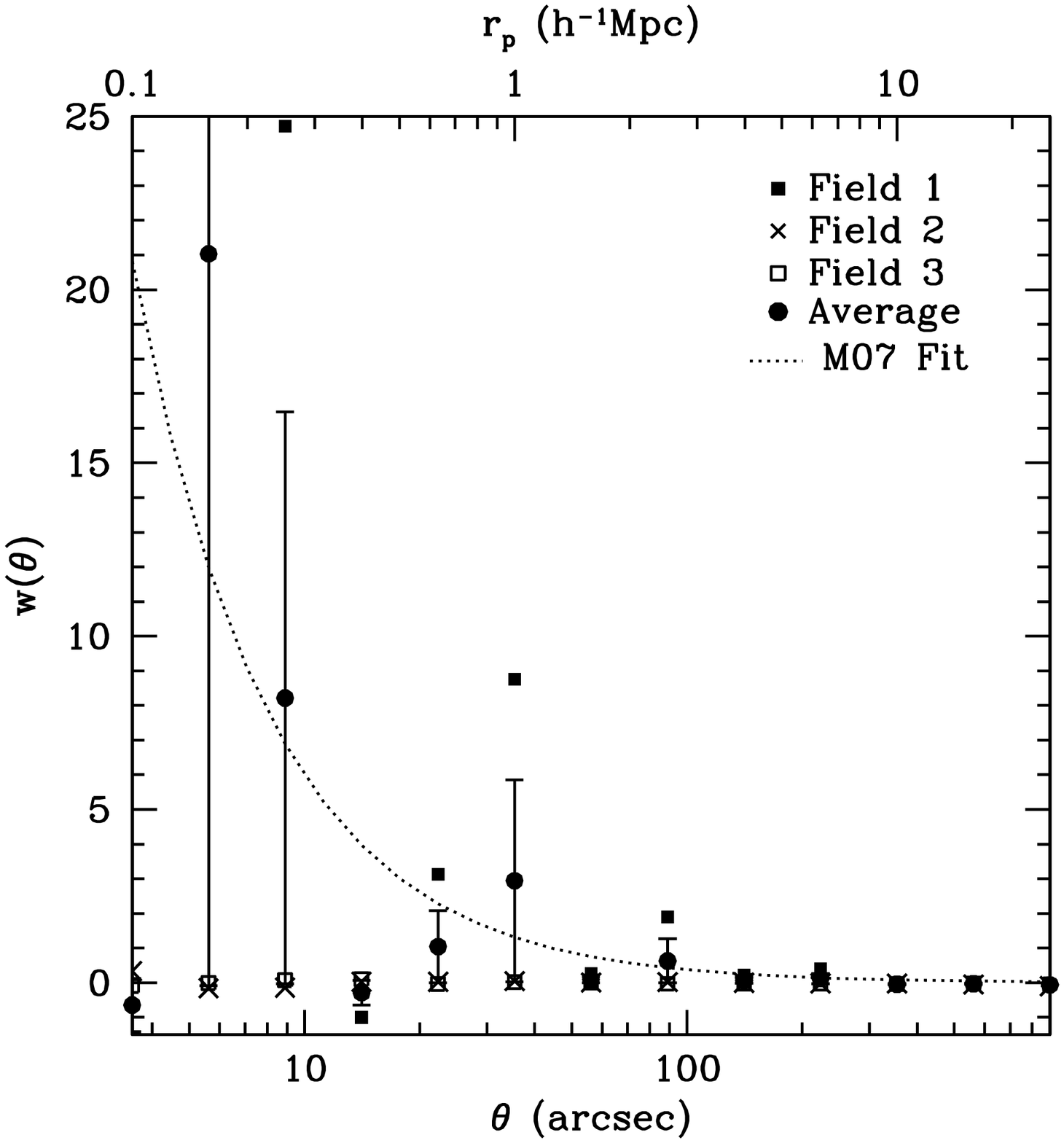}
\caption[]{
\label{fig:subaru2}
Sample variance of the angular two-point correlation function for $z\sim$5.7 
LAE sample in SXDS-like volume. The sample has a number density of 
$\sim 2\times 10^{-4}\denhMpc$, comparable to the sample in 
\citet{Murayama07}. Filled squares, crosses, and open squares are the
angular two-point correlation functions measured in three fields from the
simulation, showing a large sample variance. Filled circles with error bars
are the mean of the three measurements. The dotted curve is the best-fit
power law to the measured LAE two-point correlation function in 
\citet{Murayama07}.
}
\end{figure}

The angular 2PCFs of LAEs can be measured with LAEs identified in narrow-band 
surveys. On scales smaller than the length scale of the redshift range of
the survey, which is set by the width of the narrow-band filter, the angular 
2PCF is closely related to the projected 2PCF, or the projected power spectrum 
(Equation~(\ref{eqn:projectedP})). Here we present the projected 2PCFs of LAE 
samples and the corresponding control samples in Figure~\ref{fig:wp} 
(with the error bars in the LAE sample estimated from the jackknife method).

The \lya RT effect boosts the transverse density fluctuation, leading to an 
enhancement in the projected 2PCFs. In the simple physical model, the 
enhancement comes from the density dependence of the selection function
(Equation~(\ref{eqn:projectedP})), which is an
effective description of the complicated dependences of RT on density and 
line-of-sight and transverse velocity gradients (see 
Appendix~\ref{sec:extendedtoy}). The trends in the relative 
amplitudes between the projected 2PCFs of LAEs and halos and between the 
projected 2PCFs of LAEs and S-LAEs are similar to those seen in the 
spherically averaged 2PCFs, as shown in Figure~\ref{fig:wp}. 

To quantify the amplitude of the projected 2PCFs of LAE and control
samples, we compute an average bias factor of LAEs, S-LAEs, and halos 
for each sample number density. We measure the projected 2PCF of matter 
from the matter density field in the simulation (dashed curves in
Figure~\ref{fig:wp}). The average bias factor of LAEs/S-LAEs/halos is taken 
to be the square root of the ratio of the projected 2PCFs of LAEs/S-LAEs/halos 
and matter, averaged over the range of 1--10$\hMpc$. It turns out that the
ratio is not constant at the above scales, being smaller toward larger scales. 
The error bars we assign 
to the average bias factors largely reflect this scale dependence.

The left panel of Figure~\ref{fig:biasfactor} shows the bias factors of 
LAEs, S-LAEs, and halos, as a function of sample number density. Plotted in 
the right panel are the bias factors of LAEs/S-LAEs relative to halos.
At the same number density, S-LAEs can occupy halos with mass lower than 
the threshold mass of the halo sample, therefore S-LAE samples always show 
a lower bias factor than halos. The relative bias factor of S-LAEs and 
halos is roughly a constant ($\sim$0.8) for $n=10^{-2}$--$10^{-3}\denhMpc$.
It decreases toward lower number density and reaches $\sim$0.6 at 
$n=10^{-4}\denhMpc$. 

It is interesting that the bias factor for LAE samples does not show strong
dependence on sample number density. The value is around 10 for sample 
number density in the range of $10^{-4}$--$10^{-2}\denhMpc$ and increases 
slightly toward the low number density end. However, the bias of LAEs relative 
to halos continuously decreases toward low number density (filled circles in 
the right panel), from $\sim$1.75 at $n=10^{-2}\denhMpc$ to $\sim$0.7 at 
$n=10^{-4}\denhMpc$.

The weak dependence of LAE clustering on sample 
number density is a result of the selection effect caused by the environment 
dependent \lya RT. In the simple physical model, the amplitude of the
projected 2PCF is proportional to $(b+\alpha_1)^2$ 
(Equation~(\ref{eqn:projectedP})).
As the number density decreases, the bias factor $b$ increases. The weak 
dependence of the amplitude on number density would require that $\alpha_1$ 
drop with decreasing number density, which seems to be consistent with the
halo mass dependence seen in Figure~\ref{fig:denvelgrad_mass}. 
The prediction can be tested with LAE clustering measurements as a function of 
observed \lya luminosity. In Appendix~\ref{sec:effect_lineprof}, we show how 
the weak dependence changes if there is a large dispersion in the intrinsic 
\lya luminosity and halo mass relation.

When interpreting observational results of galaxy clustering, 
luminosity-threshold samples of galaxies are usually modeled as mass-threshold 
halo samples for a simple inference of the relation between galaxies and halos.
If the same exercise were applied to Ly$\alpha$-luminosity-threshold samples 
of LAEs,
our model would have interesting implications. For high number density samples 
($n>5\times 10^{-4}\denhMpc$), LAEs are more strongly clustered than halos in
the corresponding mass-threshold halo control sample. From the clustering 
amplitude, one would relate the LAE sample to a halo sample with higher mass 
threshold. However, the number density of these halos are less than the LAE 
sample, which would imply that there have to be multiple LAEs per halo. For 
example, the bias factor of $n=2\times 10^{-3}\denhMpc$ LAEs is similar to 
that of $n=4\times 10^{-4}\denhMpc$ halos (left panel of 
Figure~\ref{fig:biasfactor}). From the clustering amplitude, we would infer 
that these LAEs reside in halos above $2\times 10^{11}\hMsun$ 
($n=4\times 10^{-4}\denhMpc$), and comparison of the halo number density and
LAE number density would make us infer that there are on average five LAEs 
per halo. If one LAE per halo were assumed, from the number density alone, 
we would infer that these LAEs reside in halos above $7\times 10^{10}\hMsun$.
For low number density samples 
($n<5\times 10^{-4}\denhMpc$), LAEs are less clustered than halo samples,
and one would infer a duty cycle of LAEs by combining the clustering result
and the number density. We see that the simple modeling of LAEs as 
mass-threshold halo samples can lead to incorrect inference of
the relation between LAEs and halos. 
Therefore, a proper interpretation of LAE clustering
requires a careful account of the large \lya selection effect.

At present, the largest LAE sample at $z=5.7$ comes from the SXDS 
\citep{Ouchi08}.  If we divided the simulation box along our
chosen line of sight into three layers of equal size, the depth of each 
layer, 33.33$\hMpc$, is close to that from the width of the narrow-band filter 
used to search for $z\sim$5.7 LAEs in the SXDS, and the area is almost 
identical to that of the survey (1deg$^2$). Therefore, the simulation box 
gives us three SXDS-like volumes at $z\sim$5.7.

The photometric $z\sim$5.7 LAE sample in \citet{Ouchi08} corresponds to
a luminosity-threshold sample of number density 
$1.2\times 10^{-3}\denhMpc$. We construct a threshold sample of model LAEs 
that has the same number density and use the Landy--Szalay estimator 
\citep{Landy93} to measure the angular 2PCFs in the three SXDS-like volumes,
respectively. We also perform similar measurements for the halo and S-LAE
control samples. In Figure~\ref{fig:subaru1}, points connected by solid 
curves are the 2PCFs of LAEs, halos, and S-LAEs, respectively, averaged over 
the three SXDS-like volumes. Similar to the projected 2PCFs, LAEs and S-LAEs
have the highest and lowest angular clustering amplitudes, respectively.
The two dotted curves show the variance of the angular 2PCFs of LAEs from 
individual measurements in the three volumes. Even with the large variance on 
large scales, the SXDS clustering measurement is likely to reveal that LAEs are
more strongly clustered than halos of the same number density.\footnote{
\citet{Ouchi10} present the angular 2PCF of the $z\sim 5.7$ LAE sample. On 
scales above 100\arcsec, the measurement agrees with our prediction. However,
below 100\arcsec, the measured 2PCF is essentially flattened, which is not
seen in their samples at lower and higher redshifts.  
It may be caused by sample variance or may have interesting 
implications, which merits further investigations.
}

\citet{Murayama07} present an angular 2PCF measurement of $z=5.7$ LAEs in 
the Cosmic Evolution Survey (COSMOS). The sample number density is about
$2\times 10^{-4}\denhMpc$ and the effective area for the sample is 1.86deg$^2$.
They find signals at several scales, but with low significance.
The clustering results for this sample in our model are shown in 
Figure~\ref{fig:subaru2}. Apparently there is a 
large variance among samples in the three volumes. The power-law fit from 
\citet{Murayama07} (dotted curve) falls in between the three measurements.
The \citet{Murayama07} sample is only 1.86 times larger in effective area than 
our sample, so we still expect a large sample variance. Large sample
variance for low density samples in a small-area survey are reported by 
\citet{Shimasaku04} at $z\sim 5$.  They measure angular 2PCFs of LAEs at 
$z=4.79$ and $z=4.86$ in a Subaru Deep Field of $\sim$0.3deg$^2$ and find 
clustering signals in one sample and the lack of clustering in the other.
Based on the mean and uncertainty from our simulation data for the
$2\times 10^{-4}\denhMpc$ sample, we estimate that a survey about 10 times 
larger is required for a solid detection ($>3\sigma$) of the clustering 
signal, which means
a $\sim$10 deg$^2$ survey. For the (relatively small) scales considered here 
($r_p\lesssim 20\hMpc$ or $\theta\lesssim 0.2$deg), the survey can be composed 
of either several separated fields with a few square degrees each or one 
single contiguous field. With the same total area, the former case would have 
a slightly smaller number of LAE pairs on the above scales because of the 
edge effect, so a single field is slightly preferred. The upcoming Subaru 
Hyper Suprime-Cam (HSC) survey will meet such a requirement.

Based on the above investigation, we conclude that: in the SXDS-like field,
large sample variance in angular 2PCFs is expected for LAE samples with 
low number density (e.g., lower than a few times $10^{-4}\denhMpc$); for 
the LAE sample in \citet{Ouchi08}, it is likely that the enhancement in
the transverse clustering of LAEs caused by the \lya selection effect can
be detected. If future observation did not detect this effect and the null
detection could be firmly established, it would have profound implications
in our understanding of LAEs. It would mean that much stronger effects 
mask the \lya RT selection effect, which could be extremely large scatter in 
the distribution of intrinsic \lya properties caused by stochastic star 
formation and/or dust (see further discussions in Section~\ref{sec:discussion} 
and Appendix~\ref{sec:effect_lineprof}).

\section{HOD of LAE\protect\lowercase{\rm s} and the Environment Dependence}
\label{sec:hod}

\begin{figure*}
\plotone{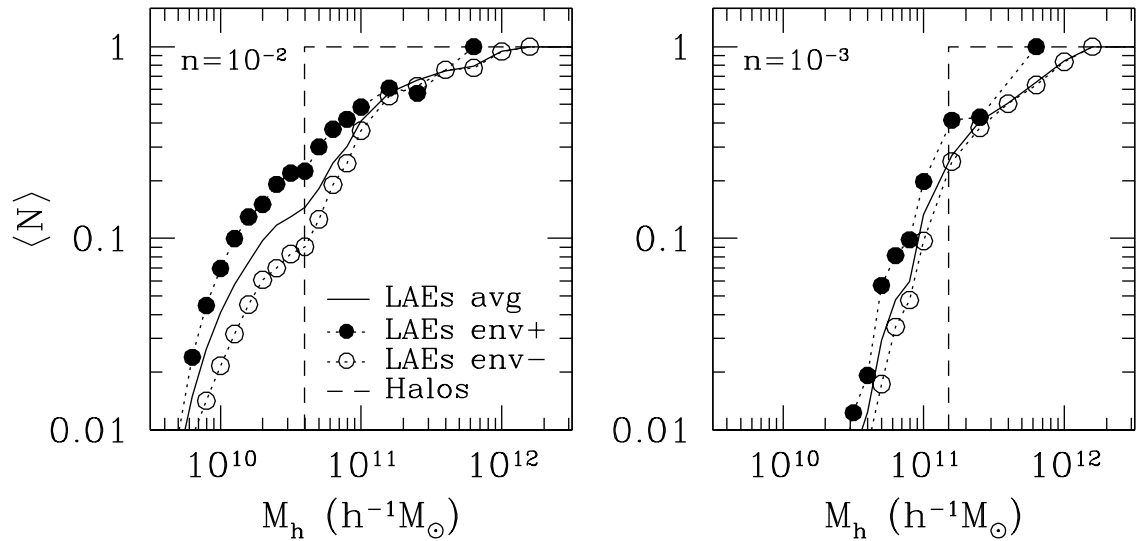}
\caption[]{
\label{fig:hod_env}
Environment dependence of the halo occupation distribution of LAEs.
LAE samples are defined by thresholds in observed \lya luminosity. The left
and right panels are for LAE samples with number density of $10^{-2}$
and $10^{-3}\denhMpc$, respectively. The line-of-sight velocity gradient
$\partial v_z/\partial z$ is used here as the environment indicator and 
halos are grouped according to the value of $\partial v_z/\partial z$. 
In each panel, filled (open) circles show the mean occupation function of 
LAEs in halos that are in the upper (lower) quartile of the 
$\partial v_z/\partial z$ distribution. The solid curve is the mean 
occupation function of LAEs over all environments. The dashed curve is the
mean occupation function of a threshold halo sample with the same number 
density.
}
\end{figure*}

The HOD framework \citep[e.g.,][]{Berlind02}
has become a powerful tool to interpret galaxy clustering. 
It describes the probability distribution
of number of galaxies of a given type in a halo as a function of halo mass,
together with the spatial and velocity distributions of galaxies inside halos.
The HOD framework recasts galaxy clustering in terms of the 
relation between galaxies and dark matter halos. The link from galaxies to dark 
matter halos enables informative tests of galaxy formation model and
tight constraints on cosmological parameters from galaxy clustering.

The basic assumption in the current version of the HOD framework is that the
statistical properties of galaxies only depend on halo mass and is independent 
of the large-scale environments. For normal galaxies (i.e., galaxies selected 
from continuum or optically-thin lines), the assumption of 
environment-independent HOD is supported by theory (e.g., \citealt{Berlind02};
but see \citealt{Zhu06}), observation \citep[e.g.,][]{Blanton06}, and 
clustering analysis \citep[e.g.,][]{Tinker08}.
Together with the properties of halo 
population of a given cosmology, the halo-mass-dependent HOD can lead to a full 
description of galaxy clustering. However, LAEs that are selected from \lya
emission suffer from environment-dependent \lya RT effect,
so the assumption of environment-independent HOD is expected to break down 
for LAEs.

Figure~\ref{fig:hod_env} shows the environment dependence of the HOD of LAEs.
Since the line-of-sight velocity gradient is the dominant variable in 
determining the observed-to-intrinsic \lya luminosity ratio, we use it
to define the environment for the HOD investigation. In each panel, the dashed
curve (step function) represents the case for the halo control sample. The 
mass threshold can be read off from the plot. The solid curve is the mean 
occupation function of LAEs, averaged over all environments. It is also the
mean occupation function of S-LAEs. 

In our model, LAEs are related to halos, not sub-halos, and there is one \lya 
emitting source per halo, which means that we do not have satellite LAEs. 
This is a good approximation, since there are few group-size halos at $z=5.7$. 
The mass of the most massive halo in the simulation box is 
$\sim 4\times 10^{12}\hMsun$, with a virial radius of $\sim 0.4\hMpc$.
Ignoring the (low) satellite fraction of LAEs only affects the clustering on
small scales (e.g., tenths of Mpc). The mean occupation functions of LAEs in
Figure~\ref{fig:hod_env} are only for central LAEs and reach unity at the 
high-mass end. 

As mentioned in Paper I, in our LAE model, the intrinsic \lya luminosity is
proportional to halo mass (Equations (1) and (2) in Paper I). 
If LAEs were selected based on intrinsic \lya
luminosity, they would be equivalent to the halo sample. Observationally, 
LAEs are selected based on their apparent (observed) \lya luminosity. We show
in Paper I that the environment-dependent \lya RT leads to a 
broad distribution of apparent \lya luminosity at a fixed intrinsic \lya
luminosity. Observation detects LAEs with observed \lya luminosity above
a threshold. This means that for sources at a fixed intrinsic \lya 
luminosity (or halo mass), only a fraction of them can be detected as
LAEs. The mean occupation function is simply this fraction.
From low to high halo mass, the mean occupation function appears as a ramp 
from zero to unity (solid curves in Figure~\ref{fig:hod_env}). With the
occupation number below unity, to maintain the same number density as the 
control halo sample, a fraction of LAEs resides in halos of mass lower
than the threshold mass of the halo sample. The cutoff profile of the 
LAE mean occupation function is similar to that of a luminosity threshold 
sample of continuum-selected galaxies \citep[e.g.,][]{Zheng05,Zheng07}, 
although the profile for LAEs are much more shallow. 
The shallow cutoffs for LAEs and for continuum-selected galaxies can be 
attributed to the same origin: for continuum-selected galaxies, the shallow 
cutoff reflects the scatter in galaxy luminosity at a fixed halo mass, while 
for LAEs the apparent \lya luminosity has a distribution at fixed halo mass.

The environment dependence of the HOD of LAEs is demonstrated by the
filled and open circles in Figure~\ref{fig:hod_env}.  The filled (open)
circles are the mean occupation function of LAEs in halos in the upper (lower)
quartile of the $\partial v_z / \partial z$ distribution. Since the 
observed-to-intrinsic \lya luminosity ratio is positively correlated with
the velocity gradient $\partial v_z / \partial z$, a higher fraction of
LAEs in halos with higher $\partial v_z / \partial z$ can be detected,
which leads to a higher mean occupation function of LAEs (filled circles) 
in these halos. The difference between the mean occupation numbers in 
the upper and lower quartiles of halos are about a factor of 2--3 for the
two samples in Figure~\ref{fig:hod_env}. 

If large-scale overdensity $\delta$ is added as another environment variable,
halos can be grouped according to the joint distribution of $\partial v_z / 
\partial z$ and $\delta$ (Figure~\ref{fig:jointdep}). We consider two extreme
environments. We divide halos in the upper quartile of 
$\partial v_z / \partial z$ according to the value of $\delta$ and keep halos 
in the upper quartile of $\delta$ as the first group. We then divide halos 
in the lower quartile of $\partial v_z / \partial z$ according to the value 
of $\delta$ and keep halos in the lower quartile of $\delta$ as the second 
group. The two groups of halos
correspond to the upper-right and lower-left corners of the shaded region
in Figure~\ref{fig:jointdep}. As the mean observed-to-intrinsic \lya 
luminosity ratios for halos in these two groups differ a lot, so does
the mean occupation functions of LAEs in the two group of halos. We find that
the mean occupation numbers can differ by more than one order of magnitude.

The strong environment dependence of the HOD of LAEs adds complexity to the
modeling of their clustering. In fact, from the difference in the clustering 
of the LAE sample and the halo control sample, we know that the simple model 
of one LAE per halo with a mass threshold does not work. The difference in
the LAE and S-LAE clustering means that the environment-independent HOD
model is not a good model, either. To fully describe the features in the LAE
clustering, the environment dependence has to be incorporated into the HOD
model.

\section{Conclusion and Discussion}
\label{sec:discussion}

We investigate the clustering of LAEs, galaxies that are selected by their 
\lya emission, within a physical model that fully accounts for the \lya 
RT. Our model of LAEs combines radiation-hydrodynamic
cosmological reionization simulations with Monte Carlo RT 
for \lya photons. It is a simple model, which assumes that RT
is the single factor in transforming intrinsic \lya emission properties 
to observed ones. As previously shown in Paper I, the simple model is able to explain 
an array of observational properties of $z=5.7$ LAEs. The model has strong
predictive power, and we predict the clustering properties of LAE clustering in this paper.

\lya RT depends on the circum-galactic and inter-galactic
environments of \lya emitting galaxies. \lya photons
emitted from a source at a halo center see a complex density and velocity 
structure and an anisotropic optical depth distribution. \lya photons tend to 
find the easiest way out, and the observed \lya emission is usually 
anisotropic. {\it We emphasize that the observed \lya flux at a given line 
of sight is not purely determined by the line-of-sight optical depth, but 
the line-of-sight optical depth relative to those in all other directions
(see Section~\ref{sec:environ}).}
Because of the resonance nature of the \lya line and the large scattering 
cross-section, the observed \lya emission is highly 
sensitive to the local environments, including the matter density, 
line-of-sight velocity, their line-of-sight gradients, and the 
velocity gradients in the transverse directions (not all of these 
environmental variables are independent). 
It is this strong coupling between observed \lya emission and environments that 
gives rise to new effects in LAE clustering that need to be taken into
account in interpreting the clustering of observed LAE samples.

The overall effects in clustering caused by the \lya selection are anisotropic 
clustering and scale-dependent galaxy bias. For density fluctuations along the 
line of sight, \lya selection leads to a higher probability in detecting LAE 
sources in underdense regions than in overdense regions, so it suppresses the 
line-of-sight density fluctuations. For density fluctuations perpendicular to 
the line of sight, the anisotropic distribution of \lya emission makes sources 
in overdense regions preferentially selected, hence the transverse density 
fluctuations are enhanced. Roughly speaking, filamentary or planar structures 
tend to be preferentially selected when they are parallel to the line of sight.

The suppression of line-of-sight fluctuations and the enhancement of 
transverse fluctuations create anisotropy in LAE clustering.
The iso-contour curves in the 3D 2PCFs, which reflect LAE pair counts as a 
function of line-of-sight and transverse separations, show a distinct 
elongation pattern along the line of sight. The elongation appears on all 
scales where we have reliable measurements of the 2PCFs, ranging from sub-Mpc 
to over 10 Mpc. The anisotropic pattern is opposite to the linear 
redshift-space distortion effect (Kaiser effect), which makes contours 
squashed along the line of sight. We emphasize that the anisotropy caused 
by \lya selection originates in real space. At $z=5.7$, the cases we consider 
in this paper, the Ly$\alpha$-selection-induced distortion in LAE clustering 
is much stronger than the linear redshift-space distortion. 
Therefore, even in redshift space the elongation pattern along the line 
of sight in the 3D 2PCFs is well preserved. 
We note that the elongation also differs from the FoG 
effect seen in galaxy clustering, which is caused by random motions of 
galaxies in virialized structures. While the FoG effect only shows up in 
redshift space and on small scales (e.g., $\lesssim$ 1Mpc), the \lya selection 
effect can appear in both real and redshift spaces and on much larger scales. 
The \lya RT induced features are not ``Fingers of God'' but ``Arms of God''.

The anisotropic clustering induced by \lya selection is a completely new
phenomenon in galaxy clustering.
Other than the usual redshift-space distortion (FoG and Kaiser effect), 
there are other forms of anisotropic clustering discussed in the literature.
\citet{Padilla05} study the cross-correlation between voids and galaxies. 
In the redshift-space two-point cross-correlation function of voids and 
galaxies, an elongation pattern along the line of sight is found on large 
scales (up to a few times the radius of voids in the sample). Unlike the
\lya selection effect, which is a real-space effect, the phenomenon in 
the void-galaxy cross-correlation is a redshift-space effect. It simply 
reflects that galaxies tend to stream out of void regions.
The closest analogy to the \lya selection effect we study is the effect of 
orientation-dependent galaxy selection investigated by \citet{Hirata09}, in 
the case that galaxies are aligned by large scale tidal fields. 
Both effects can be attributed to an environment-dependent surface brightness 
selection and act in both real and redshift spaces. 
However, the \lya RT selection effect is much stronger. While the
\citeauthor{Hirata09} effect may change the
clustering at a level of a few percent, the \lya selection can change the
clustering amplitude by a factor of a few,
owing to the high sensitivity of \lya RT to local environments of galaxies.
In addition, unlike the surface brightness distribution of stars in 
galaxies, which largely has a parity symmetry (i.e., similar surface brightness 
if viewed at opposite directions), the anisotropic distribution of \lya 
surface brightness does not necessarily have any symmetry. 

The 3D anisotropic clustering of LAEs, if measured, will be a strong test to
the RT model of LAEs. To achieve this goal, a large spectroscopic sample of 
LAEs will be needed. With current narrow-band surveys of 
LAEs, however, the clustering measurements are limited to the angular 2PCFs, 
which resemble the projected 2PCFs. Unlike the redshift-space distortion 
effect, which is largely eliminated in the projected 2PCFs, we find that the 
\lya selection effect is imprinted in the projected 2PCFs. Projection keeps 
the transverse fluctuations, which are always enhanced by the \lya selection. 
The projected 2PCF of LAEs has a higher amplitude than that of the control LAE 
sample with environment effect removed. Our model makes
the following distinctive prediction (see Figure~\ref{fig:biasfactor}): 
{\it the amplitude of the LAE 2PCF has a very weak dependence on the 
observed \lya luminosity.} 
The prediction breaks down for faint LAEs if a large dispersion (1 dex) 
between intrinsic \lya luminosity and halo mass is introduced, which may
result from stochastic star formation, but it remains valid for luminous
LAEs (see Appendix~\ref{sec:effect_lineprof}).
This is testable with large narrow-band surveys. The relation between LAEs 
and underlying halos has interesting properties. For LAE samples defined by a
threshold in observed \lya luminosity, faint LAEs appear 
to be more strongly clustered than mass threshold samples of dark matter halos 
having the same number density. The trend is reversed for very bright LAEs 
(with number density lower than $2\times 10^{-4}\denhMpc$), 
where halos are more strongly clustered than LAEs of the same number density.
If an LAE sample with low luminosity threshold 
were simply related to a mass threshold sample of halos and one used the 
observed LAE clustering amplitude to infer the halo number density, one would 
reach the conclusion that there are multiple LAEs per halo. 
This is not clearly seen from currently available clustering measurements yet, 
probably because of the small sample sizes (hence large variance) and the 
existence of contamination from low-redshift objects.

The Ly$\alpha$-selection-induced anisotropic clustering can be largely 
understood by accounting for the dependences of the selection on density and 
line-of-sight velocity gradient. Besides these dependences, \lya selection is 
also related to line-of-sight peculiar velocity and line-of-sight density 
gradient, which can give rise to scale-dependent galaxy bias, changing the 
shape of the fluctuation power spectrum with respect to matter 
(Appendices~\ref{sec:extendedtoy} and \ref{sec:powerspectrum}). We develop a 
simple physical model that incorporates all the identified environment 
variables in the selection to aid the understanding and interpretation of 
LAE clustering (Appendix~\ref{sec:extendedtoy}). Although the model is 
able to provide qualitative explanations to features in the LAE clustering,
it is an over-simplified model, which assumes linear dependence on
environment variables. A more sophisticated model will have to rely on
a detailed study of the dependence of \lya RT on environments
and an accurate description of the statistical relation between observed
\lya emission and environmental variables.

As a powerful tool, the HOD framework has been successfully applied to 
interpret galaxy clustering data from many galaxy surveys. For LAEs, 
the usual assumption in the HOD framework that galaxy properties 
are only a function of halo mass and are independent of large-scale 
environments breaks down, as a result of the selection caused by environment 
dependent \lya RT. As shown in Section~4, the HOD of LAEs is 
strongly dependent on environments. \citet{Hamana04} perform 
(environment-independent) HOD modeling of the angular 2PCF of $z=4.86$ LAEs in 
one Subaru Deep Field and find that the observed strong clustering cannot be 
reproduced by the model. Given the large sample variance in the small field 
and crude error estimate of the clustering measurement, however, it is not 
clear whether the failure of the model in this case is caused by the 
neglecting of the \lya selection effect. Our study shows that to correctly 
model LAE clustering within the HOD framework, one has to extend the framework 
to incorporate the strong environment dependence of \lya selection.

As discussed in Paper I, the uncertainties in our current model lie in the
intrinsic \lya luminosity and spectra, which deserve detailed investigation. 
By ``intrinsic'', we mean the properties of \lya 
photons after escaping the ISM. 
The intrinsic \lya luminosity is related to the SFR and the initial 
mass function of stars. 
If the model (luminosity-threshold) LAEs are matched 
to the observed LAEs in terms of the number density, the uncertainty in the 
intrinsic \lya luminosity is largely removed in studying LAE clustering. 
On the other hand, the intrinsic spectrum of \lya emission, especially the 
intrinsic line width, is an important factor in determining the strength of 
the \lya selection effect. A larger intrinsic line width would lead to a 
weaker dependence of \lya RT on environments (Paper I), which in turn  
would make the effects of \lya RT on LAE clustering weaker. For example, we 
would see a less elongated pattern in the 3D 2PCF of LAEs with a larger 
intrinsic line width. The intrinsic line shape may also be modified by
galactic winds, which can also affects the \lya RT (e.g., 
\citealt{Kunth98,Atek08}) and the effect on clustering. 
If isotropic galactic winds shift the initial \lya line by a few hundred 
$\kms$, the coupling of observed \lya emission to the environment can be 
weakened. However, galactic winds usually display a collimated bipolar 
pattern, and photons escaping in directions other than the bipolar direction 
may not achieve a large shift. High-resolution simulation of individual 
galaxies with galactic wind included 
is useful for further studying the effect of wind on the observational 
properties of LAEs. See the tests and discussions in Appendix~\ref{sec:effect_lineprof}.
Consequently, if galactic wind can have a strong effect, the 
strength of Ly$\alpha$-selection-induced clustering effects would allow us 
to potentially constrain the intrinsic \lya line width and shape of LAEs, 
which would otherwise be difficult to discern. 
For low-mass halos ($\sim 10^{10}\hMsun$) in the 
simulation, the grid for hydrodynamic calculation marginally resolves the 
virial radius. The \lya RT for sources in these halos may be limited by the 
resolution. However, the infall region (inside the turnaround radius), which 
is about 5.6 times larger and is important in shaping the \lya RT, is well 
resolved. Therefore, the \lya RT for sources in these low-mass halos is not 
expected to suffer the resolution effect significantly.

\lya RT is a physical process that must exist around LAEs. The new effects 
on LAE clustering, including the enhancement in the projected 2PCFs, are 
strong, which means that they cannot be easily masked by other effects. If
observation showed a null detection of the effects, e.g., finding no 
enhancement in the angular 2PCFs, it would have important implications in our
study of LAEs. A substantial scatter in the intrinsic \lya emission 
properties at fixed halo mass remains as a possibility to reduce or even mask
the \lya RT selection effect. In our model, the intrinsic \lya emission 
properties, especially the \lya luminosity, are tightly correlated with halo 
mass. The \lya luminosity is based on the SFR averaged over 10Myr time scale 
and the star formation prescription does not lead to significant scatter in 
the SFR at fixed halo mass and redshift \citep{Trac07}. To have a large 
scatter in the instantaneous SFR, one needs to introduce a broad distribution 
of the star formation efficiency or make the star formation stochastic in 
halos of fixed mass. Dust in the ISM could further enlarge the scatter
in the \lya luminosity. If in the end the intrinsic \lya luminosity 
distribution at fixed halo mass were much broader than that from \lya RT
effect (Paper I), the \lya RT selection could be masked. 
In Appendix~\ref{sec:effect_lineprof}, we present simple tests on the effect
of the dispersion in the relation between intrinsic \lya luminosity and halo 
mass. The dispersion can change the clustering amplitude of LAEs of fixed 
number density by including LAEs residing in lower mass halos. However, 
the anisotropic clustering pattern persists even if a large dispersion (1 dex
in luminosity) is introduced.
A broad or 
stochastic \lya luminosity distribution would give rise to an effective duty 
cycle such that only a tiny fraction of galaxies are in the \lya emitting 
phase at a given time, which means that at a given number density LAEs reside
in halos of much lower mass, compared to the case of a narrow intrinsic \lya 
luminosity distribution to start with. \citet{Nagamine10} argue that a duty 
cycle scenario provides a reasonable explanation to existing LAE clustering 
measurements. To fully address the magnitude of the distribution and 
stochasticity of star formation efficiency, the relevant processes have to 
be incorporated in the reionization simulation, which is limited by our 
understanding of baryon physics. Potential consequences, when combined 
with \lya RT selection, on LAE clustering deserves detail investigations.
Strong tests to the model and constraints on different processes are expected 
to come from the \lya LF, UV LF, and clustering of LAEs.

The completely new effects on LAE clustering due to \lya RT 
add another layer of complexity in modeling their clustering. 
This means that the existence of \lya
selection effect would complicate the inference of cosmological parameters
from LAE clustering in large LAE surveys. To map from galaxy clustering to
dark matter clustering, which directly encodes the cosmological information,
the effects that are commonly considered in current galaxy clustering analysis
include the redshift-space distortion, the nonlinear evolution of structure,
and the scale-dependent bias induced by halo biasing. For \lya selected 
galaxies, the new effects that add to the above list are the anisotropic 
clustering (opposite to and stronger than the redshift-space distortion)
and the scale-dependent bias induced by environment-dependent \lya RT. 
For our model with $z=5.7$ LAEs, the Ly$\alpha$-selection-induced 
anisotropy is overwhelming. 

At lower redshifts $z=2-3$ we expect the \lya RT induced selection effect to 
continue to operate in the regions surrounding galaxies, as at $z=5.7$ shown 
here. However, at present, it is not clear what the strength of the \lya 
selection effect will be and up to what scales the clustering of LAEs is 
affected at these redshifts. Several competing factors prevents us from having 
a simple guess. On one hand, lower matter density and higher UV background at 
$z=2-3$ (hence lower neutral hydrogen fraction in the scattering regions), 
compared to $z=5.7$, permit easier escape of \lya photons through the 
circum-galactic and inter-galactic gas. On the other hand, a lower Hubble 
velocity may provide a countering factor. We reserve a more detailed 
investigation of clustering of LAEs at $z=2-3$ for a future work.

The strong environmental dependence of the \lya selection, on the other hand,
provides a sensitive way to probe the late stage of cosmological reionization.
With simple treatments of \lya RT, it has been shown that 
\lya emission from LAEs can be used to probe the mean hydrogen neutral 
fraction as a function of redshift (e.g., from the \lya LF; 
\citealt{Malhotra04, Haiman05}). Both 
\citet{Furlanetto06} and \citet{McQuinn07} show that the clustering 
of LAEs is enhanced by the patchy reionization. Our current work is limited to 
$z=5.7$, when the reionization is completed, and we plan to perform full 
RT modeling of LAEs at higher redshifts in a subsequent paper. 
Because of the fast evolution of neutral hydrogen density as reionization 
proceeds, we would see a rapid change in the distribution of 
the observed-to-intrinsic \lya luminosity ratio, which would lead to a rapid 
change in the clustering amplitude of LAEs relative to the underlying 
population \citep{Furlanetto06}. The isolated \ion{H}{2} bubbles introduce a 
characteristic scale, which would result in a scale dependence in the bias 
factor \citep{Furlanetto06}. We also
expect that isolated \ion{H}{2} bubbles and large-scale fluctuations of
photoionization rate may introduce additional effects in the 3D anisotropic 
clustering of LAEs. As a whole, we expect that different features in
the clustering of LAEs, if detected, would 
provide a wealth of information about reionization.

\acknowledgments

We are grateful to David Weinberg for useful discussions and comments and for 
suggesting the use of shuffled samples. 
We thank Mark Dijkstra, Matt McQuinn, and Masami Ouchi for helpful discussions
and comments on an earlier draft. We thank the referee for constructive 
comments.
Z.Z. thanks Aaron Bray for useful discussions.
Z.Z. gratefully acknowledges support from Yale Center for Astronomy and 
Astrophysics through a YCAA fellowship. 
Z.Z. also thanks Aspen Center for Physics for a stimulating atmosphere
and Stuart Wyithe, Andrei Mesinger, and Paul Shapiro for interesting 
discussions. 
This work is supported in part by NASA grant NNG06GI09G and NNX08AH31G. 
H.T. is supported by an Institute for Theory and Computation Fellowship. 
J.M. is supported by the International Reintegration Grant of the
European Research Council 2006-046435 and Spanish grant AYA2009-09745.
J.M. thanks the Institute for Advanced Study for their hospitality.
Computing resources were in part provided by the NASA High-End Computing 
(HEC) Program through the NASA Advanced Supercomputing (NAS) Division at 
Ames Research Center. The \lya RT
computations were performed at the Princeton Institute for Computational
Science and Engineering (PICSciE). Z.Z. thanks Daisuke Nagai for the use of 
the chimay computer for conducting some analyses presented in this paper.

\appendix
\section{A. The Extended Version of the Simple Physical Model}
\label{sec:extendedtoy}

For LAEs, we have identified density, peculiar velocity, and their gradients 
as the main factors in transforming the 
intrinsic \lya emission properties to the observed ones. The dependence of 
\lya RT on these quantities imposes a selection function for 
the appearance of LAEs, which affects the clustering of LAEs. In the simple
model of LAE clustering presented in Section~\ref{sec:toy_model}, for 
simplicity we only 
include the density and line-of-sight velocity gradient to understand the main 
features in LAE clustering. For completeness, we add the other factors into 
the model here and follow Section~\ref{sec:toy_model} in presenting it.
 
We still consider the case that the selection function is a 
linear function of the environment variables. The real-space density of LAE 
galaxies is related to the matter density modified by the selection function,
\begin{equation}
\label{eqn:A_lae_den_sel}
\bar{n}_g(1+\delta_g) = q\bar{n}_0(1+b\delta_m) \times  
          \left[ 1 + \tilde{\alpha}_1 \delta_m
                   + \tilde{\alpha}_2 \frac{1}{Ha}\frac{\partial v_z}{\partial z}
                   + \tilde{\alpha}_3 \frac{1}{Ha}
                              \left(
                              \frac{\partial v_x}{\partial x}
                             +\frac{\partial v_y}{\partial y} 
                              \right) 
                   + \tilde{\alpha}_4 \frac{v_z}{Hr_Ha}
                   + \tilde{\alpha}_5 r_H\frac{\partial\delta_m}{\partial z}
          \right],
\end{equation}
where $b$ is the bias factor for the underlying galaxy population (with mean 
number density $\bar{n}_0$) before the \lya selection is imposed, $\bar{n}_g$ 
is the mean number density of galaxies that are selected as LAEs, and $q$ is 
the overall fraction of galaxies that are selected as LAEs, $\tilde{\alpha}_i$ 
($i$=1, 2, 3, 4, and 5) are coefficients (assumed to be constant) in the 
selection function, $r_H$ is a length scale introduced to make the coefficients
dimensionless and is chosen to be the Hubble radius $c/H$ when the scale factor
is $a$. Note that unlike the simple model presented in 
Section~\ref{sec:toy_model}, 
here we explicitly include the term (with coefficient $\tilde{\alpha}_3$) 
dependent on the transverse velocity gradient as well as that on the density 
(with coefficient $\tilde{\alpha}_1$), since they represent different physical 
effects (Section~\ref{sec:environ}). The untilded and tilded coefficients 
have a simple relation.
Since density and the divergence of velocity are 
connected from the continuity equation 
$\dot\delta+\nabla\cdot{\mathbf v}/a=0$, we have 
$\alpha_1=\tilde{\alpha}_1-\tilde{\alpha}_3 f$ and 
$\alpha_2 = \tilde{\alpha}_2-\tilde{\alpha}_3$ (see below).
In 
equation~(\ref{eqn:A_lae_den_sel}), $\tilde{\alpha}_1$ represent the selection 
effect
purely caused by density and we expect it has a negative sign. Since higher 
line-of-sight velocity gradient corresponds to higher apparent-to-intrinsic
\lya luminosity ratio, the coefficient $\tilde{\alpha}_2$ should be positive. 
For reasons discussed in Section~\ref{sec:environ}, the coefficient 
$\tilde{\alpha}_3$ is expected to be negative. 
The $\tilde{\alpha}_4$ term is related to the line-of-sight velocity, which
is not an accurate description. The selection may only depend on the 
motion of the galaxy relative to its surrounding medium that scatters the 
\lya photons, and a smoothing scale needs to be introduced in a full 
description.
Here we simply include a $\tilde{\alpha}_4$ term to gain a rough idea on the
possible effect.

Keeping the first order 
terms in Equation~(\ref{eqn:A_lae_den_sel}) and noticing that 
$\bar{n}_g=q\bar{n}_0$, we have
\begin{equation}
\delta_g  =  (b+\tilde{\alpha}_1) \delta_m 
               +\tilde{\alpha}_2 \frac{1}{Ha}\frac{\partial v_z}{\partial z}
               +\tilde{\alpha}_3 \frac{1}{Ha}
                         \left(
                               \frac{\partial v_x}{\partial x}
                              +\frac{\partial v_y}{\partial y}
                         \right) 
               +\tilde{\alpha}_4 \frac{v_z}{Hr_Ha}
               + \tilde{\alpha}_5 r_H\frac{\partial\delta_m}{\partial z}.
\end{equation}
By using 
$\delta_m = \sum_{\mathbf k} 
\delta_{m,{\mathbf k}}\exp(i{\mathbf k}\cdot{\mathbf r})$,
$\partial\delta_m/\partial z = \sum_{\mathbf k} ik_z 
\delta_{m,{\mathbf k}}\exp(i{\mathbf k}\cdot{\mathbf r})$,
$v_z=\sum_{\mathbf k} ifHak_z/k^2
\delta_{m,{\mathbf k}}\exp(i{\mathbf k}\cdot{\mathbf r})$,
and 
${\partial v_z}/{\partial z}=-\sum_{\mathbf k} fHa(k_z^2/k^2)
\delta_{m,{\mathbf k}}\exp(i{\mathbf k}\cdot{\mathbf r})$,
we can express the above relation in Fourier space as
\begin{equation}
\label{eqn:A_delta_realspace}
\delta_{g,{\mathbf k}} 
 =   \left[ (b+\tilde{\alpha}_1)-\tilde{\alpha}_2 f\mu^2-\tilde{\alpha}_3f(1-\mu^2) 
               +i \left(\tilde{\alpha}_4 \frac{f}{kr_H} + \tilde{\alpha}_5 k r_H\right)\mu
                \right]\delta_{m,{\mathbf k}}.
\end{equation} 
Redshift-space distortion contributes to the $\mu^2$ term, and in
redshift space, it reads
\begin{equation}
\label{eqn:A_delta_zspace}
\delta^s_{g,{\mathbf k}} 
 =     \left[ (b+\tilde{\alpha}_1-\tilde{\alpha}_3f)+(1-\tilde{\alpha}_2+\tilde{\alpha}_3)f\mu^2
       +i \left(\tilde{\alpha}_4 \frac{f}{kr_H} + \tilde{\alpha}_5 k r_H\right)\mu
                \right]\delta_{m,{\mathbf k}}.
\end{equation}
Although $\tilde{\alpha}_1$ is expected to be negative, the combination 
($\tilde{\alpha}_1-\tilde{\alpha}_3f$) should be positive, according to 
Figure~\ref{fig:jointdep}. Since the overall clustering effect is opposite to
the linear redshift-space distortion, we expect ($\tilde{\alpha}_2-\tilde{\alpha}_3$) to be 
positive. The dependence of the
selection on line-of-sight velocity and line-of-sight density gradient leads
to phase shifts in the fluctuation. 

With Equation~(\ref{eqn:A_delta_zspace}), the power spectrum of LAEs in 
redshift space is then
\begin{equation}
\label{eqn:A_PLAE_zspace}
P^s_g({\mathbf k}) 
  =  \left\{ \left[ \left(1+\frac{\tilde{\alpha}_1-\tilde{\alpha}_3f}{b}\right)
              +(1-\tilde{\alpha}_2+\tilde{\alpha}_3)\beta\mu^2\right]^2 
  +  \left( \tilde{\alpha}_4\beta \frac{1}{kr_H} 
                   + \frac{\tilde{\alpha}_5}{b} k r_H
                 \right)^2\mu^2
                \right\}b^2P_m({\mathbf k}).
\end{equation}
This full expression of power spectrum includes both redshift distortion and
\lya selection effects. Compared to Equation~(\ref{eqn:PLAE_zspace}), it
has more terms to represent different physical effects in the selection. 
Setting all the coefficients $\tilde{\alpha}_i$ to zero gives the usual redshift space 
linear power spectrum. The
dependence of \lya RT on the velocity gradient leads to the
same form of angular dependence as the redshift distortion in the anisotropic
power spectrum, but with an opposite sign. The dependences on line-of-sight 
velocity and line-of-sight gradient of density (the $\tilde{\alpha}_4$ and $\tilde{\alpha}_5$ 
terms in Equation~(\ref{eqn:A_PLAE_zspace}) also contribute to the anisotropy. 
Furthermore, these dependences lead to scale-dependent bias for the power
spectrum, with the velocity and density gradient terms dominant on large 
and small scales, respectively.

The monopole, quadrupole, and hexadecapole moments of the power spectrum in 
Equation~(\ref{eqn:A_PLAE_zspace}) are 
\begin{equation}
P_0(k) = 
\left[
 \left(1+\frac{\tilde{\alpha}_1-\tilde{\alpha}_3f}{b}\right)^2
+\frac{2}{3}\left(1+\frac{\tilde{\alpha}_1-\tilde{\alpha}_3f}{b}\right)(1-\tilde{\alpha}_2+\tilde{\alpha}_3)\beta
+ \frac{1}{5}(1-\tilde{\alpha}_2+\tilde{\alpha}_3)^2\beta^2
+\frac{1}{3}\left(\tilde{\alpha}_4\beta\frac{1}{kr_H}+\frac{\tilde{\alpha}_5}{b}kr_H\right)^2
 \right] b^2P_m(k),
\end{equation}
\begin{equation}
P_2(k)   =  
\left[
 \frac{4}{3}\left(1+\frac{\tilde{\alpha}_1-\tilde{\alpha}_3f}{b}\right)(1-\tilde{\alpha}_2+\tilde{\alpha}_3)\beta
+\frac{4}{7}(1-\tilde{\alpha}_2+\tilde{\alpha}_3)^2\beta^2
+\frac{2}{3}\left(\tilde{\alpha}_4\beta\frac{1}{kr_H}+\frac{\tilde{\alpha}_5}{b}kr_H\right)^2
 \right] b^2P_m(k),
\end{equation}
and
\begin{equation}
P_4(k)  = 
\frac{8}{35}(1-\tilde{\alpha}_2+\tilde{\alpha}_3)^2\beta^2
b^2P_m(k).
\end{equation}
The changes in these multipole moments with respect
to the redshift-distortion-only case ($\tilde{\alpha}_i=0$) can be clearly seen.
The monopole and quadrupole are affected by all the \lya selection factors,
while the hexadecapole only has additional contributions from the dependence
on velocity gradient.

\section{B. Power Spectrum of LAE\protect\lowercase{\rm s}}
\label{sec:powerspectrum}

\begin{figure*}
\plotone{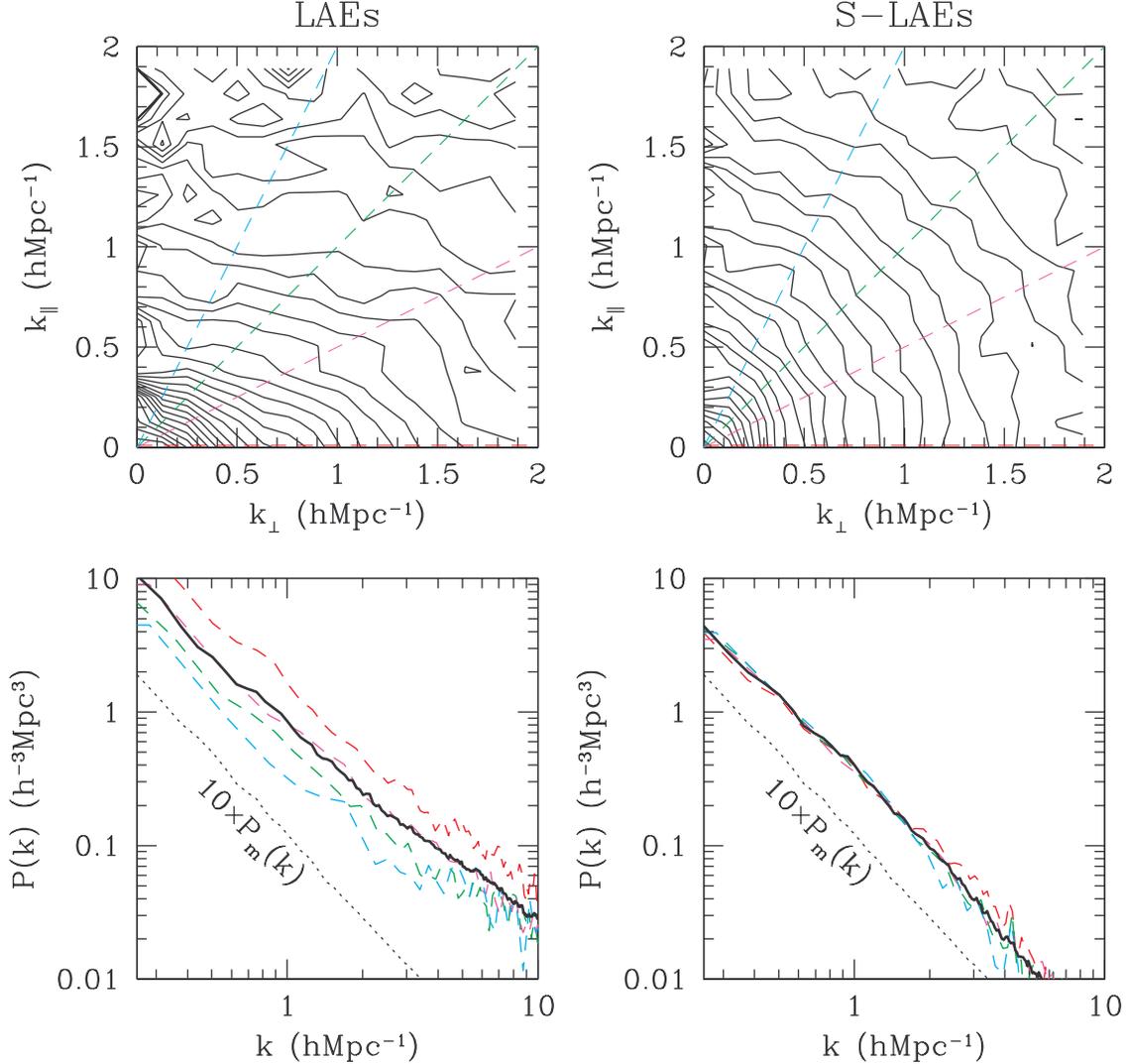}
\caption[]{
\label{fig:Pk}
Redshift-space power spectra of LAEs (left) and S-LAEs (right). The LAE and 
S-LAE samples are \lya luminosity-threshold samples of number density
$10^{-2}\denhMpc$. The S-LAE sample eliminates the environment effect
of \lya selection. For each column, the top panel shows the 3D redshift-space
power spectra. Adjacent contours differ by 0.1dex in contour levels.
The color coded dashed lines are four directions of wave vector. The
power spectra along the four directions are compared in the bottom panel
(dashed curves). The solid black curve in the bottom panel are the
spherically averaged power spectrum and the dotted curve is the matter 
power spectrum.
}
\end{figure*}

\begin{figure*}
\plotone{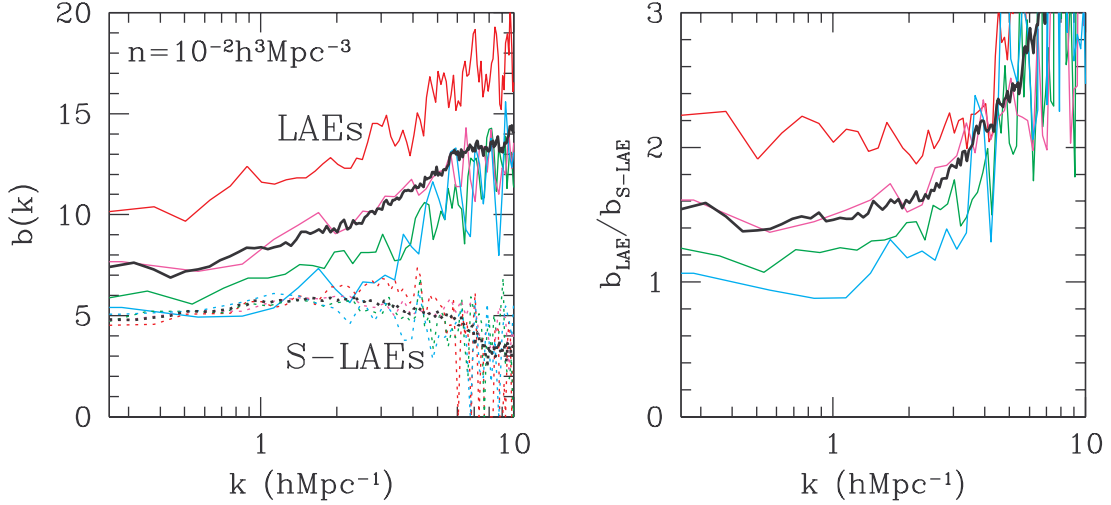}
\caption[]{
\label{fig:b_1em2}
Scale dependence of the LAE bias factor. In the left panel, the solid (dotted)
curves are the galaxy bias factors of LAEs (S-LAEs) at different directions, 
color coded in the same fashion as in Figure~\ref{fig:Pk}. The scale dependence
seen in the S-LAE bias is caused by nonlinear structure growth, while that
in the LAE bias is a combination effect of nonlinear growth and \lya selection.
In the right panel, the relative bias factors of LAEs and S-LAEs are plotted
and the scale dependence seen here is supposed to be induced only by the 
\lya selection effect. The number density of the luminosity-threshold LAE 
or S-LAE sample is $10^{-2}\denhMpc$.
}
\end{figure*}

\begin{figure*}
\plotone{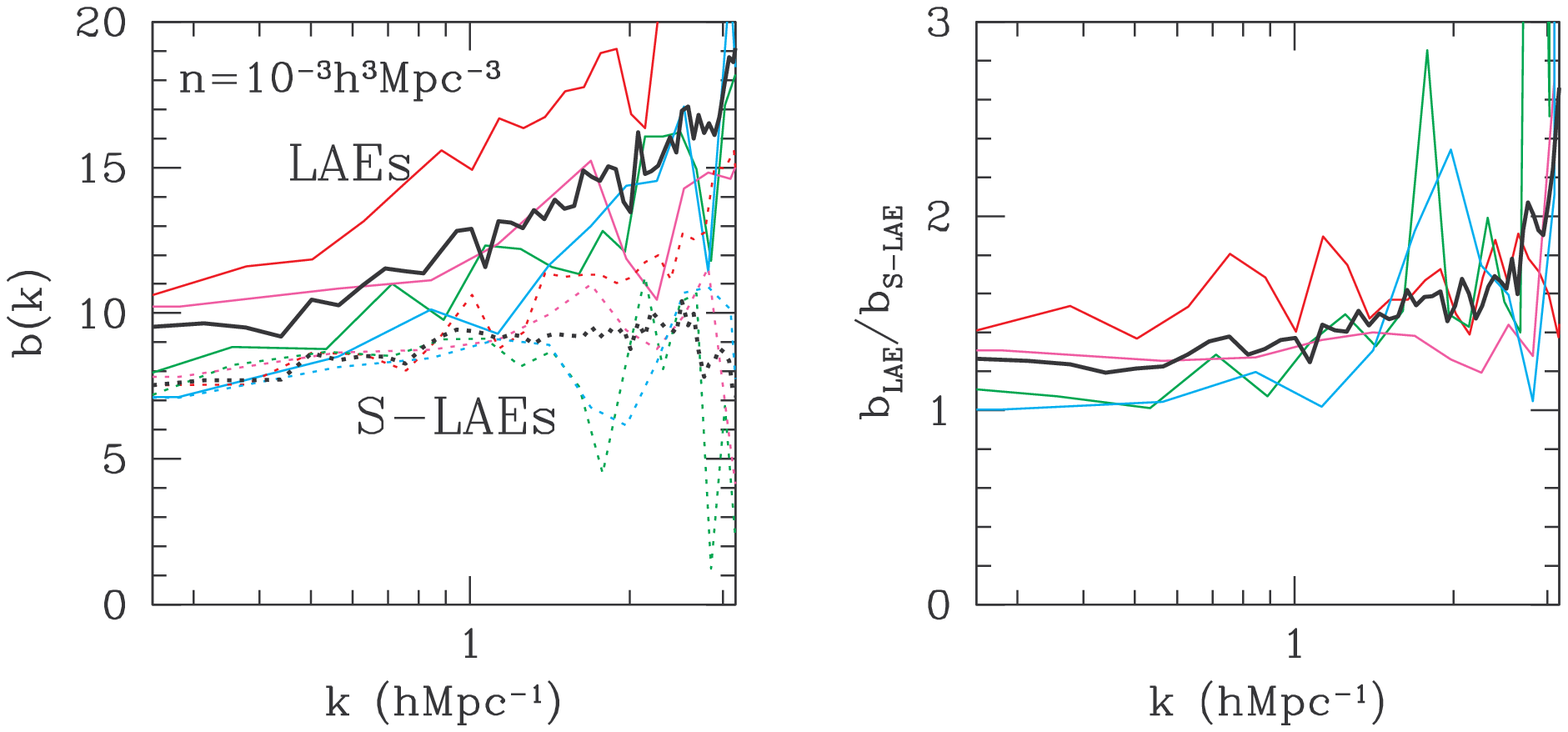}
\caption[]{
\label{fig:b_1em3}
Same as Figure~\ref{fig:b_1em2}, but for samples of number density 
$10^{-3}\denhMpc$. Note that the upper limits in $k$ differ from those
in  Figure~\ref{fig:b_1em2} to avoid plotting noisy data.
}
\end{figure*}

In Section~\ref{sec:clustering}, we present the LAE clustering results in the 
form
of 2PCFs, which are commonly measured in LAE surveys. In the simple model we
develop, however, the clustering is easily studied in terms of the power 
spectrum, the counterpart of the 2PCF in Fourier space. To be in parallel with 
the physical model and for a better comparison, we present the power spectrum 
of model LAEs measured for a couple of samples.

For a given sample of LAEs, we derive a density field from the spatial 
distribution of LAEs using cloud in cell weighting. The density field 
in the simulation box ($100\hMpc$ on a side) is put into a $768^3$ grid,
the same size as the grid used in our RT calculation. 
We then perform fast Fourier transform (FFT) with the density grid to obtain 
the fluctuation power spectrum $P({\mathbf k})$ of LAEs. To correct for 
the shot noise contribution to the power spectrum and the alias effect in the
shot noise \citep{Jing05}, we generate 16 random 
catalogs in boxes of $100\hMpc$ on a side. The number of particles in each 
random catalog is the same as that of LAEs in consideration. The shot noise
power spectrum of each random catalog is calculated from FFT. The average
shot noise spectrum over the 16 realizations is subtracted from 
the above measured power spectrum of LAEs to obtain the final power spectrum.
We also compute the power spectrum $P_m(k)$ of matter with FFT from the grid 
of matter density field.

The top-left panel of Figure~\ref{fig:Pk} shows the (apparent) redshift-space
3D power spectrum $P({\mathbf k})$ of the luminosity-threshold sample of LAEs 
with number density $10^{-2}\denhMpc$. The line-of-sight elongation pattern 
seen in the 2PCF now corresponds to the squashed contours in the power spectrum 
plot. For comparison, the S-LAE sample of the same number density only shows
a weak redshift-space distortion pattern (top-right panel of 
Figure~\ref{fig:Pk}), with contours slightly elongated along the line of 
sight. For a better comparison and to see the shape of $P({\mathbf k})$ 
more clearly, we plot the power spectra at different angles with respect to 
the line of sight in the bottom panels of Figure~\ref{fig:Pk} for the 
LAE and S-LAE samples, respectively. The curves are color coded in the same
way as the directions of wave vectors shown in the top panels. The solid black
curve in each of the bottom panels is the spherically averaged power spectrum.

For S-LAEs, the \lya selection effect is eliminated and only redshift-space
distortion effect is left. Since the sample is highly biased, the redshift-space
distortion parameter $\beta=\Omega^{0.6}/b$ is small and the redshift-space
distortion effect is weak. In the bottom-right panel, we only see small 
differences in the $P({\mathbf k})$ at different angles, or at different 
values of $\mu$ (the cosine of the angle between the line of sight and the 
wave vector). The shape of $P({\mathbf k})$ reflects that of halos. It shows 
slight difference from the matter power spectrum $P_m(k)$, which results from
nonlinear structure growth and halo biasing.

For LAEs, the anisotropy caused by \lya selection is very strong (dashed 
curves in the bottom-left panel). The amplitude of $P({\mathbf k})$ 
continuously increases as the direction of the wave vector gets closer to 
the line of sight, i.e., as the value of $\mu$ becomes larger. This is
in line with the trend from the physical model (Equation~(\ref{eqn:A_PLAE_zspace})).
The shape of $P({\mathbf k})$ shows a clear departure from that of $P_m(k)$,
which is predicted by the physical model as the results of the dependence of 
\lya selection on the line-of-sight velocity and line-of-sight density 
gradient. The departure is larger at smaller scales (larger $k$), which 
implies that the effect of line-of-sight density gradient 
(the $\tilde{\alpha}_5$ term Equation~(\ref{eqn:A_PLAE_zspace})) 
likely dominates the scale dependence.

To clearly see the scale dependence, we plot the galaxy bias factor at 
different values of $\mu$ as a function of scale in the left panel of 
Figure~\ref{fig:b_1em2}. We compute the galaxy bias factor as the
square root of $P({\mathbf k})/P_m(k)$. For S-LAEs (dotted curves), the 
deviation from a constant bias factor is a result of nonlinear structure 
growth and halo biasing. For LAEs (solid curves), the deviation is a 
combination of the above effects and the \lya selection effect. If we take 
the ratio of the bias factors of LAEs and S-LAEs at different values of 
$\mu$, we can remove the effect only caused by nonlinear structure growth 
and halo biasing. The right panel of 
Figure~\ref{fig:b_1em2} shows such relative bias factors of LAEs and S-LAEs.
The scale dependence in this panel is introduced by the selection effect
from environment-dependent \lya RT. We can see the trend 
that the scale dependence is stronger at higher values of $\mu$, in a 
qualitative agreement with the simple model 
(Equation~(\ref{eqn:A_PLAE_zspace})). 

Figure~\ref{fig:b_1em3} shows the scale dependence of the bias factors of
LAE and S-LAE samples with a lower number density ($10^{-3}\denhMpc$).
Since the samples correspond to more massive halos, it is not surprising that
the scale dependence caused by nonlinear structure growth and halo biasing
becomes stronger, as seen in the bias factors of S-LAEs (dotted curves in 
the left panel). The Ly$\alpha$-selection-induced scale dependence and the 
trend with $\mu$ can still be clearly seen (right panel). Compared with the 
$10^{-2}\denhMpc$ case, the bias factor of LAEs with the lower number density 
shows a weaker dependence on $\mu$. In the simple model, this can be largely 
explained by the larger value of $b$ in Equation~(\ref{eqn:A_PLAE_zspace}).

The overall trends seen in the power spectrum of LAEs seem to be captured
by the extended simple model. Therefore, the simple model can be used for 
qualitative understanding of LAE clustering.

\section{C. Clustering of LAE\protect\lowercase{\rm s} in the $\exp(-\tau_\nu)$ Model}
\label{sec:clustering_expmtau}

\begin{figure*}
\epsscale{1.1}
\plotone{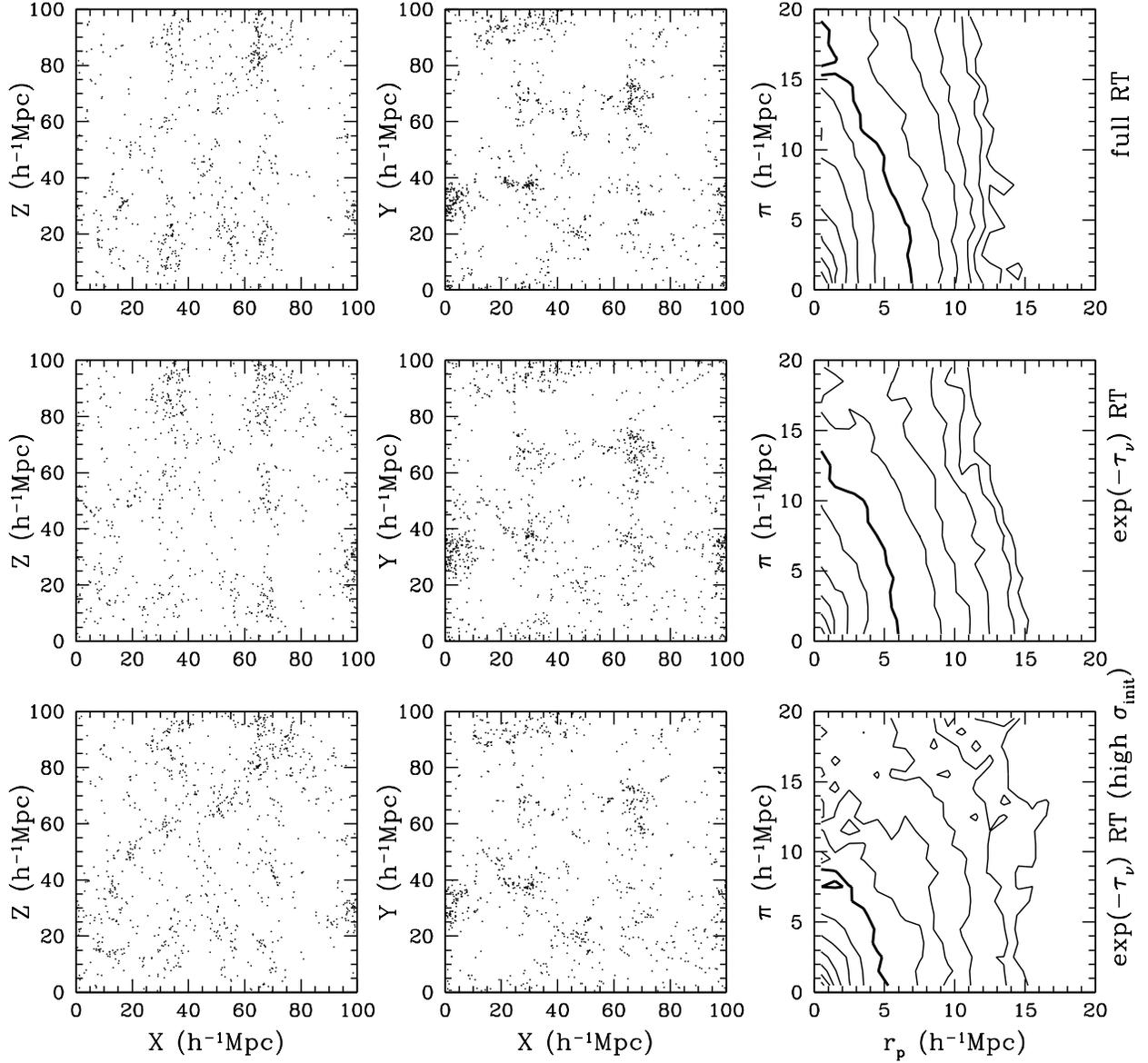}
\epsscale{1.0}
\caption[]{
\label{fig:slice_cmp}
Spatial distributions of LAEs and 3D 2PCFs in real space in the full \lya RT 
model (top row) and two $\exp(-\tau_\nu)$ RT models (middle and 
bottom rows). 
The LAE samples from the three models are defined by thresholds in observed 
\lya luminosity so that they have the same number density, 
$5\times 10^{-3}\denhMpc$.
The $\exp(-\tau_\nu)$ RT model in the middle row adopts the same intrinsic 
\lya line width as the full RT model, which cannot produce an apparent \lya LF 
that matches that from the full RT model. The $\exp(-\tau_\nu)$ RT model in the 
bottom row adopts a larger intrinsic \lya line width so that it roughly 
produces an apparent \lya LF matching that from the full RT model.
Left panels compare the distributions with one spatial direction along the 
line of sight. The observer is supposed to be on the top of the panels. 
Middle panels compare the distributions in the transverse plane perpendicular 
to the line of sight. 
The corresponding slices from the three models are from the same part of the 
box, with a thickness of 20$\hMpc$. Right panels compare the real-space 3D 
2PCFs of LAEs from the three models. The solid contour in each panel 
denotes a contour level of unity, and the adjacent contours differ by 0.2dex 
in contour levels.
}
\end{figure*}

\begin{figure*}
\epsscale{1.1}
\plotone{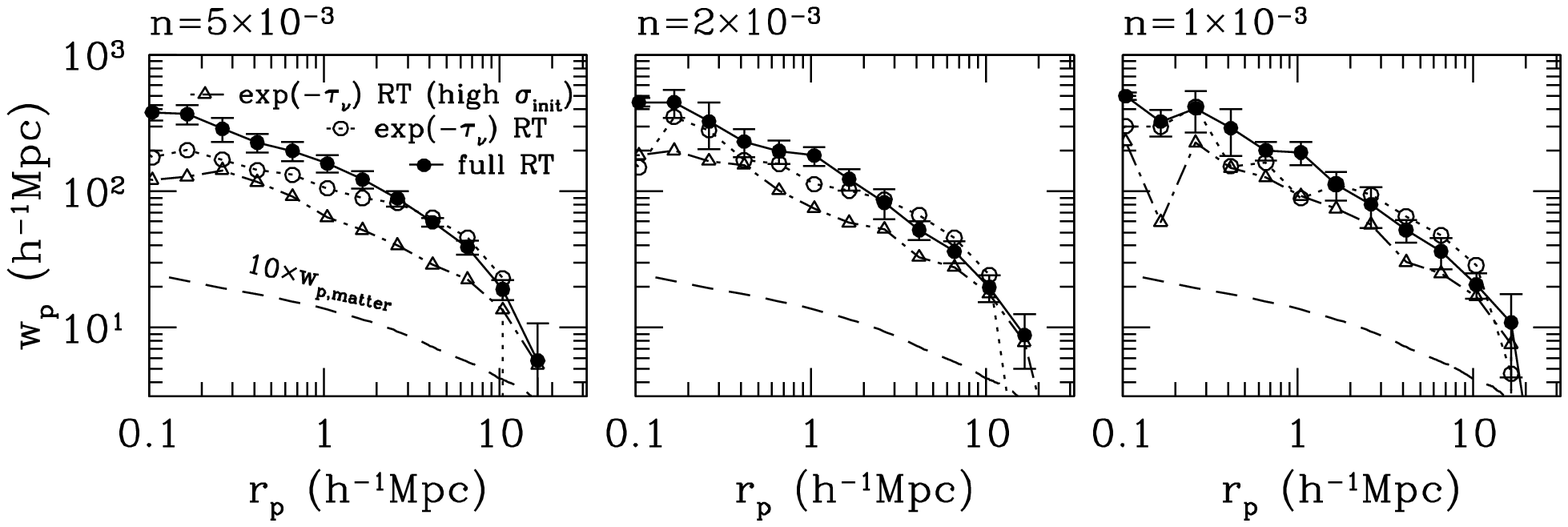}
\epsscale{1.0}
\caption[]{
\label{fig:wp_cmp}
Projected 2PCFs of LAEs in the full \lya RT model and two $\exp(-\tau_\nu)$ RT 
models.
In each panel, the LAE samples from the three models are defined by thresholds 
in observed \lya luminosity so that they have the same number density
(marked on top of the panel in units of $\denhMpc$).
The dashed curve is the projected 2PCF of matter, scaled by a factor of 10.
The $\exp(-\tau_\nu)$ RT model with open circles adopts the same intrinsic
\lya line width as the full RT model, which cannot produce an apparent \lya LF
that matches that from the full RT model. The $\exp(-\tau_\nu)$ RT model with
open triangles adopts a larger intrinsic \lya line width so that it roughly
produces an apparent \lya LF matching that from the full RT model.
}
\end{figure*}

In this Appendix, we compare the predictions of clustering of LAEs in
the full \lya RT model and the $\exp(-\tau_\nu)$ model.

The $\exp(-\tau_\nu)$ model is adopted in previous work for the \lya RT
calculation, which computes the line-of-sight optical depth $\tau_\nu$ to a 
source and modifies the intrinsic \lya emission by the $\exp(-\tau_\nu)$
factor. 
The $\exp(-\tau_\nu)$ model is reasonably accurate in describing 
\lya absorption feature for \lya photons passing through neutral 
clouds/islands, especially for absorption caused by the damping wing or 
small optical depth. In the case of studying the observed \lya emission,
as shown in Paper I, the $\exp(-\tau_\nu)$ model can qualitatively 
explain some trends in the \lya flux suppression, but it cannot provide
quantitative results because of the lack of frequency diffusion and spatial 
diffusion in the model. Previous work intends to limit the $\exp(-\tau_\nu)$ 
model to study the transfer outside of a radius much larger than halo size 
by assuming a \lya line profile at that radius. Even if this is the case, 
it is not clear what radius to use, what line profile to assume, and what 
angular distribution of \lya emission at that radius to adopt, which all rely 
on the \lya RT inside that radius. Here we take the original meaning of the
$\exp(-\tau_\nu)$ model and compute the line-of-sight optical depth $\tau_\nu$
all the way to the source.

As discussed in Section~\ref{sec:environ}, the line-of-sight optical depth is 
not the single factor in determining the \lya RT, since scatterings of \lya 
photons enable them to probe the optical depth in all directions. The 
$\exp(-\tau_\nu)$ model does not fully take into account the environment 
around the source. It fails to even qualitatively reproduce the trend of the 
\lya flux suppression with density at fixed line-of-sight gradient, as seen in 
Figure~\ref{fig:jointdep}. 

Since $\exp(-\tau_\nu)$ model still partially captures the environment 
effect on \lya RT, the new effects in LAE clustering predicted by our full 
\lya RT model are expected to be more or less seen in this model. Here, we 
compare the clustering properties of LAEs in the full RT and $\exp(-\tau_\nu)$ 
models for samples defined by thresholds in observed \lya luminosity.
With the same initial setups (i.e., intrinsic \lya luminosity and 
line profile) as in our full RT calculation, the $\exp(-\tau_\nu)$ model 
underpredicts the observed \lya luminosity (Paper I). We also include a case
of $\exp(-\tau_\nu)$ model with higher intrinsic \lya line width 
$\sigma_{\rm init}$, 
corresponding to the circular velocity at halo virial radius. The apparent
\lya LF from this $\exp(-\tau_\nu)$ model roughly match what we obtain from 
the above full RT model. We construct luminosity threshold LAE samples from 
the three models to have the same number density.

Figure~\ref{fig:slice_cmp} compares the spatial distributions and the 3D
2PCFs for LAEs of number density $5\times 10^{-3}\denhMpc$ in the three models.
The line-of-sight elongation pattern is also seen in the $\exp(-\tau_\nu)$ 
models (middle-left and bottom-left panels). However, it is not as prominent 
as in the full RT model and it is weak in the high-$\sigma_{\rm init}$ 
model (bottom-left panel). Compared to shuffled samples (not shown here), 
LAEs in the $\exp(-\tau_\nu)$ models also show enhanced clustering in the 
transverse plane. However, the clustering is not as strong as in the full RT 
model (see panels in the middle row). In terms of the reals-pace 3D 2PCFs, 
LAEs show a stronger clustering and a more prominent line-of-sight elongation 
feature in the full RT model than in the $\exp(-\tau_\nu)$ models. 

Figure~\ref{fig:wp_cmp} compares the projected 2PCFs in the three models for 
three LAE samples with different number densities. The projected 2PCF appears 
to be flatter in the $\exp(-\tau_\nu)$ models than in the full RT model. The 
shape is more parallel to the matter 2PCF, which suggests a weaker scale 
dependence in the bias factor. 
LAEs from the high-$\sigma_{\rm init}$ $\exp(-\tau_\nu)$ model shows a much 
weaker clustering than those from the full RT model.
The large scale bias factor of LAEs in the
$\exp(-\tau_\nu)$ models relative to halos has a similar trend as seen in 
Figure~\ref{fig:biasfactor}.

Our comparison shows that it may be possible to adjust the intrinsic \lya
line width in the $\exp(-\tau_\nu)$ model to approximately match the 
apparent \lya LF, but the clustering of LAEs would be far off. Alternatively, 
the intrinsic \lya line width could be adjusted to roughly match the 
clustering, but the apparent \lya LF would be far off. The $\exp(-\tau_\nu)$ 
model cannot fit both the apparent \lya LF and the clustering of LAE.

As a whole, the \lya selection effects on clustering are preserved to some 
extent in the $\exp(-\tau_\nu)$ model. The reason may lie in that the \lya 
optical depth is strongly dependent on the velocity gradient, which plays a 
major role in determining the anisotropic distribution of \lya emission of a 
LAE source (see the illustration in Figure~\ref{fig:angle_dep}) and in driving
the anisotropic clustering of LAEs (Section~\ref{sec:toy_model}). 
The line-of-sight optical depth computed in the $\exp(-\tau_\nu)$ model encodes
the information of the line-of-sight velocity gradient, so it is not surprising
that the model is able to qualitatively captures the main effects of \lya 
selection in LAE clustering. 

However, the $\exp(-\tau_\nu)$ model excludes the environment information in 
directions other than the line of sight. Furthermore, the effect of velocity 
gradient in this model is computed with the initial line profile, despite that 
the line profile changes because of frequency diffusion caused by scatterings 
around the source. The model therefore cannot give a full description of \lya 
selection effect on environment and cannot explain the observed LAE properties 
(e.g., \lya LF and clustering) quantitatively and self-consistently. 

\section{D. Effects of the Intrinsic L\protect\lowercase{\rm y}$\alpha$ Line 
Profile and the Dispersion between Intrinsic Luminosity and Halo Mass}
\label{sec:effect_lineprof}

In this appendix, we first show the effect of the intrinsic \lya line profile 
on the scattered \lya emission and discuss how it may affect the clustering 
of LAEs. We then investigate how a scatter in the intrinsic \lya (or UV) 
luminosity and halo mass relation may change the clustering effects caused by 
the \lya RT.

As mentioned in Paper I, one main uncertainty in the the RT 
model of LAEs is the initial or intrinsic \lya line profile (after \lya 
photons escape the ISM). In our current setup, the profile is assumed to be 
Gaussian with the width determined by the thermal temperature of the host halo. 
In reality, the shape and width of the intrinsic \lya line are determined by
gas dynamics in galaxies and affected by galaxy disk rotation,
galaxy merging, and galactic wind. As a result, the \lya line can be broadened,
become asymmetric, and/or have a shift in wavelength. All of these effects 
may change the strength of the coupling between observed \lya properties and
circum-galactic and inter-galactic environments, and therefore have an impact 
on the clustering effects discussed in this paper. 

A thorough investigation on the effect of the intrinsic line profile is out 
of the scope of this paper. Here we present a test to show how sensitive the
observed \lya emission is to the change in the initial \lya line wavelength.
We perform RT calculations for an individual source in a halo 
of $10^{11}\hMsun$ and consider cases with different initial values of 
\lya wavelength. For each case, the initial \lya photons all have a single 
wavelength with the luminosity normalized to be the intrinsic \lya luminosity 
of the source. That is, the line follows the Dirac $\delta$ function.

\begin{figure*}
\plotone{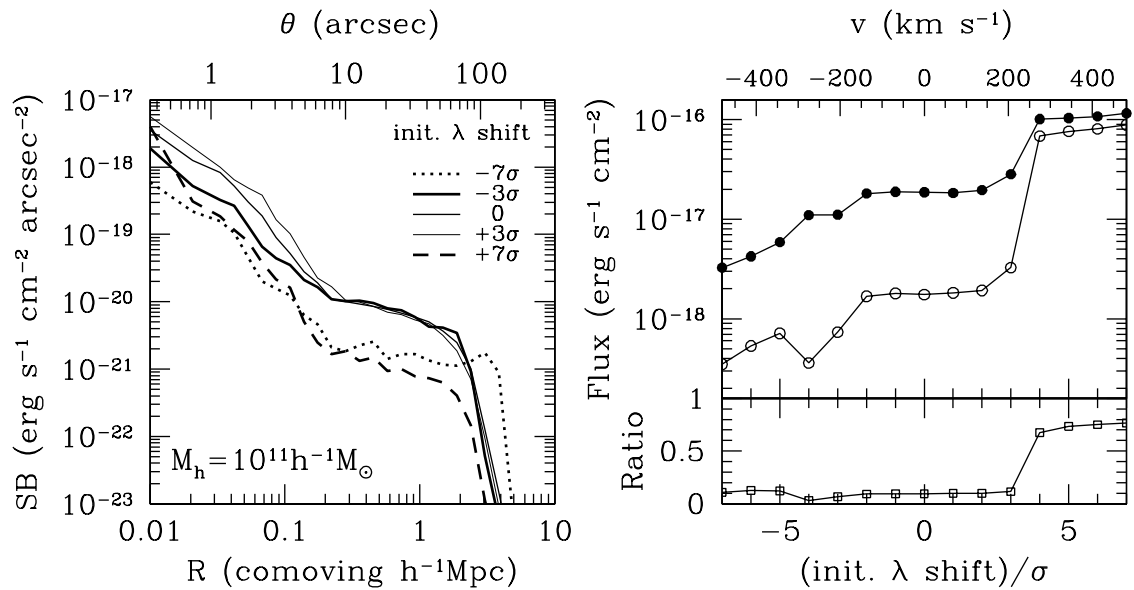}
\caption[]{
\label{fig:effect_lineprof}
Effect of initial \lya line wavelength shift on the \lya flux distribution.
The left panel shows the surface brightness profiles of the scattered \lya 
emission for a source residing in a $10^{11}\hMsun$ halo, chosen from the 
simulation box. For each profile, the initial \lya line from the central 
source has a single wavelength, denoted by the wavelength shift with respect 
to the rest-frame line center in units of the halo thermal velocity dispersion
$\sigma\simeq 70\kms$. The right panel shows the \lya flux in a central 
3\arcsec.5 diameter aperture as a function of the initial \lya wavelength 
shift. The two curves in the upper panel are for fluxes observed in 
opposite directions, and the curve in the lower panel shows their ratio.
}
\end{figure*}

In the left panel of Figure~\ref{fig:effect_lineprof}, the \lya surface 
brightness profile of the source is shown for a few cases. The initial
\lya line wavelength is denoted by the wavelength shift relative to the 
rest-frame \lya line center, in units of the halo thermal velocity dispersion 
$\sigma\simeq 70\kms$. 

On average, bluer initial \lya photons lead to less concentrated surface 
brightness profile, since these photons need to travel farther in the IGM 
before they redshift into the line center to encounter significant scatterings.
Redder initial photons make the profile more concentrated and more point-like,
A small fraction of the initial red photons can blueshift to the line center 
by encountering scatterings in the infall region around the halos, which
makes the profile still appear to be extended on large scales. The solid curves
in the left panel of Figure~\ref{fig:effect_lineprof} show that the profile is 
insensitive to the initial \lya wavelength if the initial shifts of photons 
are within the $\pm 3\sigma$ range. These photons encounter a lot of core 
scatterings and redistribute their wavelength in the $\pm 3\sigma$ range,
and therefore they lost their memory of the initial states.

In the right panel, we show the \lya flux in a central 3\arcsec.5 diameter 
aperture for different choice of the initial \lya wavelength. The two curves
in the upper panel are fluxes observed in opposite directions. The 
difference in the two curve tells us how anisotropic the scattered \lya 
emission is and provides an estimate of the strength of the 
coupling between environments and scattered \lya emission through RT.
If the initial shift of photons is smaller than $\sim 3.5\sigma$,
the flux from the central aperture is strongly suppressed. Meanwhile, there
is a large difference in the fluxes observed from the two opposite directions.
Both indicate the strong coupling between the observed \lya emission and
circum-galactic and inter-galactic environments. For photons with initial shift
larger than $\sim 3.5\sigma$, a fraction of photons can be out of resonance 
in the circum-galactic and inter-galactic media, weakening the coupling. 
However, the fluxes in opposite directions still show $\sim$20\% difference.

The above test seems to suggest that the environment-dependent RT 
effect becomes weak if the initial line shifts 
is larger than $\sim 3.5\sigma$ ($\sim 250\kms$) for halos of $10^{11}\hMsun$.
Galactic winds can have the effect of shifting \lya emission toward red
\citep[e.g.,][]{Verhamme06,Dijkstra10}. Observationally, galactic winds are
ubiquitous in star-forming galaxies. \citet{Steidel10} find that the
interstellar absorption lines and the H$\alpha$ line have a mean shift of
$\sim 160\kms$ with a large scatter for $z\sim$ 2--3 Lyman break galaxies 
with {\it baryon} mass of $10^{10}-10^{11}M_\odot$. Such a wind velocity may
be able to shift the \lya line by a few hundred $\kms$. However, the outflowing
neutral gas of the wind is observed to be usually collimated in a bipolar 
fashion \citep[e.g.,][]{Bland88,Shopbell98,Veilleux02}. \lya emission escaping 
in directions other than the wind is expected to have smaller redward shift,
which can still probe the circum-galactic and inter-galactic environments 
through scatterings. Therefore, it is likely that the clustering features we 
study in this paper persist in the presence of galactic wind. Radiative
transfer modeling of individual galaxies in high resolution simulations 
with galactic wind prescription will help to further shed light on this issue.

\begin{figure*}
\epsscale{1.1}
\plotone{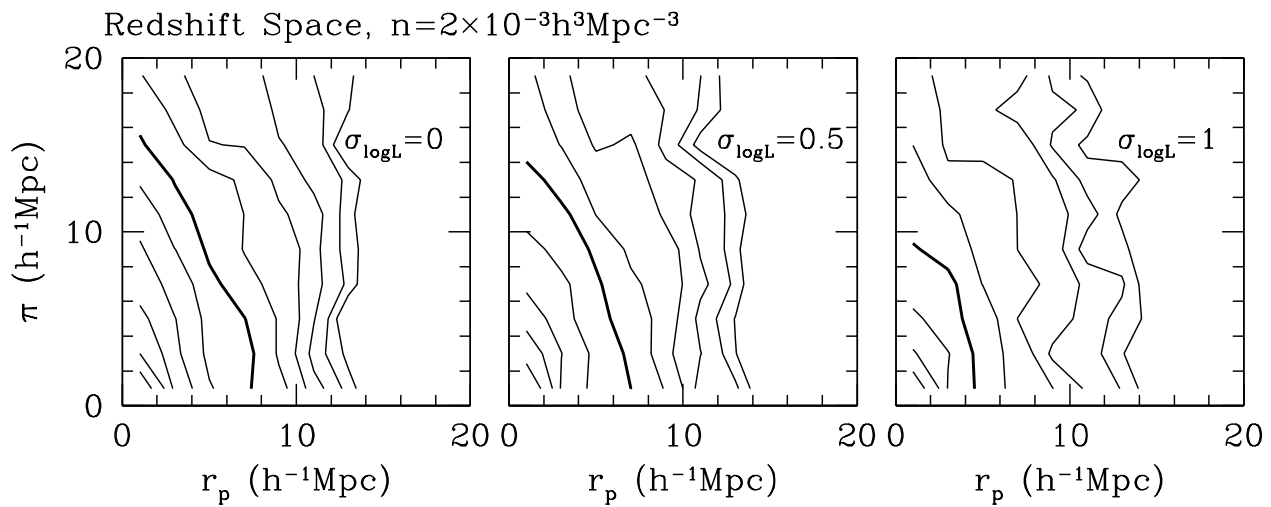}
\epsscale{1.0}
\caption[]{
\label{fig:xi3d_MLscatter}
Effect of scatter between intrinsic \lya luminosity $\Lint$ and halo mass 
on the 3D redshift-space 2PCFs of LAEs. A Gaussian distribution is introduced
for $\log\Lint$
with standard deviation $\sigma_{\log L}$ and fixed mean. The LAE sample
is constructed by selecting LAEs above a threshold in {\it observed} \lya
luminosity so that the number density is $2\times 10^{-3}\denhMpc$.
From left to right, the cases for $\sigma_{\log L}$=0, 0.5, and 1 are shown,
respectively. See the text for details.
}
\end{figure*}

\begin{figure*}
\plottwo{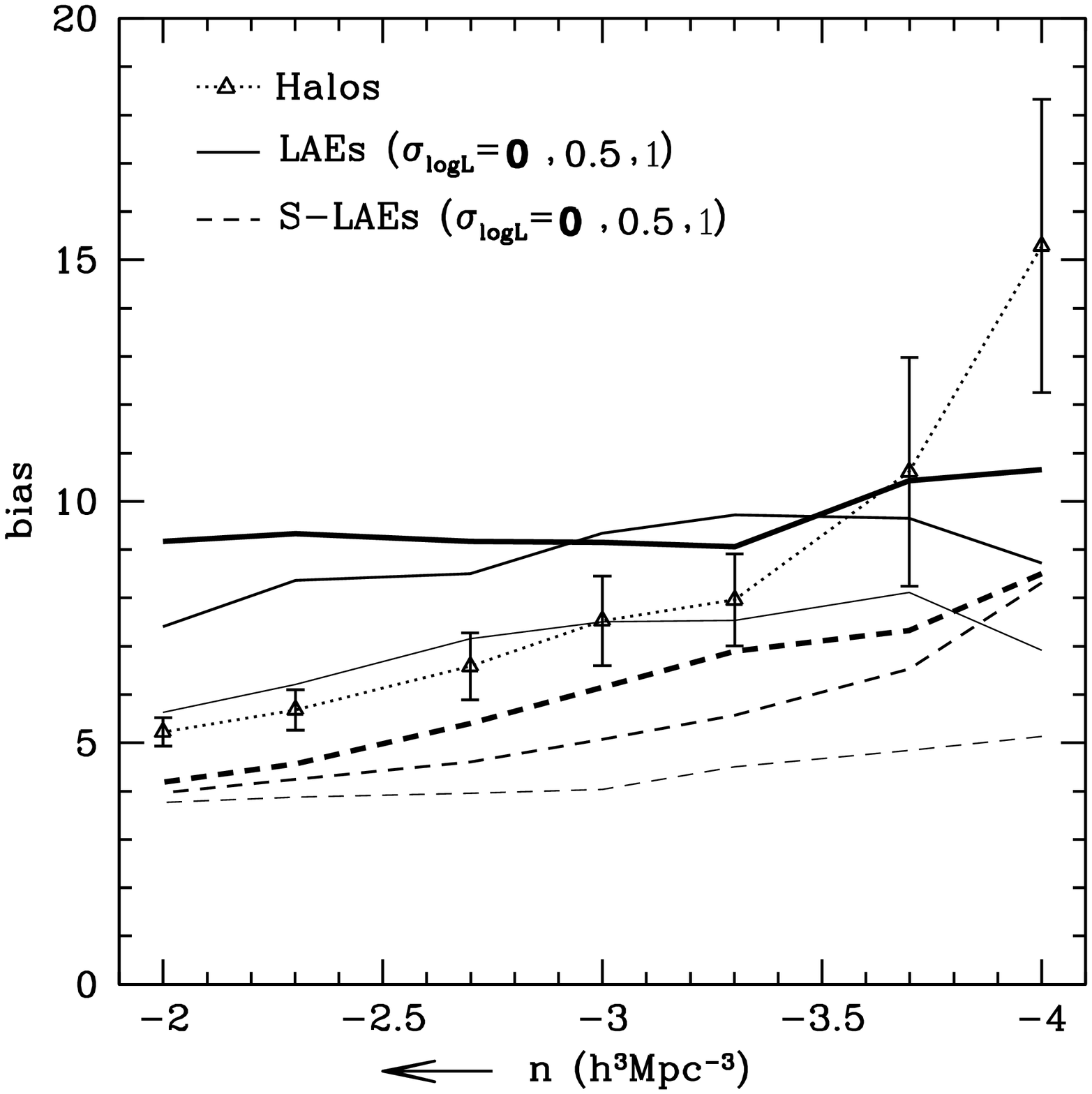}{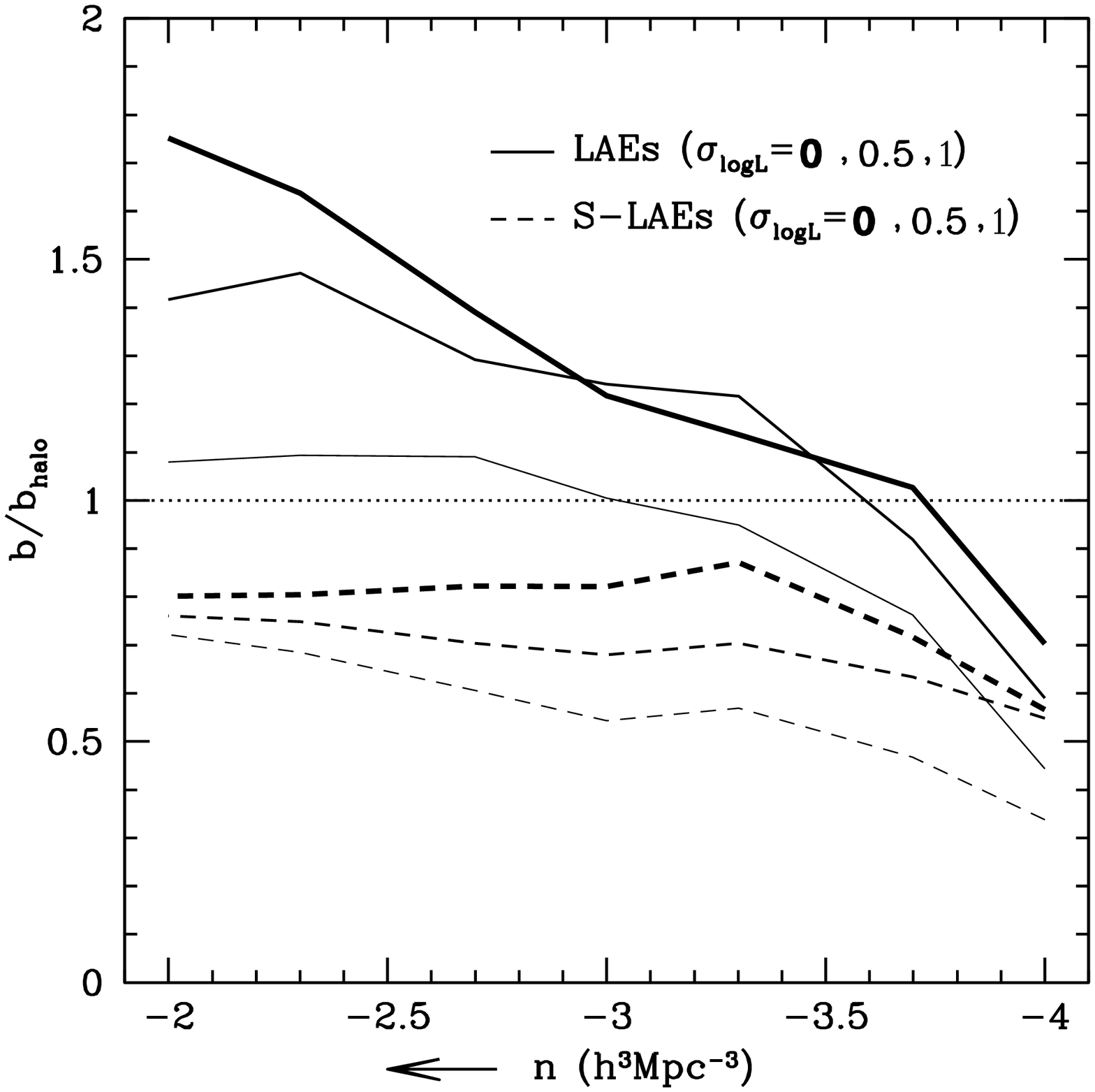}
\caption[]{
\label{fig:bias_MLscatter}
Effect of scatter between intrinsic \lya luminosity $\Lint$ and halo mass
on the bias factors of LAEs. A Gaussian distribution is introduced
for $\log\Lint$ with standard deviation $\sigma_{\log L}$ and fixed mean.
The left and right panels are for absolute bias factor and that relative to 
halos, similar to Figure~\ref{fig:biasfactor}. The bias curves are shown 
for both LAE and S-LAE samples, with $\sigma_{\log L}$=0, 0.5, and 1 (from
thick to thin curves), respectively.
}
\end{figure*}

We now turn to discuss whether a dispersion in the intrinsic \lya (UV) 
luminosity and halo mass can substantially weaken the RT caused clustering 
effects.  With the star formation recipe 
in the cosmological reionization simulation, whose output we use for the 
RT calculation, the SFR and halo mass appear
to be tightly correlated. In reality, star formation may be more stochastic
as a result of feedback, merging, gas accretion, and so on, and the 
SFR (or intrinsic \lya luminosity) at fixed halo mass may vary a lot 
from halo to halo. At a fixed observed \lya luminosity, in addition to the 
RT caused mixing of host halo masses, we would have further
mixing from the variation in the SFR. To see whether the 
stochastic star formation can wash out the clustering effects, we perform a
simple test.

We introduce a log-normal dispersion in the intrinsic \lya luminosity $\Lint$ 
and
halo mass relation, characterized by the standard deviation $\sigma_{\log L}$ 
of $\log\Lint$. To obtain the observed \lya luminosity, we assume that the 
observed-to-intrinsic \lya luminosity ratio does not change for each 
individual halo. Strictly speaking, this assumption is not accurate, since 
the radius of the contour of the detection threshold surface brightness 
(see Paper I) changes as one varies the intrinsic luminosity of a source. 
To be accurate, one needs to redo the expensive RT calculation. 
However, the assumption suffices for our purpose to investigate the effect
of the scatter. We construct LAE and S-LAE samples with different thresholds
in the observed \lya luminosity.

Figure~\ref{fig:xi3d_MLscatter} shows the 3D redshift-space 2PCFs for 
$n=2\times 10^{-3}\denhMpc$ LAE samples, with $\sigma_{\log L}$=0 (no scatter,
the default case in this paper), 0.5, and 1. At a fixed number density, a 
larger scatter brings a larger fraction of sources in low-mass halos into
the sample, so the sample of LAEs would on average have a lower bias factor. 
We see that the clustering amplitude, indicated by the position of the thick 
solid curve, decreases as $\sigma_{\log L}$ increases. However, the elongation 
pattern along the line of sight, which is an indication of the environment
dependent RT effect, remains clearly visible even for 
the case of $\sigma_{\log L}=1$ (i.e., one order of magnitude variation in 
$\Lint$).

In Figure~\ref{fig:bias_MLscatter}, we show the bias factor of LAEs and S-LAEs
as a function of sample number density. From the left panel, the range of
weak dependence of LAE clustering on sample number density tends to 
shift to lower number density for larger scatter $\sigma_{\log L}$. At higher
number density, the weak dependence breaks down, and the bias factor shows a 
similar slope as the halo samples. The right panel displays
the bias factor relative to the halo sample. It turns out to be always true 
that LAEs are more strongly clustered than the corresponding S-LAEs, which 
means that the RT caused selection effect remains strong even for
the case of large scatter. 

Our tests demonstrate that a large dispersion (up to $\sigma_{\log L}=1$) in 
the intrinsic \lya luminosity and halo mass relation cannot completely wash out 
the RT caused clustering effects studied in this paper.

{}

\end{document}